\documentclass[12 pt]{article}
\pdfoutput=1
\usepackage[hidelinks]{hyperref}

\usepackage{GatesHeader}
\usepackage{graphicx}

\usepackage[footnotesize,bf,textfont={it},margin=1 cm]{caption}
\usepackage{multirow}
\usepackage{multicol}
\usepackage{xcolor}
\usepackage{amsmath}
\usepackage{amssymb}
\usepackage[utf8]{inputenc}

\definecolor{AdinkraGreen}{rgb}{0.10196079, 0.61176473, 0.21960784 }
\definecolor{AdinkraViolet}{rgb}{0.42352942, 0.15294118, 0.4509804 }
\definecolor{AdinkraOrange}{rgb}{0.89803922, 0.57647061, 0.27450982}
\definecolor{AdinkraRed}{rgb}{0.78431374, 0, 0.12156863}
\def\gD{\mbox{\textcolor{AdinkraGreen}{${\rm D}_1$}}}
\def\vD{\mbox{\textcolor{AdinkraViolet}{${\rm D}_2$}}}
\def\oD{\mbox{\textcolor{AdinkraOrange}{${\rm D}_3$}}}
\def\rD{\mbox{\textcolor{AdinkraRed}{${\rm D}_4$}}}

\def\brL{{\bm {\rm L}}}
\def\brR{{\bm {\rm R}}}
\def\brV{{\bm {\rm V}}}
\def\brtV{{\bm{\tilde{\rm V}}}}
\def\bap{\boldsymbol{\alpha}}
\def\bbt{\boldsymbol{\beta}}
\def\bI{\boldsymbol{\rm I}}
\def\bi{\boldsymbol{i}}
\def\bj{\boldsymbol{j}}
\def\bk{\boldsymbol{k}}

\def\tQ{\tilde{Q}}

\def\Lone{\mbox{\textcolor{AdinkraGreen}{$\brL_1$}}}
\def\Ltwo{\mbox{\textcolor{AdinkraViolet}{$\brL_2$}}}
\def\Lthree{\mbox{\textcolor{AdinkraOrange}{$\brL_3$}}}
\def\Lfour{\mbox{\textcolor{AdinkraRed}{$\brL_4$}}}

\def\rI{{\rm I}}
\def\rJ{{\rm J}}
\def\rK{{\rm K}}
\def\rL{{\rm L}}
\def\rR{{\rm R}}
\def\rtL{\tilde{\rm L}}

\def\hi{{\hat\imath}}

\def\hk{{\hat{k}}}

\begin{document}
\numberwithin{equation}{section}
\setcounter{equation}{0}
\setcounter{page}{0}

\def\dt#1{\on{\hbox{\rm .}}{#1}}                
\def\Dot#1{\dt{#1}}

\def\gfrac#1#2{\frac {\scriptstyle{#1}}
        {\mbox{\raisebox{-.6ex}{$\scriptstyle{#2}$}}}}
\def\gg{{\hbox{\sc g}}}
\hspace*{25 pt}
\border
{\hbox to\hsize{\hspace*{40 pt}\today \hfill  
{Brown-HET-1721\hspace*{32 pt} } 
}}
\par {$~$ \hfill 
{arXiv:1712.07826 [hep-th]\hspace*{32 pt} } 
} 
\par

\def\adinkrawidth{0.48\textwidth}

\begin{center}
\vglue .10in
{\large\rm Generating all 36,864 Four-Color Adinkras via Signed Permutations \\ and Organizing into 
     $\ell$- and $\tilde{\ell}$-Equivalence Classes
}\\[.3in]

 S.\, James Gates, Jr.\footnote{\href{mailto:sylvester\_gates@brown.edu}{sylvester\_gates@brown.edu}}$^{a}$, Kevin Iga\footnote{kevin.iga@pepperdine.edu}$^b$, Lucas Kang\footnote{\href{mailto:lucas\_kang@alumni.brown.edu}{lucas\_kang@alumni.brown.edu}}$^a$,\\
  Vadim Korotkikh\footnote{va.korotki@gmail.com}$^c$,~and Kory Stiffler\footnote{\href{mailto:kory\_stiffler@brown.edu}{kory\_stiffler@brown.edu}}$^a$
\\[0.2in]
{\it 
\centering     
$^{a}$Department of Physics, Brown University,
\\[1pt]
Box 1843, 182 Hope Street,
Providence, RI 02912, USA 
\\[8 pt]
$^{b}$Natural Science Division, Pepperdine University,\\
24255 Pacific Coast Hwy., Malibu, CA 90263, USA
\\[8pt]    
$^c$Center for String and Particle Theory-Dept.\ of Physics,
University of Maryland, \\[-2pt]
4150 Campus Dr., College Park, MD 20472,  USA
}
\\[0.35in]

{\rm ABSTRACT}\\[.01in]
\end{center}
\begin{quotation}
{\small Recently, all 1,358,954,496 values of the gadget between the 36,864 adinkras with four colors, four bosons, and four fermions have been computed. In this paper, we further analyze these results in terms of $BC_3$, the signed permutation group of three elements, and $BC_4$, the signed permutation group of four elements. It is shown how all 36,864 adinkras can be generated via $BC_4$ boson $\times$ $BC_3$ color transformations of two quaternion adinkras that satisfy the quaternion algebra. An adinkra inner product has been used for some time, known as the \emph{gadget}, which is used to distinguish adinkras. We~show how 96 equivalence classes of adinkras that are based on the gadget emerge in terms of $BC_3$ and $BC_4$. We also comment on the importance of the gadget as it relates to separating out dynamics in terms of K\"ahler-like potentials. Thus, on the basis of the complete analysis of the supersymmetrical representations achieved in the preparatory first four sections, the final comprehensive achievement of this work is the construction of the universal $BC_4$ non-linear $\sigma$-model.
}
\endtitle

\setcounter{equation}{0}

\section{Introduction}
Adinkras are diagrams that encode supersymmetric (SUSY) transformation laws with complete fidelity in one spacetime dimension and in two spacetime dimensions~\cite{Iga:2015ysa}. Different colored lines in adinkra diagrams encode the action of distinct one-dimensional supercharges on the field variables of a supermultiplet. The lines connect nodes that encode fields related by the supersymmetry transformation. Adinkras are useful theoretical tools for many reasons. First, adinkras are elegant and concise classification tools that encode a plethora of mathematics similar to Dynkin diagrams and Feynman diagrams. Second, adinkras have proven useful in discovering previously-unknown supersymmetric multiplets~\cite{Doran:2007bx}. We seek to further develop adinkras as a search tool to uncover finite realizations of off-shell supersymmetric representations: most notably 4D, $\mathcal{N}=4$ super Yang--Mills theory and 10D and 11D supergravity. Such representations lie outside the no-go theorem of~\cite{Siegel:1981dx}. A~finite representation of the off-shell superconformal hypermultiplet has recently been uncovered~\cite{Faux:2016ygh}. This is promising evidence pointing toward the possibility of a finite realization of 4D, $\mathcal{N}=4$ super Yang--Mills theory. The utility of adinkras in analyzing extended supersymmetric systems was demonstrated in \cite{Gates:2014vxa} where the adinkra parameter $\chi_0$ was used to classify which 4D, $\mathcal{N}=2$ off-shell supersymmetric systems can be represented with finite numbers of auxiliary field and which cannot. Third, adinkras relate supersymmetric systems that exist in different dimensions, possibly providing a holographic path to uncover unknown representations. Adinkras can be ``shadows'' of higher dimensional supersymmetry where an adinkra can be drawn that encodes the entire transformation laws when the system is considered to depend only on one or two of the spacetime dimensions. 

Classifying supersymmetric systems in terms of which adinkras they reduce to is known as supersymmetric genomics~\cite{Gates:2009me,Gates:2011aa,Chappell:2012qf}, whereas building higher dimensional supersymmetry from lower dimensional supersymmetry (dimensional enhancement) is known as supersymmetric holography~\cite{Faux:2009wd,Faux:2009rz,Gates:2011mu,Gates:2012xb,Chappell:2012ms,Calkins:2014exa,Gates:2014npa,Calkins:2014sma,Gates:2016bzr}. As there are generally more low-dimensional than high-dimensional supersymmetric systems, a~classification scheme is necessary to sort out which lower dimensional systems are related to which higher dimensional systems. Defining equivalence classes is essential to the adinkra sorting process. 

In this paper, we report the discovery of 96 equivalence classes of four-color, four-boson, and four-fermion adinkras. These equivalence classes can be thought of as classes of inner products within an orthogonal basis for adinkras: two adinkras can be thought of as equivalent if they decompose the same way in this basis, thus having a normalized inner product of one. As shown in~\cite{Gates:2012xb,Calkins:2014sma}, a set of holoraumy (the word ``holoraumy'' was defined in~\cite{Gates:2014npa} as a combination of the Greek word \emph{holos} (complete) and the German word \emph{raum} (space)) 
matrices $\boldsymbol{\tilde{\rm V}}_{\rm IJ}$ can be constructed from the transformation laws encoded by the adinkra. These holoraumy matrices exist in a space spanned by six basis elements $\boldsymbol{\alpha}^{\hat{a}}$ and $\boldsymbol{\beta}^{\hat{a}}$. The basis elements $\boldsymbol{\alpha}^{\hat{a}}$ and $\boldsymbol{\beta}^{\hat{a}}$ form mutually-commuting su(2) algebras. An inner product was first defined for this basis in~\cite{Gates:2012xb}. This inner product has been called the \emph{gadget} in many subsequent works such as \cite{Calkins:2014sma,Gdgt1,Gdgt2}. The gadget between two adinkra representations equals one if the adinkras have identical $\boldsymbol{\tilde{\rm V}}_{IJ}$'s. Following~\cite{Gdgt2}, we define holoraumy-equivalence classes of adinkras whose inner products equal one. We organize our results in terms of $BC_3$ and $BC_4$: the signed permutation groups of three and four elements, respectively. The relationships between the different holoraumy-equivalence classes are presented in terms of $BC_3$ color transformations that map one equivalence class to another. The main results of this paper are

\begin{enumerate}
	\item the discovery of the $BC_3$ mappings to all $\boldsymbol{\tilde{\rm V}}$-equivalence classes,
	\item the generation of all 36,864 four-color, four-boson, and four-fermion adinkras in terms of $BC_4$ boson $\times$ $BC_3$ color transformations of two $\boldsymbol{\tilde{\rm V}}$-inequivalent adinkras, dubbed the quaternion adinkras,
	\item the presentation of a formula that encodes \emph{all} 36,864 four-color, four-boson, and four-fermion adinkras in terms of their $\brtV$-equivalence classes, 
	\item the explanation of the count of all possible gadget values in terms of equivalence classes, and
	\item the connections between the gadget, holoraumy, and dynamics through K\"ahler-like potentials.
\end{enumerate} 

The second and third results will elucidate \emph{why} there are 36,864 four-color, four-boson, and four-fermion adinkras, as well as encode all such adinkras in two succinct equations. The matrix representations of the quaternion adinkras we define satisfy the quaternion algebra. Though lesser discussed at present, describing supersymmetry in the language of quaternions has been investigated before~\cite{Kugo:1982bn}.

\begin{quote}
\emph{The fifth result is the major achievement of this work. For the first time, a universal non-linear $\s$-model is defined over the entirety of the 36,864 adinkras that provide the basis for all linear representations of 1D, $N$ = 4 supersymmetry.}
\end{quote}

The vast majority of our previous efforts in the study of adinkras has concentrated
on the issue of building a rigorous representation theory. However, there have
been two exceptions. In the work of \cite{DFGHIL1}, it was shown how to couple 
one-dimensional SUSY models with arbitrary extensions of the numbers of worldline supersymmetries to external magnetic fields. The existence of a ``super 
Zeeman effect'' was noted. The other exception \cite{DFGHIL2} explored the 
question of the compatibility of the traditional superfield approach of one-dimensional 
supersymmetrical theories with the approach of using adinkras in these same 
theories. Compatibility was shown, and this opens the path for explaining how 
the adinkra approach leads to a uniform and universal formalism for describing 
the ``model space'' of 1D, $N$ = 4 supersymmetric $\s$-models.


The topic of one-dimensional $N$ = 4 $\sigma$-models, which began in 1991~\cite{Ivanov:1988it,Ivanov:1990cz,Ivanov:1990jn}, developed 
into a substantial literature~\cite{T1,T2,T3,T4,T5,Ivanov:2003nk,E1,E2,E3,E4,E5,E6,E7,B1,B2,B3,B4,B5,B6,L1,L2,L3,L4,L5,L6,L7}, some even prior to the work in~\cite{DFGHIL2}.
Parts of this
work have been empowered by some of the insights (e.g., ``root superfields''
provide one example) gained from adinkras. The efficacy of these models
can be seen by the range of concepts to which they connect such as:
\newline \indent (a.) 
Hopf maps, 
\newline \indent (b.)
superconformal mechanics,
\newline \indent (c.)
supersymmetric Calogero models,
\newline \indent (d.)
supersymmetric CP(n) mechanics,
\newline \indent (e.)
superconformal mechanics and black holes,
\newline \indent (f.)
supersymmetric WDVV equations and roots, and
\newline \indent (g.)
 Hyper-K\" ahler, and Clifford K\" ahler geometries with torsion.
\newline
The reader should keep in mind that the recitation and citations of this paragraph
constitute only a very small slice of the literature. If one looks at the cited work 
of this paragraph, it can be noted there was an effort to establish a universal 
formalism, via the use of harmonic superspace techniques \cite{E5,E6,E7}, to~describe all such models. Such a universal formalism is what we mean by the
use of the term ``model space.'' We~next review the issue of the control of the
model space in the more familiar context of 4D, $\cal N$ = 1, 2D, $\cal N$ 
= 2, and 2D, $\cal N$ = (2, 0) $\s$-models. These are domains in which these 
issues are well understood and settled. We use this discussion as the basis
for the construction of a universal formalism for describing the universal
$BC_4$ Coxeter group non-linear $\sigma$-model.

This paper is organized as follows. In Section~\ref{s:AR}, we briefly review adinkras and how they can describe higher dimensional systems, as well as show how $\brtV$-equivalence classes have already been used to distinguish some of these systems. We illustrate the process of SUSY holography, demonstrating the missing steps and commenting on possible solutions. In Section~\ref{s:QA}, we introduce the quaternion adinkras, and in Section~\ref{s:96}, we report the 96 $\brtV$-equivalence classes that are generated from the quaternion adinkras and span all 36,864 four-color, four-boson, and four-fermion adinkras. We use these equivalence classes to explain the counts of the four different gadget values calculated in~\cite{Gdgt2} and present the formula that encodes all 36,864 four-color, four-boson, and four-fermion adinkras. In Sections~\ref{s:Weapon} and~\ref{s:Kahler}, we demonstrate connections between holoraumy and dynamics. Specifically, in Section~\ref{s:Weapon}, we demonstrate how the holoraumy for each of four different 2D SUSY sigma models is different from the others, thus having gadgets differ from unity. In Section~\ref{s:Kahler}, we demonstrate how the holoraumy matrices appear in a set of 1D SUSY actions generated by a K\"ahler-like potential and demonstrate how a particular choice of K\"ahler-like potential yields the common action for all 36,864 1D SUSY models investigated throughout the rest of the paper. The connections shown in Sections~\ref{s:Weapon} and~\ref{s:Kahler} will be particularly important as we continue our quest to develop SUSY holography.
 
\section{Adinkra Review}\label{s:AR}
In this section, we review how adinkras are used to encode supersymmetry 
entirely in one-dimension and partially in four-dimensions. We demonstrate the (at present incomplete) holographic map from 1D adinkras to 4D SUSY. For a more complete review, we refer the reader to~\cite{Gates:2009me,Gates:2011aa,Chappell:2012qf}, whose conventions we follow for four-dimensional gamma matrices $(\g^\m)_{ab}$. 
\subsection{Adinkras and One-Dimensional Supersymmetry}
Consider four bosonic fields $\Phi_i$ and four fermionic fields $\Psi_{\hat{j}}$ in one dimension, that of time. The~Lagrangian for these fields is: 
\begin{align}\label{e:L0}
	\mathcal{L} = \frac{1}{2} \delta^{ij}\dot{\Phi}_i \dot{\Phi}_j - i \frac{1}{2} \delta^{\hat{i}\hat{j}}\Psi_{\hat{i}} \dot{\Psi}_{\hat{j}}
\end{align}
where a dot above a field indicates a time derivative, $\dot{\Phi}_i = d\Phi_i/d\tau$ for example.
We now explain how the two adinkras shown in Figure~\ref{f:ctAdinkra} each encode a set of SUSY transformation laws that leave the Lagrangian~(\ref{e:L0}) invariant. 
\begin{figure}[!htbp]
\centering
	\includegraphics[width = \adinkrawidth]{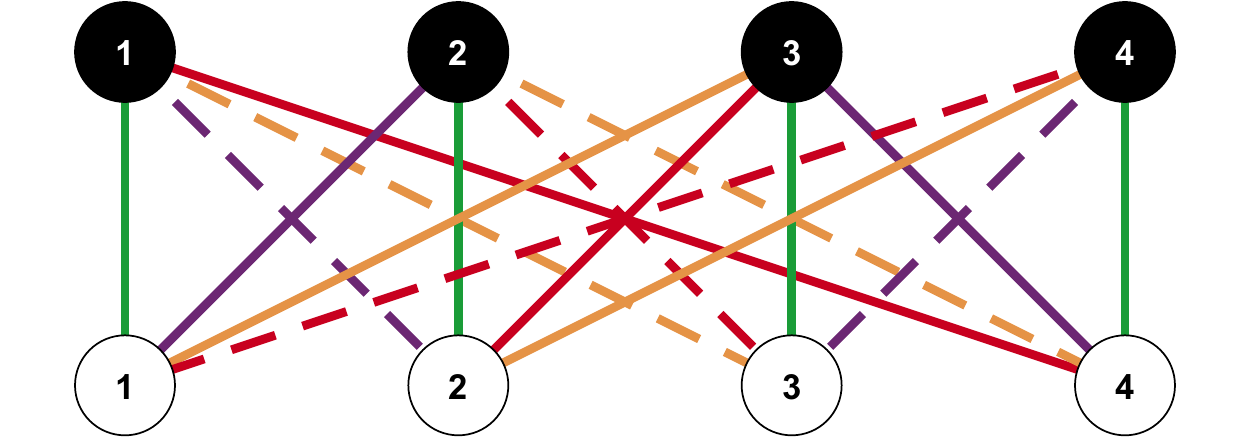} \quad \includegraphics[width = \adinkrawidth]{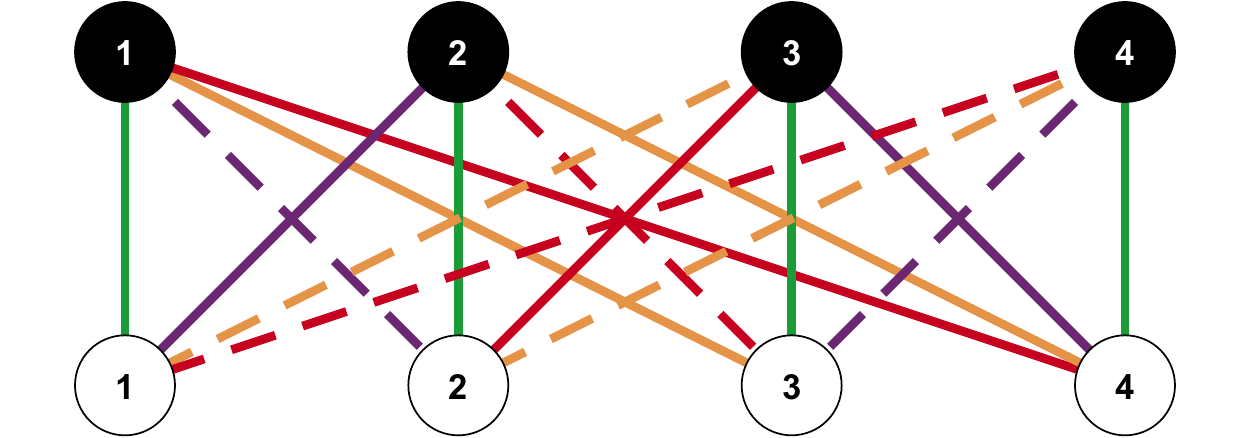}
	\caption{The \emph{cis}- ({\textbf{left}}) and \emph{trans}- ({\textbf{right}}) adinkra. The black nodes are the fermions {$\Psi_{\hat{i}}$}, and the white nodes are the bosons {$\Phi_i$} in the transformation laws.} 
	\label{f:ctAdinkra}
\end{figure}
The white nodes of the adinkras encode the bosons $\Phi_i$, and the black nodes encode the fermions multiplied by the imaginary number $ i \Psi_{\hat{j}}$. A line connecting two nodes indicates a SUSY transformation law between the corresponding fields. Each of the $N=4$ colors encodes a different SUSY transformation. A solid (dashed) line indicates a plus (minus) sign in SUSY transformations. In~transforming from a higher node to a lower node (higher mass dimension field to one-half lower mass dimension field), a time derivative appears on the field of the lower node. 

For the two adinkras in Figure~\ref{f:ctAdinkra}, we define $\chi_0 = +1$ for the leftmost adinkra (the \emph{cis}-adinkra) and $\chi_0 = -1$ for the rightmost adinkra (the \emph{trans}-adinkra), which allows us to simultaneously write the transformation laws encoded by each adinkra as follows. For boson to fermion, we have:
\begin{subequations}
\label{e:DPhi}
\begin{align}
	\label{e:DPhi1}
	\text{\gD} \Phi_1 = & i \Psi_1,~~~\text{\vD} \Phi_1 = i \Psi_2,~~~\text{\oD} \Phi_1 = \chi_0 i \Psi_3,~~~\text{\rD} \Phi_1 = -i \Psi_4 \\ 
	\label{e:DPhi2}
	\text{\gD} \Phi_2 = & i \Psi_2,~~~\text{\vD} \Phi_2 = -i \Psi_1,~~~\text{\oD} \Phi_2 = \chi_0 i \Psi_4,~~~\text{\rD} \Phi_2 = i \Psi_3 \\
	\label{e:DPhi3}
	\text{\gD} \Phi_3 = & i \Psi_3,~~~\text{\vD} \Phi_3 = -i \Psi_4,~~~\text{\oD} \Phi_3 = -\chi_0 i \Psi_1,~~~\text{\rD} \Phi_3 = -i \Psi_2 \\
	\label{e:DPhi4}
	\text{\gD} \Phi_4 = & i \Psi_4,~~~\text{\vD} \Phi_4 = i \Psi_3,~~~\text{\oD} \Phi_4 = -\chi_0 i \Psi_2,~~~\text{\rD} \Phi_4 = i \Psi_1 
\end{align} 
\end{subequations}

For fermion to boson, we have:
\begin{subequations}
\label{e:DPsi}
\begin{align}
	\label{e:DPsi1}
	\text{\gD} \Psi_1 = & \dot{\Phi}_1,~~~\text{\vD} \Psi_1 = -\dot{\Phi}_2,~~~\text{\oD} \Psi_1 = -\chi_0 \dot{\Phi}_3,~~~\text{\rD} \Psi_1 = \dot{\Phi}_4 \\
	\label{e:DPsi2}
	\text{\gD} \Psi_2 = & \dot{\Phi}_2,~~~\text{\vD} \Psi_2 = \dot{\Phi}_1,~~~\text{\oD} \Psi_2 = -\chi_0\dot{\Phi}_4,~~~\text{\rD} \Psi_2 = -\dot{\Phi}_3 \\
	\label{e:DPsi3}
	\text{\gD} \Psi_3 = & \dot{\Phi}_3,~~~\text{\vD} \Psi_3 = \dot{\Phi}_4,~~~\text{\oD} \Psi_3 = \chi_0 \dot{\Phi}_1,~~~\text{\rD} \Psi_3 = \dot{\Phi}_2 \\
	\label{e:DPsi4}
	\text{\gD} \Psi_4 = & \dot{\Phi}_4,~~~\text{\vD} \Psi_4 = -\dot{\Phi}_3,~~~\text{\oD} \Psi_4 = \chi_0 \dot{\Phi}_2,~~~\text{\rD} \Psi_4 = -\dot{\Phi}_1 
\end{align} 
\end{subequations}

Adinkras such as those in Figure~\ref{f:ctAdinkra} that have all bosons at the same height and all fermions at the same height are known as valise adinkras. The SUSY transformation laws encoded by a valise adinkra can be succinctly written as:
\begin{align}\label{e:D}
	{\rm D}_{\rm I} \Phi = i \brL_{\rm I} \Psi,~~~{\rm D}_{\rm I} \Psi = \brR_{\rm I} \dot{\Phi},
\end{align}
with $\brR_\rI$ and $\brL_\rI$ inverses and transposes of each other:
\begin{align}\label{e:adinkraic}
	\brR_\rI = \brL_{\rm I}^T = \brL_{\rm I}^{-1}.
\end{align}
 
The adinkra matrices $\brL_\rI$ and $\brR_\rI$ satisfy the $\mathcal{GR}(d,N)$ algebra, also known as the garden algebra:
\begin{align}\label{e:GRdN}
	\brL_\rI \brR_\rJ + \brL_\rJ \brR_\rI = 2 \delta_{\rm IJ} {\bm {\rm I}}_4,~~~\brR_\rI \brL_\rJ + \brR_\rJ \brL_\rI = 2 \delta_{\rm IJ} {\bm {\rm I}}_4.
\end{align} 

The $\mathcal{GR}(d,N)$ algebra is the algebra of general, real matrices encoding the supersymmetry transformation laws between $d$ bosons, $d$ fermions, and $N$ supersymmetries. The adinkras in Figure~\ref{f:ctAdinkra} each have $d=4$ and $N=4$. Hitherto, the word adinkra shall refer to $d=4$, $N=4$ valise adinkras, unless otherwise specified. With the bosonic ($i, j, k, \dots$) and fermionic ($\hat{i}, \hat{j}, \hat{k},\dots$) indices exposed, Equations~(\ref{e:D}) and (\ref{e:GRdN}) are:
\begin{align}\label{e:Dexposed}
	&{\rm D}_{\rm I} \Phi_i = i (\rL_{\rm I})_i^{~\hat{j}} \Psi_{\hat{j}},~~~{\rm D}_{\rm I} \Psi_{\hat{j}} = (\rR_{\rm I})_{\hat{j}}^{~i} \dot{\Phi}_i, \\
\label{e:GRdNLRexposed}
	&(\rL_\rI)_i^{~\hat{j}} (\rR_\rJ)_{\hat{j}}^{~k} + (\rL_\rJ)_i^{~\hat{j}} (\rR_\rI)_{\hat{j}}^{~k} = 2 \delta_{\rm IJ} \delta_i^{~k}, \\
\label{e:GRdNRLexposed}
	&(\rR_\rI)_{\hat{j}}^{~i} (\rL_\rJ)_i^{~\hat{k}} + (\rR_\rJ)_{\hat{j}}^{~i} (\rL_\rI)_i^{~\hat{k}} = 2 \delta_{\rm IJ} \delta_{\hat{j}}^{~\hat{k}}.
\end{align} 

Using Equation~(\ref{e:Dexposed}), we concisely write the matrices $\brL_\rI$ and $\brR_\rI$ for the transformation laws in Equations~(\ref{e:DPhi}) and~(\ref{e:DPsi}) for the \emph{cis}- and \emph{trans}-adinkra as:\begin{equation}
\begin{aligned}\label{e:Lmatrices}
\mbox{\textcolor{AdinkraGreen}
{
$	\brL_1 = \left(
\begin{array}{cccc}
 1 & 0 & 0 & 0 \\
 0 & 1 & 0 & 0 \\
 0 & 0 & 1 & 0 \\
 0 & 0 & 0 & 1 \\
\end{array}
\right)
$}}
&~~~,~~~
\mbox{\textcolor{AdinkraViolet}
{
$\brL_2 = \left(
\begin{array}{cccc}
 0 & 1 & 0 & 0 \\
 -1 & 0 & 0 & 0 \\
 0 & 0 & 0 & -1 \\
 0 & 0 & 1 & 0 \\
\end{array}
\right)\
$}}
~~~,\cr
\mbox{\textcolor{AdinkraOrange}
{
$
\brL_3 = \chi_0 \left(
\begin{array}{cccc}
 0 & 0 & 1 & 0 \\
 0 & 0 & 0 & 1 \\
 -1 & 0 & 0 & 0 \\
 0 & -1 & 0 & 0 \\
\end{array}
\right)
$}}
&~~~,~~~
\mbox{\textcolor{AdinkraRed}
{
$
\brL_4 = \left(
\begin{array}{cccc}
 0 & 0 & 0 & -1 \\
 0 & 0 & 1 & 0 \\
 0 & -1 & 0 & 0 \\
 1 & 0 & 0 & 0 \\
\end{array}
\right)
$}}
\end{aligned}\end{equation}\begin{equation}
\begin{aligned}\label{e:Rmatrices}
\mbox{\textcolor{AdinkraGreen}
{
$
	\brR_1 = \left(
\begin{array}{cccc}
 1 & 0 & 0 & 0 \\
 0 & 1 & 0 & 0 \\
 0 & 0 & 1 & 0 \\
 0 & 0 & 0 & 1 \\
\end{array}
\right)
$}}
&~~~,~~~
\mbox{\textcolor{AdinkraViolet}
{
$
\brR_2 = \left(
\begin{array}{cccc}
 0 & -1 & 0 & 0 \\
 1 & 0 & 0 & 0 \\
 0 & 0 & 0 & 1 \\
 0 & 0 & -1 & 0 \\
\end{array}
\right)
$}}
~~~,\cr
\mbox{\textcolor{AdinkraOrange}
{
$
\brR_3 = \chi_0 \left(
\begin{array}{cccc}
 0 & 0 & -1 & 0 \\
 0 & 0 & 0 & -1 \\
 1 & 0 & 0 & 0 \\
 0 & 1 & 0 & 0 \\
\end{array}
\right)\
$}}
&~~~,~~~
\mbox{\textcolor{AdinkraRed}
{
$
\brR_4 = \left(
\begin{array}{cccc}
 0 & 0 & 0 & 1 \\
 0 & 0 & -1 & 0 \\
 0 & 1 & 0 & 0 \\
 -1 & 0 & 0 & 0 \\
\end{array}
\right)
$}}~~~.
\end{aligned}\end{equation}

A SUSY transformation between two fields is encoded as a nonzero entry in the corresponding row and column of each of the corresponding $\brL$ and $\brR$ matrices. A solid (dashed) line in an adinkra encodes a plus (minus) one in the corresponding matrix entry. The only difference between the \emph{cis}- and \emph{trans}-valise 
adinkras in Figure~\ref{f:ctAdinkra} is the orange lines, where they are dashed in one adinkra, solid in the other; hence, matrices $\Lone$, $\Ltwo$, and $\Lfour$ are identical, but $\Lthree$ is opposite for the \emph{cis}- and \emph{trans}-valise adinkras. For an arbitrary $d=4$, $N=4$ valise adinkra, we define $\chi_0$ from the following chromocharacter equation~\cite{Gates:2009me}:
\begin{align}\label{e:chromo}
	Tr(\brL_{\rm I} \brL_\rJ^T \brL_\rK \brL_\rL^T) = 4 (\delta_{\rI \rJ}\delta_{\rK \rL} - \delta_{\rI \rL}\delta_{\rJ \rK} + \delta_{\rI \rJ}\delta_{\rK \rL} + \chi_0~\epsilon_{\rI \rJ \rK \rL})~~~
\end{align}
We have for \emph{all} $d=4$, $N=4$ valise adinkras, $\chi_0 = \pm 1$. Here, we define the $\chi_0$-equivalence class:
\begin{center}
\emph{ Two adinkras are $\chi_0$-equivalent if they share the same value of $\chi_0$.}
\end{center}

\subsection{Adinkras and Four-Dimensional Supersymmetry}

In this section, we will describe how adinkras can partially encode SUSY transformations in four-dimensions. All valise adinkras belong to either the $\chi_0 = +1$ or $\chi_0 = -1$ equivalence class. In~\cite{Gates:2009me}, it was shown that the \emph{cis}- and \emph{trans}-valise adinkras can encode the zero-brane reduced transformation laws (the transformation laws when only temporal dependence of the fields is considered) of the 4D, $\mathcal{N} = 1$ chiral multiplet ($CM$), tensor multiplet ($TM$), and vector multiplet ($VM$). In terms of $\chi_0$ equivalence classes, we have specifically that $\chi_0 = +1$ for $CM$ and $\chi_0= -1$ for $VM$ and $TM$. This~first step of simply identifying which type of fields are described by adinkras nodes is at present unknown simply given the adinkra. 

Given that starting point, however, we now demonstrate how the 4D transformation laws can be seen to arise from the adinkra representations that exist on the zero-brane, as shown in Figures~\ref{f:CMVal1}--\ref{f:VMVal1} for $CM$, $VM$, and $TM$. Along with the nodal definitions, the other missing piece of such a full holographic map is how to fill in the spatial dependence. This will be illustrated in the following. While showing the parts of the map from the zero-brane to 4D that exist (namely that time-dependent parts), we will also demonstrate that the $\chi_0$-equivalence class can be used to separate partially which adinkras correspond to which higher dimensional multiplets. Holoraumy allows us to separate more completely, which we discuss in the next section.

\begin{figure}[!htbp]
\centering
	\includegraphics[width = \adinkrawidth]{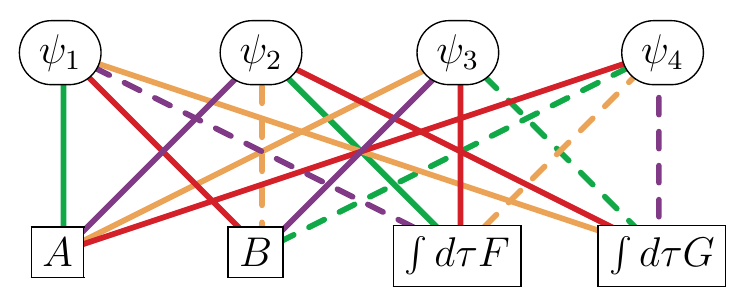}
	\vspace{-10 pt}
\caption{A valise adinkra for the chiral multiplet $CM$.}
\label{f:CMVal1}
\end{figure} 

\begin{figure}[!htbp]
\centering
	\includegraphics[width = \adinkrawidth]{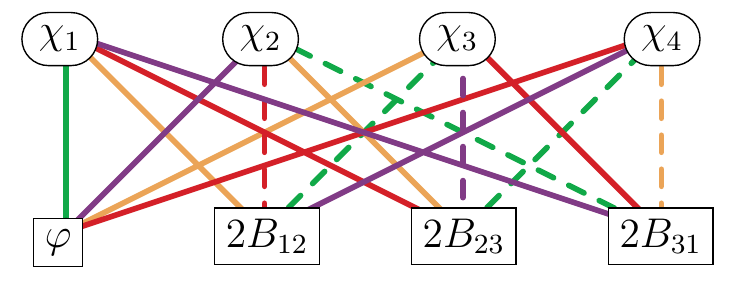}
	\vspace{-10 pt}
\caption{A valise adinkra for the tensor multiplet $TM$.}
\label{f:TMVal1}
\end{figure} 

\begin{figure}[!htbp]
\centering
	\includegraphics[width = \adinkrawidth]{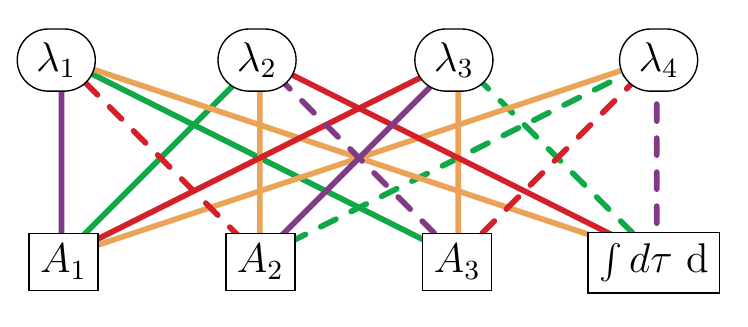}
	\vspace{-10 pt}
\caption{A valise adinkra for the vector multiplet $VM$.}
\label{f:VMVal1}
\end{figure}

The adinkras in Figures~\ref{f:CMVal1}--\ref{f:VMVal1} encode the transformation laws in Equation~(\ref{e:D}) with the following node identifications and adinkra matrices $\brL_\rI$ and $\brR_\rI$. The node identifications for each adinkra are: 
\begin{subequations}
\label{e:CMnodes}
\begin{align}
	\label{e:CMfermions}
	 CM: \hspace{ 15 pt} & i\Psi_1 = \psi_1~~~,~~~i\Psi_2 = \psi_2~~~,~~~i\Psi_3 = \psi_3~~~,~~~i\Psi_4 = \psi_4~~~\\
	\label{e:CMbosons}
	&\Phi_1 = A~~~,~~~\Phi_2 = B~~~,~~~\Phi_3 = \int d\tau F~~~,~~~\Phi_4 = \int d\tau G~~~
\end{align}
\end{subequations}
\begin{subequations}
\label{e:TMnodes}
\begin{align}
	\label{e:TMfermions}
	 TM: \hspace{ 15 pt} & i\Psi_1 = \chi_1~~~,~~~i\Psi_2 = \chi_2~~~,~~~i\Psi_3 = \chi_3~~~,~~~i\Psi_4 = \chi_4~~~\\
	\label{e:TMbosons}
	&\Phi_1 = \varphi~~~,~~~\Phi_2 = 2 B_{12}~~~,~~~\Phi_3 = 2 B_{23}~~~,~~~\Phi_4 = 2 B_{31}~~~
\end{align}
\end{subequations}
\begin{subequations}
\label{e:VMnodes}
\begin{align}
	\label{e:VMfermions}
	 TM: \hspace{ 15 pt} & i\Psi_1 = \lambda_1~~~,~~~i\Psi_2 = \lambda_2~~~,~~~i\Psi_3 = \lambda_3~~~,~~~i\Psi_4 = \lambda_4~~~\\
	\label{e:VMbosons}
	&\Phi_1 = A_1~~~,~~~\Phi_2 = A_2~~~,~~~\Phi_3 = A_3~~~,~~~\Phi_4 = \int d\tau~{\rm d}~~~
\end{align}
\end{subequations}

The $\brL_\rI$ for the $CM$, $TM$, and $VM$ adinkras are shown below. Note that the $\brR_\rI$ are the transposes of these as in Equation~(\ref{e:adinkraic}).\begin{equation}
\begin{aligned}\label{e:CM}
\mbox{\textcolor{AdinkraGreen}
{
$
	\brL^{(CM)}_1 = \left(
\begin{array}{cccc}
 1 & 0 & 0 & 0 \\
 0 & 0 & 0 & -1 \\
 0 & 1 & 0 & 0 \\
 0 & 0 & -1 & 0 \\
\end{array}
\right)
$}}
&~~~,~~~
\mbox{\textcolor{AdinkraViolet}
{
$
\brL_2^{(CM)} =\left(
\begin{array}{cccc}
 0 & 1 & 0 & 0 \\
 0 & 0 & 1 & 0 \\
 -1 & 0 & 0 & 0 \\
 0 & 0 & 0 & -1 \\
\end{array}
\right)
$}}
~~~,\cr
\mbox{\textcolor{AdinkraOrange}
{
$
\brL_3^{(CM)} =\left(
\begin{array}{cccc}
 0 & 0 & 1 & 0 \\
 0 & -1 & 0 & 0 \\
 0 & 0 & 0 & -1 \\
 1 & 0 & 0 & 0 \\
\end{array}
\right)
$}}
&~~~,~~~
\mbox{\textcolor{AdinkraRed}
{
$
\brL_4^{(CM)} =\left(
\begin{array}{cccc}
 0 & 0 & 0 & 1 \\
 1 & 0 & 0 & 0 \\
 0 & 0 & 1 & 0 \\
 0 & 1 & 0 & 0 \\
\end{array}
\right)
$}}~~~.
\end{aligned}\end{equation}
\begin{equation}
\begin{aligned}\label{e:TM}
\mbox{\textcolor{AdinkraGreen}
{
$	\brL_1^{(TM)} =\left(
\begin{array}{cccc}
 1 & 0 & 0 & 0 \\
 0 & 0 & -1 & 0 \\
 0 & 0 & 0 & -1 \\
 0 & -1 & 0 & 0 \\
\end{array}
\right)
$}}
&~~~,~~~
\mbox{\textcolor{AdinkraViolet}
{
$
\brL_2^{(TM)} = \left(
\begin{array}{cccc}
 0 & 1 & 0 & 0 \\
 0 & 0 & 0 & 1 \\
 0 & 0 & -1 & 0 \\
 1 & 0 & 0 & 0 \\
\end{array}
\right)
$}}
~~~,\cr
\mbox{\textcolor{AdinkraOrange}
{
$
\brL_3^{(TM)} = \left(
\begin{array}{cccc}
 0 & 0 & 1 & 0 \\
 1 & 0 & 0 & 0 \\
 0 & 1 & 0 & 0 \\
 0 & 0 & 0 & -1 \\
\end{array}
\right)
$}}
&~~~,~~~
\mbox{\textcolor{AdinkraRed}
{
$
\brL_4^{(TM)} =\left(
\begin{array}{cccc}
 0 & 0 & 0 & 1 \\
 0 & -1 & 0 & 0 \\
 1 & 0 & 0 & 0 \\
 0 & 0 & 1 & 0 \\
\end{array}
\right)
$}}
\end{aligned}\end{equation}\begin{equation}
\begin{aligned}\label{e:VM}
\mbox{\textcolor{AdinkraGreen}
{
$	\brL_1^{(VM)} =\left(
\begin{array}{cccc}
 0 & 1 & 0 & 0 \\
 0 & 0 & 0 & -1 \\
 1 & 0 & 0 & 0 \\
 0 & 0 & -1 & 0 \\
\end{array}
\right)
$}}
&~~~,~~~
\mbox{\textcolor{AdinkraViolet}
{
$
\brL_2^{(VM)} =\left(
\begin{array}{cccc}
 1 & 0 & 0 & 0 \\
 0 & 0 & 1 & 0 \\
 0 & -1 & 0 & 0 \\
 0 & 0 & 0 & -1 \\
\end{array}
\right)
$}}
~~~,\cr
\mbox{\textcolor{AdinkraOrange}
{
$
\brL_3^{(VM)} =\left(
\begin{array}{cccc}
 0 & 0 & 0 & 1 \\
 0 & 1 & 0 & 0 \\
 0 & 0 & 1 & 0 \\
 1 & 0 & 0 & 0 \\
\end{array}
\right)
$}}
&~~~,~~~
\mbox{\textcolor{AdinkraRed}
{
$
\brL_4^{(VM)} =\left(
\begin{array}{cccc}
 0 & 0 & 1 & 0 \\
 -1 & 0 & 0 & 0 \\
 0 & 0 & 0 & -1 \\
 0 & 1 & 0 & 0 \\
\end{array}
\right)
$}}
\end{aligned}\end{equation}

The Lagrangian for the $CM$, $VM$, and $TM$ are, in terms of their $\Phi_i$ and $\Psi_i$, all the same: Equation~(\ref{e:L0}). Substituting in the nodal definitions Equations~(\ref{e:CMnodes})--(\ref{e:VMnodes}), we arrive at their respective Lagrangians and transformation laws, where we start to see differences:
\be\label{eq:CMLagrangian0}\eqalign{
 {\mathcal L}_{\text{$CM$}}^{(0)}= & \frac{1}{2}\dot{A}^2 +\frac{1}{2}\dot{B}^2 +i\frac{1}{2}(\g^0)^{ab}\psi_{a}\dot{\psi}_{b} +\frac{1}{2} F^{2}+\frac{1}{2} G^{2}~~~
}\ee
\begin{align}\label{eq:TMLagrangian0}
 \mathcal{L}^{(0)}_{\text{TM}} = & 2\left( \dot{B}_{12}^2 + \dot{B}_{23}^2 + \dot{B}_{31}^2\right) + \frac{1}{2} \dot{\varphi}^2 + \frac{1}{2} i (\g^0)^{ab} \chi_a \dot{\chi}_b~~~
\end{align}
\be\label{eq:VMLagrangian0}\eqalign{
 {\mathcal L}_{VM}^{(0)} = &\frac{1}{2}(\dot{A}_1^2 + \dot{A}_2^2 + \dot{A}_3^2) +\frac{1}{2}i (\g^0)^{ab}\lambda_{a}\dot{\lambda}_{b}+\frac{1}{2}{\rm d}^2
}\ee 

Again, we are using the gamma matrix conventions of~\cite{Gates:2009me}. The corresponding transformation laws for the CM
 are:
\be
\eqalign{
{\rm D}_a A~&=~\psi_a ,~~~~~~~~~~~~~\,~~~~
{\rm D}_a B~=~i \, ( \gamma^5 )_a{}^b \psi_b ,~~~~~~~~~~~~~\cr
{\rm D}_a F~&=~( \gamma^0)_a{}^b \, \partial_{\tau} \psi_b ,~~~~~~~\,~~
{\rm D}_a G~=~i\, ( \gamma^5 \gamma^0 )_a{}^b \, \partial_{\tau} \psi_b 
,~~~~\,~~\cr
{\rm D}_a \psi_b~&=~i\, ( \gamma^0 )_{ab} \left( \,\partial_{
\tau} A \, \right) - ( \gamma^5 \gamma^0)_{ab} \left( \, 
\partial_{\tau} B \, \right) \cr
& {~~~~~} - i C_{ab} F + 
( \gamma^5 )_{ab} G 
.~~~~~~~~~~~~~~\,~~~
}
 \label{BV2}
\ee

The corresponding transformation laws for the $TM$ are:
\be
\eqalign{
{\rm D}{}_a \varphi~&=~\chi_a ,~~~~~~
{\rm D}{}_a B_{m \, n}~=~- \tfrac{1}{4} ([\gamma_m,\, 
\gamma_n])_a{}^b \chi_b, \cr
{\rm D}{}_a \chi_b~&=~i (\gamma^0)_{ab} \, \partial_{\tau}
\varphi - i \tfrac{1}{2} (\gamma^0 \, [\gamma^m,\, \gamma^n]
)_{ab} \, \partial_\tau B_{m \, n} 
.~~~~
} \label{BV6}
\ee
where early Latin indices $a,b,\dots = 1,2,3,4$ 
and late Latin indices $m,n,\dots = 1,2,3$. Furthermore, we define $B_{mn} = - B_{nm}$.
The corresponding transformation laws for the $VM$ are:
\be
\eqalign{
{\rm D}{}_a A_m &~=~(\gamma_m)_a{}^b \lambda_b,~~~{\rm D}_a {\rm d} 
 \, =\, i (\gamma^5 \gamma^0)_a{}^b \, \partial_\tau \lambda_b 
 ,~~~~~~~~~\cr
{\rm D}{}_a \lambda_b &~=~- i \, ( \gamma^0 
\gamma^m )_{ab} \, \left( \, \pa_{\tau} A_m \, \right)~+~(\gamma^5
)_{ab} \, {\rm d}
.~~~
} \label{BV4}
\ee

So far, the only missing piece in the holographic map has been the nodal definitions. The next step, where we insert the spatial dependence, is the other unknown piece. Said another way, a main open question in SUSY holography is:
\begin{quote} 
\emph{What needs to be added to zero-brane Lagrangians such as~(\ref{eq:CMLagrangian0}), (\ref{eq:TMLagrangian0}), (\ref{eq:VMLagrangian0}), and their corresponding transformation laws (\ref{BV2}), (\ref{BV6}), and (\ref{BV4}) to arrive at full 4D, $\mathcal{N} =1$ off-shell SUSY representations?} 
\end{quote}

Though we do not generally know a holographic procedure, from dimensional reduction, we know the answer~\cite{Gates:2009me}. Below are the 4D transformation laws and corresponding Lagrangians that \emph{reduce} to the zero-brane transformation laws and Lagrangians. We introduce Greek indices $\mu,\nu,\dots = 0,1,2,3$ where for the $CM$ we have merely to introduce more derivatives on the gauge fields and more derivatives and gamma matrices for the fermions:
\be\label{eq:CMLagrangian}\eqalign{
 {\mathcal L}_{\text{CM}} = & -\frac{1}{2}(\partial_{\mu}A)(\partial^{\mu}A) -\frac{1}{2}(\partial_{\mu} B)(\partial^{\mu}B)+i\frac{1}{2}(\gamma^{\mu})^{ab}\psi_{a}\partial_{\mu}\psi_{b} +\frac{1}{2} F^{2}+\frac{1}{2} G^{2} 
}\ee
\be
\eqalign{
{\rm D}_a A~&=~\psi_a ,~~~~~~~~~~~~~\,~~~~~~~~
{\rm D}_a B~=~i\, ( \gamma^5 )_a{}^b \psi_b~~~~~~~~\,
 \cr
{\rm D}_a \psi_b~&=~i\, ( \gamma^{\mu} )_{ab} \left( \,\partial_{
\mu} A \, \right) - ( \gamma^5 \gamma^{\mu})_{ab} \left( \, 
\partial_{\mu} B \, \right) \cr
& {~~~~~} - i C_{ab} F~+~
( \gamma^5 )_{ab} \, G 
~~~~~~~~~~~~~~\,~~~~~~~~~~~~~~~~\,\cr
{\rm D}_a F~&=~( \gamma^{\mu})_a{}^b \, \pa_{\mu} \psi_b ,~~~~~~~\,~~
{\rm D}_a G~=~i\, ( \gamma^5 \gamma^{\mu} )_a{}^b \, \pa_{\mu} \psi_b 
~~\,~~~~~~~~~~~~~
} \label{BV1}
\ee

For the $TM$ and $VM$, in addition to adding more derivatives and more gamma matrices, we must also introduce components that were gauge fixed to zero at the adinkra level: $B_{01}$, $B_{02}$, and $B_{03}$ for the TM and $A_0$ for the VM. For the TM, we have:
\be\eqalign{
 \mathcal{L}_{\text{TM}} = & - \frac{1}{3}H_{\mu\nu\alpha}H^{\mu\nu\alpha}- \frac{1}2 \partial_\mu \varphi \partial^\mu \varphi + \frac{1}{2} i (\gamma^\mu)^{bc} \chi_b \partial_\mu \chi_c 
}\ee
where:
\be
 H_{\mu\nu\alpha} \equiv \partial_\m B_{\n\a} + \partial_\n B_{\a\m} + \partial_\a B_{\m\n} 
\ee
\be
 \eqalign{
{\rm D}_a \varphi~&=~\chi_a \cr
{\rm D}_a B{}_{\mu \, \nu }~&=~-\, \fracm 14 ( [\, \gamma_{\mu}
\, , \, \gamma_{\nu} \,]){}_a{}^b \, \chi_b \cr
{\rm D}_a \chi_b~&=~i\, (\gamma^\mu){}_{a \,b}\, \partial_\mu \varphi 
~-~(\gamma^5\gamma^\mu){}_{a \,b} \, \e{}_{\mu}{}^{\r \, \s \, \t}
\partial_\r B {}_{\s \, \t}
} \label{BV5}
\ee
For the $VM$, we have:

\be\label{eq:VMLagrangian}\eqalign{
 {\mathcal L}_{\text{VM}} = &-\frac{1}{4}F_{\mu\nu}F^{\mu\nu} +\frac{1}{2}i(\gamma^{\mu})^{ab}\lambda_{a}\partial_{\mu}\lambda_{b}+\frac{1}{2}{\rm d}^2
}\ee 
\be \eqalign{
{\rm D}{}_a A_{\mu} &~=~(\gamma_{\mu})_a{}^b \lambda_b 
~~~~~~~~~~~~~~~~~~~~~~~~~~~~~\, \cr
{\rm D}{}_a \lambda_b &~=~- i \, \fracm 12 \, ( \gamma^{\mu}
\gamma^{\nu} )_{ab} \, F{}_{\mu \, \nu} \,~+~(\gamma^5
)_{ab} \, {\rm d} \cr 
{\rm D}_a {\rm d} &~=~i (\gamma^5 \gamma^{\mu})_a{}^b \, \pa_{\mu} \lambda_b 
~~~~~~~~~~~~~~\,~~~~~~~\, 
} \label{BV3}
\ee
where:
\begin{align}
	F_{\m\n} = \partial_\m A_\n - \partial_\n A_\m
\end{align}
The preceding illustrates four missing steps in the SUSY holography procedure. 
\begin{enumerate}
	\item Nodal field definitions and which adinkras encode which particle spins.
	\item How to augment with spatial derivatives.
	\item How to introduce gamma matrices.
	\item How to introduce more degrees of freedom for the gauge fields.
\end{enumerate}

The second has arguably a simple solution as applied to the previous examples: simply replace $\gamma^0\partial_\tau \to \gamma^\mu \partial_\mu$. This is related to the third point about $\gamma^\mu$ matrices, which was addressed in~\cite{Gates:2014npa,Calkins:2014sma}, where a correspondence between the zero-brane holoraumy matrices and the 4D gamma matrices was shown. The first and fourth are the least understood, but likely will have a common solution: knowing which particle spins are encoded by a given adinkra should lead to the knowledge of which gauge degrees of freedom must be added and vice versa. In the remainder of this paper, we continue to develop holoraumy as tool to be applied to the first and fourth steps in future works.

We conclude this section with some comments on the $\chi_0$-equivalence class. Performing the calculation in Equation~(\ref{e:chromo}) on the adinkra $\brL_\rI$-matrices for the $CM$, $VM$, and $TM$ results in $\chi_0 = +1$ for the $CM$ and $\chi_0 = -1$ for the $VM$ and $TM$. The $CM$ is therefore in a different $\chi_0$-equivalence class from the $VM$ and $TM$. Furthermore, the nodes of the chiral, tensor, and vector valise adinkras in Figures~\ref{f:CMVal1}--\ref{f:VMVal1} can be rearranged to arrive at different adinkras that equivalently encode the SUSY transformations. We define permutations of the nodes as \emph{flops} and sign flips of the nodes as \emph{flips}. Flipping and flopping the nodes of a representation $\mathcal{R}$ to a new representation $\mathcal{R}'$ corresponds to multiplication by matrices $\boldsymbol{\mathcal{X}}$ and $\boldsymbol{\mathcal{Y}}$ as in:\begin{equation}
\begin{aligned}\label{e:XY}
	\Phi^{(\mathcal{R}')} = \boldsymbol{\mathcal{X}} \Phi^{(\mathcal{R})},~~~&~~~\Psi^{(\mathcal{R}')} = \boldsymbol{\mathcal{Y}}^T \Psi^{(\mathcal{R})}~~~\cr
	\brL_\rI^{(\mathcal{R}')} = \boldsymbol{\mathcal{X}} \brL_\rI^{(\mathcal{R})} \boldsymbol{\mathcal{Y}},~~~&~~~\brR_\rI^{(\mathcal{R}')} = \boldsymbol{\mathcal{Y}^T} \brR_\rI^{(\mathcal{R})} \boldsymbol{\mathcal{X}}^T. 
\end{aligned} \end{equation}
The matrices $\boldsymbol{\mathcal{X}}$ and $\boldsymbol{\mathcal{Y}}$ can in principle be any non-singular matrix; however, in this paper, they will be elements of $BC_4$.

As Equation~(\ref{e:chromo}) is invariant with respect to such nodal flips and flops, they leave the adinkra in the same $\chi_0$-equivalence class. Both the tensor and vector adinkras can be node flopped and flipped to the \emph{trans}-adinkra ($\chi_0 = -1$) in Figure~\ref{f:ctAdinkra}, and the chiral adinkra can be node flopped and flipped to the \emph{cis}-adinkra ($\chi_0 = +1$) in Figure~\ref{f:ctAdinkra}~\cite{Gates:2009me}. The $\chi_0$-equivalence class can therefore be used to distinguish adinkras that are representative of the chiral multiplet from adinkras that are representative of the tensor or vector multiplet. 

\subsection{\texorpdfstring{$\brV$}{V}- and \texorpdfstring{$\brtV$}{V}-equivalence Classes}
Once $\chi_0$-equivalence classes are seen to separate the chiral multiplet from both the tensor and vector multiplets, the natural questions is: Do other equivalence classes exist that separate the vector and tensor multiplets? In~\cite{Gates:2012xb,Calkins:2014sma,Gdgt1,Gdgt2}, the holoraumy matrices $\brV_{\rm IJ}$ and $\brtV_{\rm IJ}$ were seen to do just that\footnote{these are the conventions of~\cite{Gdgt2}, which we use throughout this paper}:
\begin{align}\label{e:V}
\brV_{\rm IJ} = -i \frac{1}{2} \brL_{[\rI} \brR_{\rJ]} =& \sum_{\hat{a} = 1}^3 (\kappa^{\hat{a}}_{\rm IJ} \bap^{\hat{a}} + \tilde{\kappa}^{\hat{a}}_{\rm IJ} \bbt^{\hat{a}}) \\
\label{e:Vtilde}
	\brtV_{\rm IJ} = -i \frac{1}{2} \brR_{[\rI} \brL_{\rJ]} =& \sum_{\hat{a} = 1}^3 (\ell^{\hat{a}}_{\rm IJ} \bap^{\hat{a}} + \tilde{\ell}^{\hat{a}}_{\rm IJ} \bbt^{\hat{a}})~~~.
\end{align}
The matrices $\bap^{\hat{a}}$ and $\bbt^{\hat{a}}$ are given in Appendix~\ref{a:ab}.

Note that by construction $\ell^{\hat{a}}_{IJ} = -\ell^{\hat{a}}_{JI}$ and similarly for $\tilde{\ell}$, $\kappa$, and $\tilde{\kappa}$. For the $CM$, $VM$, and $TM$ adinkras in Equations~(\ref{e:CM})--(\ref{e:VM}), the only independent, non-vanishing $\brtV_{\rI\rJ}$ coefficients are:
\begin{subequations}
\begin{align}
 &CM: \ell^2_{12} = \ell^3_{13} = \ell^1_{14}= \ell^1_{23} = -\ell^3_{24} = \ell^2_{34} = 1\\
&TM: \tilde{\ell}^3_{12} = \tilde{\ell}^2_{13} = \tilde{\ell}^1_{14}= -\tilde{\ell}^1_{23} = \tilde{\ell}^2_{24} = -\tilde{\ell}^3_{34} = 1\\
 &VM: -\tilde{\ell}^3_{12} = \tilde{\ell}^2_{13} = -\tilde{\ell}^1_{14}= \tilde{\ell}^1_{23} = \tilde{\ell}^2_{24} = \tilde{\ell}^3_{34} = 1
\end{align}
\end{subequations}
Once $\brV$ and $\brtV$ are calculated for two adinkra representations $\mathcal{R}$ and $\mathcal{R}'$, their {gadgets} can be computed as\footnote{these are the conventions of~\cite{Gates:2012xb,Gdgt2}, which we use throughout this paper}:
\begin{align}\label{e:gadgetboson}
	\mathcal{G}_B[(\mathcal{R}),(\mathcal{R}')] = \frac{1}{48} \sum_{\rm IJ} Tr\left( \brV_{\rm IJ}^{(\mathcal{R})} \brV_{\rm IJ}^{(\mathcal{R}')} \right) = \frac{1}{12} \sum_{\hat{a}\rm IJ} [\kappa^{(\mathcal{R})\hat{a}}_{\rm IJ} \kappa^{(\mathcal{R}')\hat{a}}_{\rm IJ} + \tilde{\kappa}^{(\mathcal{R})\hat{a}}_{\rm IJ} \tilde{\kappa}^{(\mathcal{R}')\hat{a}}_{\rm IJ} ]\\
	\label{e:gadget}
	\mathcal{G}[(\mathcal{R}),(\mathcal{R}')] = \frac{1}{48} \sum_{\rm IJ} Tr \left( \brtV_{\rm IJ}^{(\mathcal{R})} \brtV_{\rm IJ}^{(\mathcal{R}')} \right) = \frac{1}{12} \sum_{\hat{a}\rm IJ} [\ell^{(\mathcal{R})\hat{a}}_{\rm IJ} \ell^{(\mathcal{R}')\hat{a}}_{\rm IJ} + \tilde{\ell}^{(\mathcal{R})\hat{a}}_{\rm IJ} \tilde{\ell}^{(\mathcal{R}')\hat{a}}_{\rm IJ} ]~~~
\end{align}
 For the chiral, tensor, and vector multiplets, we have the gadgets~\cite{Gates:2012xb,Calkins:2014sma,Gdgt1,Gdgt2}:
\begin{align}\label{e:gadgetsCMTMVM}
	\mathcal{G}[(CM),(CM)] =& \mathcal{G}[(VM),(VM)] =\mathcal{G}[(TM),(TM)] =1~~~,\cr
	\mathcal{G}[(CM),(VM)] =& \mathcal{G}[(CM),(TM)] = 0,~~~~~~\mathcal{G}[(VM),(TM)] = - \frac{1}{3}.~~~
\end{align}
Gadgets are thought of as inner products in a $\kappa$, $\tilde{\kappa}$, $\ell$, and $\tilde{\ell}$ space that is normalized to one for identical $\brV$ and $\brtV$'s. We conclude that the chiral, vector, and tensor multiplets are all in separate $\brtV$-equivalence classes.

Generalizing, we define two adinkras to be $\brV$-equivalent or $\brtV$-equivalent if they have identical $\brV$'s or $\brtV$'s, respectively. The set of all adinkras that are $\brtV$-equivalent ($\brV$-equivalent) comprise a $\brtV$-equivalence class ($\brV$-equivalence class). We separate $\brV$-equivalence classes into $\kappa$- and $\tilde{\kappa}$-equivalence classes and separate $\brtV$-equivalence classes into $\ell$ and $\tilde{\ell}$-equivalence classes. Two adinkras are $\kappa$-equivalent if and only if their $\kappa$ coefficients are identical and their $\tilde{\kappa}$ coefficients vanish. Similarly, we define two adinkras to be $\tilde{\kappa}$-equivalent if and only if their $\tilde{\kappa}$ coefficients are identical, and their $\kappa$ coefficients vanish. Two adinkras are $\ell$-equivalent if and only if their $\ell$ coefficients are identical and~their $\tilde{\ell}$ coefficients vanish. Similarly, we define two adinkras to be $\tilde{\ell}$-equivalent if and only if their $\tilde{\ell}$ coefficients are identical and their $\ell$ coefficients vanish.

 In this paper, we focus on the gadget $\mathcal{G}$, hitherto referred to as simply the gadget, and so focus on $\ell$- and $\tilde{\ell}$-equivalence classes. It has been known for some time~\cite{Chappell:2012ms,Gates:2016bzr} that any adinkra has either vanishing $\ell$ or vanishing $\tilde{\ell}$ coefficients. Therefore, the adinkras that belong to the $\ell$- and $\tilde{\ell}$-equivalence classes constitute all possible adinkras. The main result of this paper is to explain the results of~\cite{Gdgt2}, where all gadgets between all adinkras were presented, in terms of their $\ell$- and $\tilde{\ell}$-equivalence classes. 
Note that gadgets between different $\ell$-equivalence classes are always less than one: they will either be $+1/3$, 0, or $-1/3$. If a gadget between two adinkras is one, then they necessarily belong to the same $\ell$- or $\tilde{\ell}$-equivalence class.

\section{Quaternion Adinkras}\label{s:QA}
We denote the group of signed permutations of three elements as $BC_3$. Any element of $BC_3$ can be expressed as a sign flip element $H^a$ times a permutation element $S_3^\mu$. The indices take the values $a,b,c,\dots = 1,\dots 8$ and $\mu, \nu,\dots = 1, \dots, 6$.
\begin{align}
	H^{a} =& \{(), (\overline{12}), (\overline{13}), (\overline{23}), (\overline{1}), (\overline{2}), (\overline{3}), (\overline{123}) \} \\ 
	S_3^\mu =& \{ (), (12), (13), (23), (123), (132) \}
\end{align} 
A line over a number indicates a sign flip for that element. The explicit matrix forms for these elements are given in Appendix~\ref{a:FlipFlop}. Hitherto, we shall refer to permutation elements as flops and sign flips as simply flips. 

A general element of $BC_3$ is given by:
\begin{align}\label{e:BC3}
	BC_3^{a\mu} =& H^{a} S_3^\mu 
\end{align}
The Vierergruppe $\mathcal{V}^A$, also known as the Klein four-group, is a subgroup of the permutation group of four elements $S_4$:
\begin{align}
	\mathcal{V}^A &= \{ (), (12)(34), (13)(24), (14)(23) \} 
\end{align}
where $A, B, \dots = 1,2,3,4$. A general element of $BC_4$, the group of signed permutations of four elements, can be expressed as plus or minus one times an $BC_3$ element times an element of the Vierergruppe.
\begin{align}\label{e:BC4}
	BC_4^{\pm a\mu A} = & \pm H^{a} S_3^\mu \mathcal{V}^A
\end{align}

Left cosets of the Vierergruppe via $S_3$ generate all elements of $S_4$ as follows~\cite{Chappell:2012ms, Gates:2016kck}:\begin{equation}
\begin{aligned}\label{e:cosets}
	\mathcal{V} &= \{ (), (12)(34), (13)(24), (14)(23) \} \cr
	(12) \mathcal{V} &= \{ (12), (34), (1324), (1423) \} \cr
	(13) \mathcal{V} &= \{ (13),(1234) , (24), (1432) \}\cr
	(23) \mathcal{V} &= \{ (23), (1342), (1243), (14)\}\cr
	(123) \mathcal{V} &= \{(123),(134) , (243), (142) \} \cr
	(132) \mathcal{V} &= \{(132) , (234), (124) , (143) \}~~~.
\end{aligned}\end{equation}

Consider the adinkras in Figure~\ref{f:Q}, dubbed the quaternion adinkras $Q$ and $\tQ$:
\begin{figure}[!htbp]
	\centering
	\includegraphics[width = \adinkrawidth]{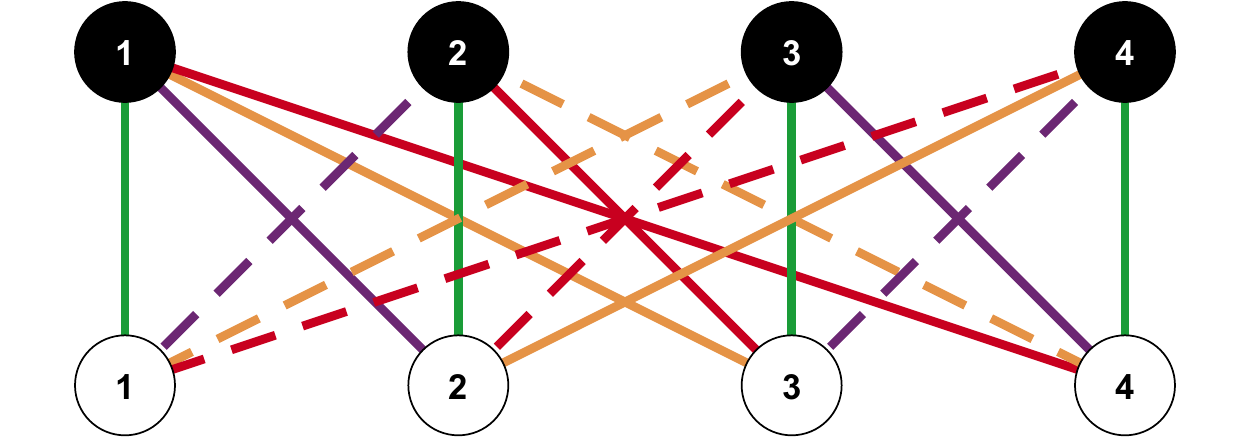}\quad \includegraphics[width = \adinkrawidth]{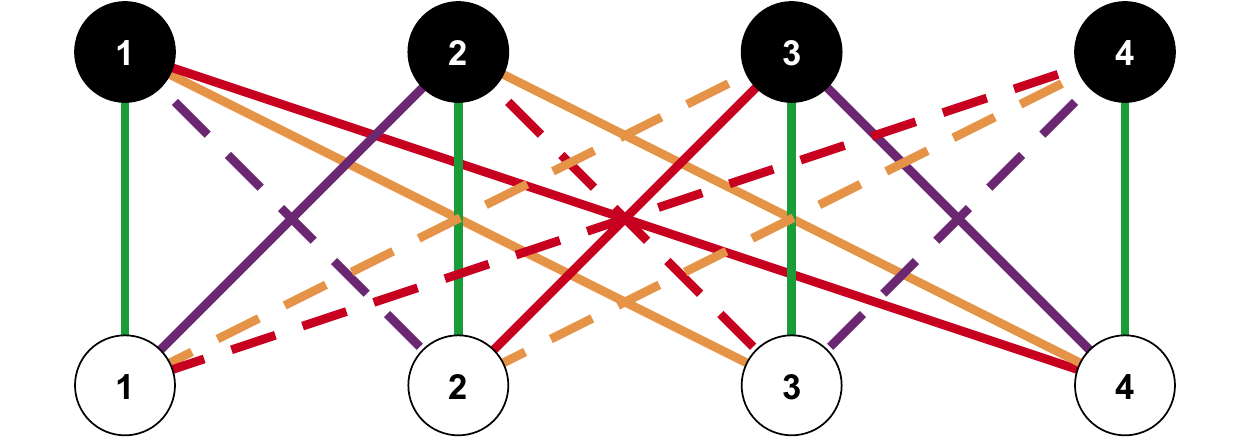}
	\caption{The quaternion adinkras $Q$ ({\textbf{left}}) and $\tQ$ ({\textbf{right}}).}
	\label{f:Q}
\end{figure}

These have the matrix representations:\begin{equation}
\begin{aligned}\label{e:Qmatrices}
\mbox{\textcolor{AdinkraGreen}
{
$	\brL_1^{(Q)} = \left(
\begin{array}{cccc}
 1 & 0 & 0 & 0 \\
 0 & 1 & 0 & 0 \\
 0 & 0 & 1 & 0 \\
 0 & 0 & 0 & 1 \\
\end{array}
\right)
$}}
&~~~,~~~
\mbox{\textcolor{AdinkraViolet}
{
$
\brL_2^{(Q)} = \left(
\begin{array}{cccc}
 0 & -1 & 0 & 0 \\
 1 & 0 & 0 & 0 \\
 0 & 0 & 0 & -1 \\
 0 & 0 & 1 & 0 \\
\end{array}
\right)
$}}
~~~,\cr
\mbox{\textcolor{AdinkraOrange}
{
$
\brL_3^{(Q)} =\left(
\begin{array}{cccc}
 0 & 0 & -1 & 0 \\
 0 & 0 & 0 & 1 \\
 1 & 0 & 0 & 0 \\
 0 & -1 & 0 & 0 \\
\end{array}
\right)
$}}
&~~~,~~~
\mbox{\textcolor{AdinkraRed}
{
$
\brL_4^{(Q)} =\left(
\begin{array}{cccc}
 0 & 0 & 0 & -1 \\
 0 & 0 & -1 & 0 \\
 0 & 1 & 0 & 0 \\
 1 & 0 & 0 & 0 \\
\end{array}
\right)
$}}~~~,
\end{aligned}\end{equation}\begin{equation}
\begin{aligned}\label{e:Qtildematrices}
	\mbox{\textcolor{AdinkraGreen}
{
$	\tilde{\brL}_1^{(\tilde{Q})} =\left(
\begin{array}{cccc}
 1 & 0 & 0 & 0 \\
 0 & 1 & 0 & 0 \\
 0 & 0 & 1 & 0 \\
 0 & 0 & 0 & 1 \\
\end{array}
\right)
$}}
&~~~,~~~
\mbox{\textcolor{AdinkraViolet}
{
$\tilde{\brL}_2^{(\tilde{Q})} = \left(
\begin{array}{cccc}
 0 & 1 & 0 & 0 \\
 -1 & 0 & 0 & 0 \\
 0 & 0 & 0 & -1 \\
 0 & 0 & 1 & 0 \\
\end{array}
\right)
$}}
~~~,\cr
\mbox{\textcolor{AdinkraOrange}
{
$
\tilde{\brL}_3^{(\tilde{Q})} =\left(
\begin{array}{cccc}
 0 & 0 & -1 & 0 \\
 0 & 0 & 0 & -1 \\
 1 & 0 & 0 & 0 \\
 0 & 1 & 0 & 0 \\
\end{array}
\right)
$}}
&~~~,~~~
\mbox{\textcolor{AdinkraRed}
{
$
\tilde{\brL}_4^{(\tilde{Q})} =
\left(
\begin{array}{cccc}
 0 & 0 & 0 & -1 \\
 0 & 0 & 1 & 0 \\
 0 & -1 & 0 & 0 \\
 1 & 0 & 0 & 0 \\
\end{array}
\right)
$}}~~~.
\end{aligned}\end{equation}
Notice that $\tilde{Q}$ is the same as the \emph{trans} adinkra in Figure~\ref{f:ctAdinkra}.

We can express any adinkra matrices as elements of $BC_4$. For the quaternion adinkras, we have:\begin{equation}
\begin{aligned}\label{e:Q}
	\brL_\rI^{(Q)} =& \{ (), (\overline{13})(12)(34), -(\overline{23})(13)(24), (\overline{12})(14)(23) \} \cr
	=& \{\bI, -i \bap^2, - i \bap^3, -i \bap^1 \} = \{ \rm I, \bi, \bj, \bk \} \end{aligned}\end{equation}
	\begin{equation}
\begin{aligned}\label{e:Qtilde}
	\tilde{\brL}_\rI^{(\tQ)} =& \{ (), (\overline{23})(12)(34), (\overline{12})(13)(24), (\overline{13})(14)(23) \} \cr
	=& \{\bI, i \bbt^3, - i \bbt^2, -i \bbt^1 \} = \{ \rm I, \tilde{\bi}, \tilde{\bj}, \tilde{\bk} \}
\end{aligned}\end{equation}
\noindent with $\rI = 1,2,3,4$. The matrices $\bap$ and $\bbt$ are given in Appendix~\ref{a:ab}, and $\bI$ is the $4 \times 4$ identity matrix. Forgetting the bosonic and fermionic nature of the rows and columns of $\brL_\rI^{(Q)}$ and $\tilde{\brL}_\rI^{(\tQ)}$, they satisfy the quaternion multiplication rules\footnote{Technically, two $\brL$-matrices can not be multiplied together. Here, it is meant that $\brL_\rI^2 = - \brL_\rI \brR_\rI$; no $\rI$ 
 sum and for $\rI = 2,~3,~\text{or}~4$.}:
\begin{align}
	\brL_2^2 =& \brL_3^2 = \brL_4^2 = \brL_2 \brL_3 \brL_4 = -\bI \\
	\tilde{\brL}_2^2 =& \tilde{\brL}_3^2 = \tilde{\brL}_4^2 = \tilde{\brL}_2 \tilde{\brL}_3 \tilde{\brL}_4 = -\bI~~~.
\end{align}
They are also mutually commuting:
\begin{align*}
	[\brL_\rI^{(Q)}, \tilde{\brL}_\rJ^{(\tQ)} ] = 0~~~\text{for}~\rI = 2,3,4
\end{align*}

The quaternion adinkras $Q$ and $\tQ$ belong to separate $\ell$ and $\tilde{\ell}$-equivalence classes:
\begin{align}\label{e:Qell}
	Q:&~~\ell_{23}^1=\ell_{41}^1 =\ell_{21}^2 = \ell_{34}^2 = \ell_{31}^3 = \ell_{42}^3 = +1~~~,~~~\tilde{\ell}^{\hat{a}}_{\rI\rJ} = 0 \\
	\label{e:Qtildeelltilde}
	\tQ:&~~\tilde{\ell}_{23}^1=\tilde{\ell}_{41}^1 =\tilde{\ell}_{31}^2 = \tilde{\ell}_{42}^2 = \tilde{\ell}_{12}^3 = \tilde{\ell}_{43}^3 = +1~~~,~~~\ell^{\hat{a}}_{\rI\rJ} = 0
\end{align}
These values of $\ell$ and $\tilde{\ell}$ can be succinctly written as $\bbt$ matrices:
\begin{align}
	Q:~~~i \ell^1_{\rI\rJ} &= \beta ^1_{\rI\rJ}~~~,~~~i \ell^2_{\rI\rJ} = \beta ^3_{\rI\rJ}~~~,~~~i \ell^3_{\rI\rJ} =\beta ^2_{\rI\rJ}~~~,~~~\tilde{\ell}^{\hat{a}}_{\rI\rJ} = 0 \\
	\tQ:~~~i \tilde{\ell}^1_{\rI\rJ} &= \beta ^1_{\rI\rJ}~~~,~~~i \tilde{\ell}^2_{\rI\rJ} = \beta ^2_{\rI\rJ}~~~,~~~i \tilde{\ell}^3_{\rI\rJ} = -\beta ^3_{\rI\rJ}~~~,~~~\ell^{\hat{a}}_{\rI\rJ} = 0.
\end{align}

The gadget between the quaternion adinkras is therefore zero: $\mathcal{G}[(Q),(\tQ)] = 0$. Both quaternion adinkras belong to the $\chi_0$-equivalence class $\chi_0= -1$.

\newpage
\section{The 96 \texorpdfstring{$\brtV$}{V}-Equivalence Classes of Adinkras}\label{s:96}

We define $BC_4$ color transformations as $BC_4$ elements acting on the color indices $\rI, \rJ, \dots$ of the adinkras, $BC_4$ boson transformations as $BC_4$ elements acting on the bosonic indices $i,j,\dots$ of the adinkras, and $BC_4$ fermion transformations as $BC_4$ elements acting on the fermionic indices $\hat{i},\hat{j},\dots$ of the adinkras.
\begin{align}\label{e:BC4Color}
	BC_4~\text{color:}~~(BC_4^{\pm a\mu A})_{\rI}^{~\rJ} = & \pm \left(H^a S_3^\mu \mathcal{V}^A\right)_{\rI}^{~\rJ} \\
	\label{e:BC4Boson}
	BC_4~\text{boson:}~~(BC_4^{\pm a\mu A})_{i}^{~j} = & \pm\left(H^a S_3^\mu \mathcal{V}^A\right)_{i}^{~j}
	\\
	\label{e:BC4Fermion}
	BC_4~\text{fermion:}~~(BC_4^{\pm a\mu A})_{\hat{i}}^{~\hat{j}} = & \pm\left( H^a S_3^\mu \mathcal{V}^A\right)_{\hat{i}}^{~\hat{j}}
\end{align}

The $BC_4$ boson (fermion) transformations act on the adinkra matrices as the $\boldsymbol{\mathcal{X}}$ ($\boldsymbol{\mathcal{Y}}$) matrices in Equation~(\ref{e:XY}).
To generate adinkras starting from the quaternion adinkras, any combination of these transformation laws can be used. As the gadget is invariant with respect to $BC_4$ boson transformations, we will use these transformation laws and exclude the fermion transformation laws so that the boson transformations will take us around the orbit of the equivalence classes. We will also have to use $BC_3$ color transformations to be able to generate all adinkras as the $BC_4$ boson does not have enough elements to do so alone.

\subsection{Examples}
We now show how the representations discussed thus far can be generated from the quaternion adinkras. Consider the following set of calculations that generate the $CM$:\begin{equation}
\begin{aligned}
(BC_4^{+362})_{i}^{~j} (-L_1^{(Q)})_{j}^{~\hat{j}} &= [(\overline{13})(234)]_{i}^{~j}[-()]_{j}^{~\hat{j}} \cr
 &= \left(
\begin{array}{cccc}
 -1 & 0 & 0 & 0 \\
 0 & 1 & 0 & 0 \\
 0 & 0 & -1 & 0 \\
 0 & 0 & 0 & 1 \\
\end{array}
\right) \left(
\begin{array}{cccc}
 1 & 0 & 0 & 0 \\
 0 & 0 & 0 & 1 \\
 0 & 1 & 0 & 0 \\
 0 & 0 & 1 & 0 \\
\end{array}
\right)\left(
\begin{array}{cccc}
 -1 & 0 & 0 & 0 \\
 0 & -1 & 0 & 0 \\
 0 & 0 & -1 & 0 \\
 0 & 0 & 0 & -1 \\
\end{array}
\right) \cr
&= \left(
\begin{array}{cccc}
 1 & 0 & 0 & 0 \\
 0 & 0 & 0 & -1 \\
 0 & 1 & 0 & 0 \\
 0 & 0 & -1 & 0 \\
\end{array}
\right) = [(\overline{24})(234)]_i^{~\hat{j}} \cr
 &=(L_1^{(CM)})_{i}^{~\hat{j}} 
\end{aligned}\end{equation}
\begin{equation}
\begin{aligned}	
(BC_4^{+362})_{i}^{~j} (L_2^{(Q)})_{j}^{~\hat{j}} &= [(\overline{13})(234)]_{i}^{~j}(L_2^{(Q)})_{j}^{~\hat{j}} \cr
 &= \left(
\begin{array}{cccc}
 -1 & 0 & 0 & 0 \\
 0 & 1 & 0 & 0 \\
 0 & 0 & -1 & 0 \\
 0 & 0 & 0 & 1 \\
\end{array}
\right) \left(
\begin{array}{cccc}
 1 & 0 & 0 & 0 \\
 0 & 0 & 0 & 1 \\
 0 & 1 & 0 & 0 \\
 0 & 0 & 1 & 0 \\
\end{array}
\right) \left(
\begin{array}{cccc}
 0 & -1 & 0 & 0 \\
 1 & 0 & 0 & 0 \\
 0 & 0 & 0 & -1 \\
 0 & 0 & 1 & 0 \\
\end{array}
\right)\cr
&= \left(
\begin{array}{cccc}
 0 & 1 & 0 & 0 \\
 0 & 0 & 1 & 0 \\
 -1 & 0 & 0 & 0 \\
 0 & 0 & 0 & -1 \\
\end{array}
\right) = [(\overline{34})(132)]_i^{~\hat{j}} \cr
&=(L_2^{(CM)})_{i}^{~\hat{j}} 
\end{aligned}\end{equation}\begin{equation}
\begin{aligned} 
(BC_4^{+362})_{i}^{~j} (L_3^{(Q)})_{j}^{~\hat{j}} &= [(\overline{13})(234)]_{i}^{~j}(L_3^{(Q)})_{j}^{~\hat{j}} \cr
 &= \left(
\begin{array}{cccc}
 -1 & 0 & 0 & 0 \\
 0 & 1 & 0 & 0 \\
 0 & 0 & -1 & 0 \\
 0 & 0 & 0 & 1 \\
\end{array}
\right) \left(
\begin{array}{cccc}
 1 & 0 & 0 & 0 \\
 0 & 0 & 0 & 1 \\
 0 & 1 & 0 & 0 \\
 0 & 0 & 1 & 0 \\
\end{array}
\right) \left(
\begin{array}{cccc}
 0 & 0 & -1 & 0 \\
 0 & 0 & 0 & 1 \\
 1 & 0 & 0 & 0 \\
 0 & -1 & 0 & 0 \\
\end{array}
\right)\cr
&= \left(
\begin{array}{cccc}
 0 & 0 & 1 & 0 \\
 0 & -1 & 0 & 0 \\
 0 & 0 & 0 & -1 \\
 1 & 0 & 0 & 0 \\
\end{array}
\right) = [(\overline{23})(143)]_i^{~\hat{j}} \cr
&= (L_3^{(CM)})_{i}^{~\hat{j}}
\end{aligned}\end{equation}\begin{equation}
\begin{aligned}
(BC_4^{+362})_{i}^{~j} (L_4^{(Q)})_{j}^{~\hat{j}}&= [(\overline{13})(234)]_{i}^{~j}(L_4^{(Q)})_{j}^{~\hat{j}} \cr
 &= \left(
\begin{array}{cccc}
 -1 & 0 & 0 & 0 \\
 0 & 1 & 0 & 0 \\
 0 & 0 & -1 & 0 \\
 0 & 0 & 0 & 1 \\
\end{array}
\right) \left(
\begin{array}{cccc}
 1 & 0 & 0 & 0 \\
 0 & 0 & 0 & 1 \\
 0 & 1 & 0 & 0 \\
 0 & 0 & 1 & 0 \\
\end{array}
\right) \left(
\begin{array}{cccc}
 0 & 0 & 0 & -1 \\
 0 & 0 & -1 & 0 \\
 0 & 1 & 0 & 0 \\
 1 & 0 & 0 & 0 \\
\end{array}
\right)\cr
&=\left(
\begin{array}{cccc}
 0 & 0 & 0 & 1 \\
 1 & 0 & 0 & 0 \\
 0 & 0 & 1 & 0 \\
 0 & 1 & 0 & 0 \\
\end{array}
\right) = (124)_i^{~\hat{j}}\cr
&=(L_4^{(CM)})_{i}^{~\hat{j}} 
\end{aligned}\end{equation}

Putting this all together, we find that $CM$ is generated from $Q$ via the 
$(BC_4^{+362})_i^{~j} = [(\overline{13})(234)]_i^{~j}$ boson flip and flop and the $(BC_3^{51})_\rI^{~\rJ} = (\overline{1})_\rI^{~\rJ}$ color flip:
\begin{align}\label{e:CMgen}
	(\rL_\rI^{(CM)})_i^{~\hat{j}} =& (BC_4^{+362})_i^{~j}(BC_3^{51})_\rI^{~\rJ}(\rL_\rJ^{(Q)})_j^{~\hat{j}}~~~.
\end{align}

A more succinct derivation of 
{Equation~(\ref{e:CMgen})}
is shown below in permutation notation, where colored text/numbers encode the corresponding free color index. For instance, a row of {\color{AdinkraGreen}green} in a matrix indicates that that row multiplying a column vector yields the {\color{AdinkraGreen} $\rI = 1$} element, {\color{AdinkraViolet}violet $\rI = 2$}, {\color{AdinkraOrange}orange $\rI = 3$}, and {\color{AdinkraRed}red $\rI = 4$}. 

\begin{equation}
\begin{aligned}
 (BC_4^{+362})_i^{~k}(BC_3^{51})_\rI^{~\rJ} (\rL_\rJ^{(Q)})_{k}^{~\hat{j}} &= [(\overline{13})(234)]_i^{~k}(\overline{1})_\rI^{~\rJ} (L_\rJ^{(Q)})_k^{~\hat{j}} \cr
 &= [(\overline{13})(234)]_i^{~k} 
 \left( \begin{array}{cccc} 
 \color{AdinkraGreen}-1 & \color{AdinkraGreen}0 &\color{AdinkraGreen} 0 & \color{AdinkraGreen}0 \\
 \color{AdinkraViolet}0 & \color{AdinkraViolet}1 & \color{AdinkraViolet} 0 & \color{AdinkraViolet} 0 \\
 \color{AdinkraOrange}0 & \color{AdinkraOrange} 0 &\color{AdinkraOrange} 1 & \color{AdinkraOrange}0 \\
 \color{AdinkraRed}0 &\color{AdinkraRed} 0 &\color{AdinkraRed} 0 &\color{AdinkraRed} 1
 \end{array}
 \right)
\left(
\begin{array}{c}
 	 ()_k^{~\hat{j}} \\
 	{[}(\overline{13})(12)(34){]}_k^{~\hat{j}} \\
	 -{[}(\overline{23})(13)(24){]}_k^{~\hat{j}} \\
 	{[}(\overline{12})(14)(23){]}_k^{~\hat{j}}
\end{array}
\right) \cr
 &= [(\overline{13})(234)]_i^{~k} 
\left(
\begin{array}{c}
 	\color{AdinkraGreen}-()_k^{~\hat{j}} \\
 	\color{AdinkraViolet}{[}(\overline{13})(12)(34){]}_k^{~\hat{j}} \\
	 \color{AdinkraOrange}-{[}(\overline{23})(13)(24){]}_k^{~\hat{j}} \\
 	\color{AdinkraRed}{[}(\overline{12})(14)(23){]}_k^{~\hat{j}}
\end{array}
\right) \cr
 &= 
\left(
\begin{array}{c}
 	\color{AdinkraGreen}-{[}(\overline{13})(234){]}_i^{~\hat{j}} \\
 	\color{AdinkraViolet}{[}(\overline{13})(234)(\overline{13})(12)(34){]}_i^{~\hat{j}} \\
	 \color{AdinkraOrange}-{[}(\overline{13})(234)(\overline{23})(13)(24){]}_i^{~\hat{j}} \\
 	\color{AdinkraRed}{[}(\overline{13})(234)(\overline{12})(14)(23){]}_i^{~\hat{j}}
\end{array}
\right) \cr
 &= 
\left(
\begin{array}{c}
 	\color{AdinkraGreen}{[}(\overline{24})(234){]}_i^{~\hat{j}} \\
 	\color{AdinkraViolet}{[}(\overline{34})(132){]}_i^{~\hat{j}} \\
	 \color{AdinkraOrange}{[}(\overline{23})(143){]}_i^{~\hat{j}} \\
 	\color{AdinkraRed}(124)_i^{~\hat{j}}
\end{array}
\right) \cr
&= (\rL_\rI^{(CM)})_i^{~\hat{j}}~~~.
\end{aligned}\end{equation}

Similarly, we can calculate how other representations are generated from $Q$ or $\tilde{Q}$:
\begin{subequations}\label{e:Lgen}
\begin{align}
	(\rL_\rI^{(TM)})_i^{~\hat{j}} &= (BC_4^{+553})_i^{~j}(BC_3^{21})_\rI^{~\rJ}(\tilde{\rL}_J^{(\tilde{Q})})_j^{~\hat{j}}~~~,\\
	(\rL_\rI^{(VM)})_i^{~\hat{j}} &= (BC_4^{-342})_i^{~j}(BC_3^{41})_\rI^{~\rJ}(\tilde{\rL}_J^{(\tilde{Q})})_j^{~\hat{j}}~~~,\\
	(\rL_\rI^{(cis)})_i^{~\hat{j}} &= (BC_4^{+111})_i^{~j}(BC_3^{71})_\rI^{~\rJ}(\tilde{\rL}_J^{(\tilde{Q})})_j^{~\hat{j}}~~~,\\
	(\rL_\rI^{(trans)})_i^{~\hat{j}} &= (BC_4^{+111})_i^{~j}(BC_3^{11})_\rI^{~\rJ}(\tilde{\rL}_J^{(\tilde{Q})})_j^{~\hat{j}}~~~.
\end{align}
\end{subequations}

Notice that the representations in Equations~(\ref{e:CMgen}) and (\ref{e:Lgen}) all utilize color flips, but not flops. For an example of color flips and flops and boson flips and flops, we define the representation $(\tilde{\rL}_\rI^{-73426})_i^{~\hat{j}} $ that is generated as:
\begin{align}\label{e:tm73426gen}
	(\tilde{\rL}_\rI^{-73426})_i^{~\hat{j}} \equiv & (BC_4^{-734})_i^{~j}(BC_3^{26})_\rI^{~\rJ}(\tilde{\rL}_\rJ^{(\tilde{Q})})_j^{~\hat{j}}~~~.
\end{align}

The details of this calculation are shown below:\begin{equation}
\begin{aligned}
 (BC_4^{-734})_i^{~k}(BC_3^{26})_\rI^{~\rJ} (\tilde{\rL}_\rJ^{(\tilde{Q})})_{k}^{~\hat{j}} &= -[(\overline{3})(1432)]_i^{~k}[(\overline{12})(132)]_\rI^{~\rJ} (\tilde{\rL}_\rJ^{(\tilde{Q})})_k^{~\hat{j}} \cr
 &= -[(\overline{3})(1432)]_i^{~k}
 \left( \begin{array}{cccc} 
 \color{AdinkraGreen}0 &\color{AdinkraGreen} -1 &\color{AdinkraGreen} 0 &\color{AdinkraGreen} 0 \\
\color{AdinkraViolet} 0 & \color{AdinkraViolet} 0 & \color{AdinkraViolet}-1 &\color{AdinkraViolet} 0 \\
\color{AdinkraOrange} 1 &\color{AdinkraOrange} 0 &\color{AdinkraOrange} 0 &\color{AdinkraOrange} 0 \\
 \color{AdinkraRed} 0 &\color{AdinkraRed} 0 &\color{AdinkraRed} 0 &\color{AdinkraRed} 1
 \end{array}
 \right)
\left(
\begin{array}{c}
 	()_k^{~\hat{j}} \\
 	{[}(\overline{23})(12)(34){]}_k^{~\hat{j}} \\
	{[}(\overline{12})(13)(24){]}_k^{~\hat{j}} \\
 	{[}(\overline{13})(14)(23){]}_k^{~\hat{j}}
\end{array}
\right) \cr
 &= -[(\overline{3})(1432)]_i^{~k} 
\left(
\begin{array}{c}
 	\color{AdinkraGreen}-{[}(\overline{23})(12)(34){]}_k^{~\hat{j}} \\
 	\color{AdinkraViolet}-{[}(\overline{12})(13)(24){]}_k^{~\hat{j}} \\
	\color{AdinkraOrange}()_k^{~\hat{j}} \\
 	\color{AdinkraRed}{[}(\overline{13})(14)(23){]}_k^{~\hat{j}}
\end{array}
\right) \cr
 &=
 \left(
\begin{array}{c}
 	\color{AdinkraGreen}{[}(\overline{3})(1432)(\overline{23})(12)(34){]}_k^{~\hat{j}} \\
 	\color{AdinkraViolet}{[}(\overline{3})(1432)(\overline{12})(13)(24){]}_k^{~\hat{j}} \\
	\color{AdinkraOrange}-[(\overline{3})(1432)]_k^{~\hat{j}} \\
 	\color{AdinkraRed}-{[}(\overline{3})(1432)(\overline{13})(14)(23){]}_k^{~\hat{j}}
\end{array}
\right) \cr
 &= 
\left(
\begin{array}{c}
 	\color{AdinkraGreen}{[}(\overline{123})(24){]}_i^{~\hat{j}} \\
 	\color{AdinkraViolet}{[}(\overline{134})(1234){]}_i^{~\hat{j}} \\
	\color{AdinkraOrange}{[}(\overline{124})(1432){]}_i^{~\hat{j}} \\
 	\color{AdinkraRed}{[}(\overline{1})(13){]}_i^{~\hat{j}}
\end{array}
\right) \cr
&= (\tilde{\rL}_\rI^{-73426})_i^{~\hat{j}}~~~.
\end{aligned}\end{equation}

The explicit matrices for this representation are\begin{equation}
\begin{aligned}\label{e:tm73426}
	\mbox{\textcolor{AdinkraGreen}
{
$	\tilde{\brL}_1^{-73426} = \left(
\begin{array}{cccc}
 -1 & 0 & 0 & 0 \\
 0 & 0 & 0 & -1 \\
 0 & 0 & -1 & 0 \\
 0 & 1 & 0 & 0 \\
\end{array}
\right)
$}}
&~~~,~~~
\mbox{\textcolor{AdinkraViolet}
{
$\tilde{\brL}_2^{-73426} =\left(
\begin{array}{cccc}
 0 & 0 & 0 & -1 \\
 1 & 0 & 0 & 0 \\
 0 & -1 & 0 & 0 \\
 0 & 0 & -1 & 0 \\
\end{array}
\right)
$}}
~~~,\cr
\mbox{\textcolor{AdinkraOrange}
{
$
\tilde{\brL}_3^{-73426} =\left(
\begin{array}{cccc}
 0 & -1 & 0 & 0 \\
 0 & 0 & -1 & 0 \\
 0 & 0 & 0 & 1 \\
 -1 & 0 & 0 & 0 \\
\end{array}
\right)
$}}
&~~~,~~~
\mbox{\textcolor{AdinkraRed}
{
$
\tilde{\brL}_4^{-73426} =\left(
\begin{array}{cccc}
 0 & 0 & -1 & 0 \\
 0 & 1 & 0 & 0 \\
 1 & 0 & 0 & 0 \\
 0 & 0 & 0 & 1 \\
\end{array}
\right)
$}}~~~.
\end{aligned}\end{equation}

The adinkra for $\tilde{\brL}_\rI^{-73426}$ is as in Figure~\ref{f:tm73426}.

\begin{figure}[!htbp]
\centering
	\includegraphics[width =\adinkrawidth]{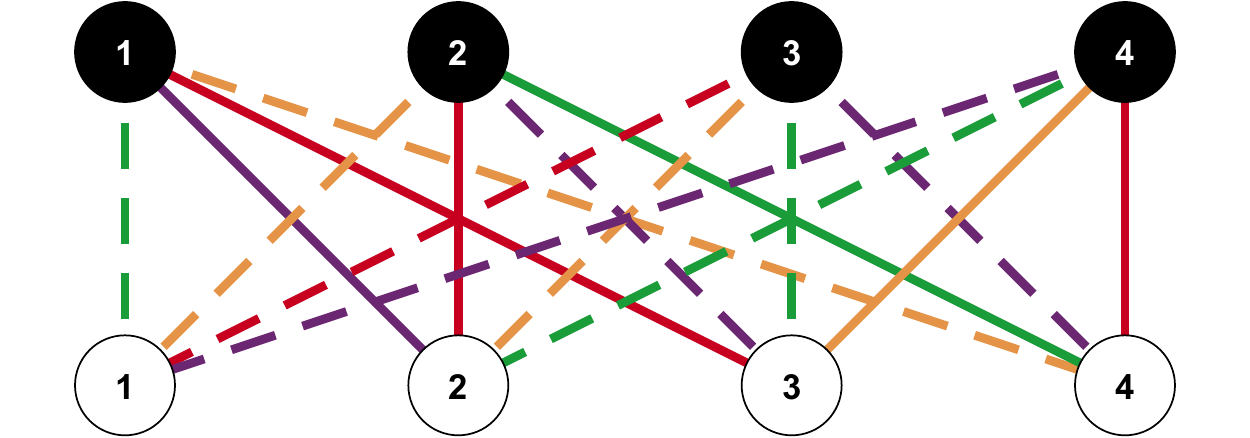}
\caption{A valise adinkra for the representation $\tilde{\brL}_\rI^{-73426}$.}
\label{f:tm73426}
\end{figure} 

Now that we have seen how $BC_4$ boson $\times$ $BC_3$ color can generate different adinkra representations from the quaternion adinkras, we will next discuss which transformations leave $\brL_\rI^{(Q)}$ and $\tilde{\brL}_\rI^{(\tilde{Q})}$ invariant. Said another way, we will discuss the isometries of $\brL_\rI^{(Q)}$ and $\tilde{\brL}_\rI^{(\tilde{Q})}$. One isometry is demonstrated in the following example.
\begin{align}
 (BC_4^{-312})_i^{~k}(BC_4^{-312})_\rI^{~\rJ} (\rL_\rJ^{(Q)})_{k}^{~\hat{j}}  &= [(\overline{24})(12)(34)]_i^{~k}[(\overline{24})(12)(34)]_\rI^{~\rJ} (L_\rJ^{(Q)})_k^{~\hat{j}}  \cr
 &= [(\overline{24})(12)(34)]_i^{~k}
 \left( \begin{array}{cccc} 
 \color{AdinkraGreen}0 & \color{AdinkraGreen}1 & \color{AdinkraGreen}0 & \color{AdinkraGreen}0 \\
 \color{AdinkraViolet}-1 &\color{AdinkraViolet} 0 &\color{AdinkraViolet} 0 &\color{AdinkraViolet} 0 \\
\color{AdinkraOrange} 0 &\color{AdinkraOrange} 0 &\color{AdinkraOrange} 0 &\color{AdinkraOrange} 1 \\
 \color{AdinkraRed}0 &\color{AdinkraRed} 0 &\color{AdinkraRed} -1 & \color{AdinkraRed}0
 \end{array}
 \right)
\left(
\begin{array}{c}
 	()_k^{~\hat{j}} \\
 	{[}(\overline{13})(12)(34){]}_k^{~\hat{j}}  \\
	-{[}(\overline{23})(13)(24){]}_k^{~\hat{j}} \\
 	{[}(\overline{12})(14)(23){]}_k^{~\hat{j}}
\end{array}
\right)  \cr
 &= [(\overline{24})(12)(34)]_i^{~k}
\left(
\begin{array}{c}
 	\color{AdinkraGreen}{[}(\overline{13})(12)(34){]}_k^{~\hat{j}}  \\
 	\color{AdinkraViolet}-()_k^{~\hat{j}} \\
 	\color{AdinkraOrange}{[}(\overline{12})(14)(23){]}_k^{~\hat{j}}\\
	\color{AdinkraRed}{[}(\overline{23})(13)(24){]}_k^{~\hat{j}} \\
\end{array}
\right)  \cr
 &=  
\left(
\begin{array}{c}
 	\color{AdinkraGreen}{[}(\overline{24})(12)(34)(\overline{13})(12)(34){]}_i^{~\hat{j}}  \\
 	\color{AdinkraViolet}-{[}(\overline{24})(12)(34) {]}_i^{~\hat{j}} \\
 \color{AdinkraOrange}	{[}(\overline{24})(12)(34)(\overline{12})(14)(23){]}_i^{~\hat{j}}\\
	\color{AdinkraRed}{[}(\overline{24})(12)(34)(\overline{23})(13)(24){]}_i^{~\hat{j}} \\
\end{array}
\right)  \cr
 &=  
\left(
\begin{array}{c}
 	\color{AdinkraGreen}()_k^{~\hat{j}} \\
 	\color{AdinkraViolet}{[}(\overline{13})(12)(34){]}_k^{~\hat{j}}  \\
	\color{AdinkraOrange}-{[}(\overline{23})(13)(24){]}_k^{~\hat{j}} \\
 	\color{AdinkraRed}{[}(\overline{12})(14)(23){]}_k^{~\hat{j}}
\end{array}
\right)  \cr
&= (\rL_\rI^{(Q)})_i^{~\hat{j}}~~~.
\end{align}

There is a two-fold degenerate list of eight such isometry transformations for each quaternion adinkra. The isometries of $Q$ are:
\begin{subequations}\label{e:Qiso}
\begin{align}
	&(BC_4^{+312})_{i}^{~j}(BC_4^{+312})_{\rI}^{~\rJ} = (BC_4^{-312})_{i}^{~j}(BC_4^{-312})_{\rI}^{~\rJ}\cr
	&~~~~~~=\text{((}\overline{13})\text{(12)(34)})_i{}^j \text{((}\overline{13})\text{(12)(34)})_\rI{}^\rJ = \text{((}\overline{24})\text{(12)(34)})_i{}^j \text{((}\overline{24})\text{(12)(34)})_\rI{}^\rJ   \\
	&(BC_4^{+413})_{i}^{~j}(BC_4^{+413})_{\rI}^{~\rJ} = (BC_4^{-413})_{i}^{~j}(BC_4^{-413})_{\rI}^{~\rJ} =\cr
 &~~~~~~= \text{((}\overline{23})\text{(13)(24)})_i{}^j \text{((}\overline{23})\text{(13)(24)})_\rI{}^\rJ = \text{((}\overline{14})\text{(13)(24)})_i{}^j \text{((}\overline{14})\text{(13)(24)})_\rI{}^\rJ\\
	&(BC_4^{+214})_{i}^{~j}(BC_4^{+214})_{\rI}^{~\rJ} = (BC_4^{-214})_{i}^{~j}(BC_4^{-214})_{\rI}^{~\rJ} = \cr
	&~~~~~~= \text{((}\overline{12})\text{(14)(23)})_i{}^j \text{((}\overline{12})\text{(14)(23)})_\rI{}^\rJ = \text{((}\overline{34})\text{(14)(23)})_i{}^j \text{((}\overline{34})\text{(14)(23)})_\rI{}^\rJ \\
	&(BC_4^{+111})_{i}^{~j}(BC_4^{+111})_{\rI}^{~\rJ} = (BC_4^{-111})_{i}^{~j}(BC_4^{-111})_{\rI}^{~\rJ} =\cr
	&~~~~~~= ()_i{}^j ()_\rI{}^\rJ = (\overline{1234})_i{}^j (\overline{1234})_\rI{}^\rJ~~~.
\end{align}
\end{subequations}

The isometries of $\tQ$ are:
\begin{subequations}\label{e:Qtiso}
\begin{align}
&(BC_4^{+412})_{i}^{~j}(BC_4^{+312})_{\rI}^{~\rJ} = (BC_4^{-412})_{i}^{~j}(BC_4^{-312})_{\rI}^{~\rJ} = \cr 
&~~~~~~=\text{((}\overline{23})\text{(12)(34)})_i{}^j \text{((}\overline{13})\text{(12)(34)})_\rI{}^\rJ=\text{((}\overline{14})\text{(12)(34)})_i{}^j \text{((}\overline{24})\text{(12)(34)})_\rI{}^\rJ\\
&(BC_4^{+213})_{i}^{~j}(BC_4^{-413})_{\rI}^{~\rJ} = (BC_4^{-213})_{i}^{~j}(BC_4^{+413})_{\rI}^{~\rJ} =\cr
&~~~~~~= \text{((}\overline{12})\text{(13)(24)})_i{}^j \text{((}\overline{14})\text{(13)(24)})_\rI{}^\rJ=\text{((}\overline{34})\text{(13)(24)})_i{}^j \text{((}\overline{23})\text{(13)(24)})_\rI{}^\rJ\\
&(BC_4^{+314})_{i}^{~j}(BC_4^{+214})_{\rI}^{~\rJ} = (BC_4^{-314})_{i}^{~j}(BC_4^{-214})_{\rI}^{~\rJ} =\cr
&~~~~~~=\text{((}\overline{13})\text{(14)(23)})_i{}^j \text{((}\overline{12})\text{(14)(23)})_\rI{}^\rJ=\text{((}\overline{24})\text{(14)(23)})_i{}^j \text{((}\overline{34})\text{(14)(23)})_\rI{}^\rJ \\
&(BC_4^{+111})_{i}^{~j}(BC_4^{+111})_{\rI}^{~\rJ} = (BC_4^{-111})_{i}^{~j}(BC_4^{-111})_{\rI}^{~\rJ} = \cr
	&~~~~~~= ()_i{}^j ()_\rI{}^\rJ=(\overline{1234})_i{}^j (\overline{1234})_\rI{}^\rJ~~~.
\end{align}
\end{subequations}

Notice the two-fold degeneracy in $BC_4~\text{boson}\times BC_4~\text{color}$ that stems from the fact that for instance $(\overline{13}) = - (\overline{24})$. Furthermore, the isometry transformations for both $Q$ and $\tQ$ are all signed elements of the Vierergruppe. This has consequences for the equivalence classes as explained in the next section.

\subsection{Flipping and Flopping via \texorpdfstring{$BC_4$}{BC4}~Boson \texorpdfstring{$\times$}{X}~\texorpdfstring{$BC_3$}{BC3}~Color: The Group Theory Reason for 36,864 Adinkras}
As $\tilde{\brV}_{\rI\rJ}$ has only color indices and fermionic indices, explicitly $(\Tilde{\rm V}_{\rm IJ})_{\hat{i}}{}^{\hat{j}}$, it is invariant with respect to $BC_4$ boson transformations. Owing to its color indices $\rI, \rJ$, $(\Tilde{\rm V}_{\rm IJ})_{\hat{i}}{}^{\hat{j}}$ is not invariant with respect to all $BC_4$ color transformations, but it will be invariant with respect to some $BC_4$ color transformations. For a given adinkra, $(\Tilde{\rm V}_{\rm IJ})_{\hat{i}}{}^{\hat{j}}$ will have an eight-fold set of $BC_4$ color isometries stemming from the isometries of the quaternion adinkras, Equations~(\ref{e:Qiso}) and (\ref{e:Qtiso}). 
Recall from the previous section that the isometry transformations for $Q$ and $\tQ$ were all elements of the signed Vierergruppe. Since the 48-element group $BC_3$ is $BC_4$ with $\pm$Vierergruppe removed, Equations~(\ref{e:BC3}) and (\ref{e:BC4}), there is a total of 96 $\brtV$-equivalence classes: 48 $BC_3$ color transformations of $Q$ and 48 $BC_3$ color transformations of $\tQ$. The 384 $BC_4$ boson transformations then fill each $\brtV$-equivalence classes with its distinct adinkras. This explains the total number of distinct adinkras as $BC_4$ boson $\times$ $BC_3$ color $\times$ 2 quaternions: 
{36,864 $= 384 \times 48 \times 2$.}

\subsubsection{Flipping and Flopping between Different \texorpdfstring{$\brtV$}{V}-Equivalence Classes via \texorpdfstring{$BC_3$}{BC3}~Color}
Color flipping and flopping $Q$ and $\tQ$ via $BC_3$ color leaves us with the following set of 96 adinkras:\vspace{-12pt}
\begin{subequations}\label{e:LLtequiv}
\begin{align}\label{e:Lequiv}
	(\rL_{\rI}^{b\nu})_i{}^{\hat{j}} &= (BC_3^{b\nu})_\rI{}^{\rJ}(\rL^{(Q)}_\rJ)_i{}^{\hat{j}} \\
	\label{e:Ltequiv}
	(\rtL_{\rI}^{b\nu})_i{}^{\hat{j}} &= (BC_3^{b\nu})_\rI{}^{\rJ}(\tilde{\rL}^{(\tQ)}_\rJ)_i{}^{\hat{j}} 
\end{align}
\end{subequations}
Each is a member of a distinct $\ell$- or $\tilde{\ell}$-equivalence class as summarized in Tables~\ref{t:ellcis}--\ref{t:elltildetrans}. Each table below describes 24 $\ell$- or $\tilde{\ell}$-equivalence classes of the same $\chi_0$-equivalence class. We define the 24 classes in each $\ell$-equivalence class table to comprise an isomer-equivalence class (either a \emph{trans}- or \emph{cis}-equivalence class), and we define the 24 classes in each $\tilde{\ell}$-equivalence class table to comprise an isometry-tilde-equivalence class (either a \emph{trans}-tilde- or \emph{cis}-tilde-equivalence class). We label these isomer- and isomer-tilde-equivalence classes as $A$, $B$, $\tilde{A}$, and $\tilde{B}$ for $\boldsymbol{l} \sim \bap$, $\boldsymbol{l} \sim \bbt$, $\tilde{\boldsymbol{l}} \sim \bap$, or $\tilde{\boldsymbol{l}} \sim \bbt$, respectively. 

\begin{table}[!h]
\centering
	\begin{tabular}{|c|c|c|c|}
	\hline
			Class & $b\nu$& $(BC_3^{b\nu})_\rI{}^{\rJ}$ & $i\ell_{IJ}^1, i\ell_{IJ}^2, i\ell_{IJ}^3 $\\
			\hline
			$A_{1}$ & 51&$(\bar{1})$&$-\alpha ^1,-\alpha ^2,-\alpha ^3$\\
			\hline
			$A_{2}$ & 61&$(\bar{2})$&$\alpha ^1,-\alpha ^2,\alpha ^3$\\
			\hline
			$A_{3}$ & 71&$(\bar{3})$&$\alpha ^1,\alpha ^2,-\alpha ^3$\\
			\hline
			$A_{4}$ & 81&$(\overline{123})$&$-\alpha ^1,\alpha ^2,\alpha ^3$\\
			\hline
			\hline
			$A_{5}$ & 55 &$(\bar{1}\text{)(123)}$&$-\alpha ^3,\alpha ^1,\alpha ^2$\\
			\hline
			$A_{6}$ & 65&$(\bar{2}\text{)(123)}$&$\alpha ^3,-\alpha ^1,\alpha ^2$\\
			\hline
			$A_{7}$ & 75&$(\bar{3}\text{)(123)}$&$-\alpha ^3,-\alpha ^1,-\alpha ^2$\\
			\hline
			$A_{8}$ & 85&$(\overline{123}\text{)(123)}$&$\alpha ^3,\alpha ^1,-\alpha ^2$\\
			\hline
			\hline
			$A_{9}$ & 56&$(\bar{1}\text{)(132)}$&$\alpha ^2,\alpha ^3,-\alpha ^1$\\
			\hline
			$A_{10}$ & 66&$(\bar{2}\text{)(132)}$&$\alpha ^2,-\alpha ^3,\alpha ^1$\\
			\hline
			$A_{11}$ & 76 &$(\bar{3}\text{)(132)}$&$-\alpha ^2,\alpha ^3,\alpha ^1$\\
			\hline
			$A_{12}$ & 86 &$(\overline{123}\text{)(132)}$&$-\alpha ^2,-\alpha ^3,-\alpha ^1$ \\
			\hline
	\end{tabular}
	\quad
	\begin{tabular}{|c|c|c|c|}
		\hline
			Class & $b\nu$& $(BC_3^{b\nu})_\rI{}^{\rJ}$ & $i\ell_{IJ}^1, i\ell_{IJ}^2, i\ell_{IJ}^3 $\\
			\hline
			$A_{13}$ &12&$\text{(12)}$&$-\alpha ^3,-\alpha ^2,\alpha ^1$\\
			\hline
			$A_{14}$ &22&$(\overline{12}\text{)(12)}$&$\alpha ^3,-\alpha ^2,-\alpha ^1$\\
			\hline
			$A_{15}$ &32&$(\overline{13}\text{)(12)}$&$-\alpha ^3,\alpha ^2,-\alpha ^1$\\
			\hline
			$A_{16}$ &42&$(\overline{23}\text{)(12)}$&$\alpha ^3,\alpha ^2,\alpha ^1$\\
			\hline
			\hline
			$A_{17}$ &13&$\text{(13)}$&$\alpha ^2,-\alpha ^1,-\alpha ^3$\\
			\hline
			$A_{18}$ &23&$(\overline{12}\text{)(13)}$&$\alpha ^2,\alpha ^1,\alpha ^3$\\
			\hline
			$A_{19}$ &33&$(\overline{13}\text{)(13)}$&$-\alpha ^2,\alpha ^1,-\alpha ^3$\\
			\hline
			$A_{20}$ &43&$(\overline{23}\text{)(13)}$&$-\alpha ^2,-\alpha ^1,\alpha ^3$\\
			\hline
			\hline
			$A_{21}$ &14&$\text{(23)}$&$\alpha ^1,\alpha ^3,\alpha ^2$\\
			\hline
			$A_{22}$ &24&$(\overline{12}\text{)(23)}$&$-\alpha ^1,-\alpha ^3,\alpha ^2$\\
			\hline
			$A_{23}$ &34&$(\overline{13}\text{)(23)}$&$-\alpha ^1,\alpha ^3,-\alpha ^2$\\
			\hline
			$A_{24}$ &44 &$(\overline{23}\text{)(23)}$&$\alpha ^1,-\alpha ^3,-\alpha ^2$\\
			\hline
	\end{tabular}
	\caption{The 24 $\ell$-equivalence classes that constitute the cis-equivalence class $A$ with $\chi_0 = +1$. Shown  are the values of $\ell$ that constitute $\brtV_{IJ}$ for $\brL_{\rI}^{b\nu}$. }
	\label{t:ellcis}
\end{table}

\begin{table}[!h]
\centering
	\begin{tabular}{|c|c|c|c|}
	\hline
			Class &$b\nu$& $(BC_3^{b\nu})_\rI{}^{\rJ}$ & $i\ell_{IJ}^1, i\ell_{IJ}^2, i\ell_{IJ}^3 $\\
			\hline
			$B_1$ &11& () & $\beta ^1,\beta ^3,\beta ^2$\\
			\hline
			$B_{2}$ &21& $(\overline{12})$&$-\beta ^1,\beta ^3,-\beta ^2$ \\
			\hline
			$B_{3}$ &31&$(\overline{13})$&$-\beta ^1,-\beta ^3,\beta ^2$ \\
			 \hline
			$B_{4}$ & 41&$(\overline{23})$&$\beta ^1,-\beta ^3,-\beta ^2$ \\
			\hline
			\hline
			$B_{5}$ &15&\text{(123)}&$\beta ^2,-\beta ^1,-\beta ^3 $\\
			\hline
			 $B_{6}$ &25& $(\overline{12}\text{)(123)}$&$-\beta ^2,\beta ^1,-\beta ^3$ \\
			\hline
			 $B_{7}$ &35&$(\overline{13}\text{)(123)}$&$\beta ^2,\beta ^1,\beta ^3$ \\
			 \hline
			$B_{8}$ & 45&$(\overline{23}\text{)(123)}$&$-\beta ^2,-\beta ^1,\beta ^3$ \\
			\hline
			\hline
			$B_{9}$ &16&$\text{(132)}$&$-\beta ^3,-\beta ^2,\beta ^1$\\
			\hline
			$B_{10}$ & 26 &$(\overline{12}\text{)(132)}$&$-\beta ^3,\beta ^2,-\beta ^1$ \\
			\hline
			$B_{11}$ & 36 &$(\overline{13}\text{)(132)}$&$\beta ^3,-\beta ^2,-\beta ^1$ \\
			 \hline
			$B_{12}$ & 46&$(\overline{23}\text{)(132)}$&$\beta ^3,\beta ^2,\beta ^1$ \\
			\hline
	\end{tabular}
	\quad
	\begin{tabular}{|c|c|c|c|}
		\hline
		Class &	$b\nu$& $(BC_3^{b\nu})_\rI{}^{\rJ}$ & $i\ell_{IJ}^1, i\ell_{IJ}^2, i\ell_{IJ}^3 $\\
			\hline
			$B_{13}$ &52&$(\bar{1}\text{)(12)}$&$\beta ^2,\beta ^3,-\beta ^1$\\
			\hline
		$B_{14}$ &	 62 &$(\bar{2}\text{)(12)}$&$-\beta ^2,\beta ^3,\beta ^1$ \\
			\hline
			$B_{15}$ & 72&$(\bar{3}\text{)(12)}$&$\beta ^2,-\beta ^3,\beta ^1$ \\
			 \hline
		$B_{16}$ &	 82&$(\overline{123}\text{)(12)}$&$-\beta ^2,-\beta ^3,-\beta ^1$ \\
			\hline
			\hline
			$B_{17}$ &53&$(\bar{1}\text{)(13)}$&$-\beta ^3,\beta ^1,\beta ^2$\\
			\hline
		$B_{18}$ &	 63&$(\bar{2}\text{)(13)}$&$-\beta ^3,-\beta ^1,-\beta ^2$ \\
			\hline
			 $B_{19}$ &73&$(\bar{3}\text{)(13)}$&$\beta ^3,-\beta ^1,\beta ^2$ \\
			 \hline
			$B_{20}$ & 83&$(\overline{123}\text{)(13)}$&$\beta ^3,\beta ^1,-\beta ^2$ \\
			\hline
			\hline
		$B_{21}$ &	54&$(\bar{1}\text{)(23)}$&$-\beta ^1,-\beta ^2,-\beta ^3$\\
			\hline
		$B_{22}$ &	 64&$(\bar{2}\text{)(23)}$&$\beta ^1,\beta ^2,-\beta ^3$ \\
			\hline
		$B_{23}$ &	 74 &$(\bar{3}\text{)(23)}$&$\beta ^1,-\beta ^2,\beta ^3$ \\
			 \hline
		$B_{24}$ &	 84 &$(\overline{123}\text{)(23)}$&$-\beta ^1,\beta ^2,\beta ^3$ \\
			\hline
	\end{tabular}
	\caption{The 24 $\ell$-equivalence classes that constitute the trans-equivalence class $B$ with $\chi_0 = -1$. Shown are the values of $\ell$ that constitute $\brtV_{IJ}$ for $\brL_{\rI}^{b\nu}$. }
	\label{t:elltrans}
\end{table}

\begin{table}[!h]
\centering
	\begin{tabular}{|c|c|c|c|}
	\hline
			Class & $b\nu$& $(BC_3^{b\nu})_\rI{}^{\rJ}$ & $i\tilde{\ell}_{IJ}^1, i\tilde{\ell}_{IJ}^2, i\tilde{\ell}_{IJ}^3 $\\
			\hline
			$\tilde{A}_{1}$ &51&$(\bar{1})$&$-\alpha ^1,-\alpha ^3,\alpha ^2$\\
			\hline
			$\tilde{A}_{2}$ &61&$(\bar{2})$&$\alpha ^1,\alpha ^3,\alpha ^2$\\
			\hline
			$\tilde{A}_{3}$ &71&$(\bar{3})$&$\alpha ^1,-\alpha ^3,-\alpha ^2$\\
			\hline
			$\tilde{A}_{4}$ &81&$(\overline{123})$&$-\alpha ^1,\alpha ^3,-\alpha ^2$\\
			\hline
			\hline
			$\tilde{A}_{5}$ &55&$(\bar{1}\text{)(123)}$&$-\alpha ^3,\alpha ^2,-\alpha ^1$\\
			\hline
			$\tilde{A}_{6}$ &65&$(\bar{2}\text{)(123)}$&$\alpha ^3,\alpha ^2,\alpha ^1$\\
			\hline
			$\tilde{A}_{7}$ &75&$(\bar{3}\text{)(123)}$&$-\alpha ^3,-\alpha ^2,\alpha ^1$\\
			\hline
			$\tilde{A}_{8}$ &85&$(\overline{123}\text{)(123)}$&$\alpha ^3,-\alpha ^2,-\alpha ^1$\\
			\hline
			\hline
			$\tilde{A}_{9}$ &56&$(\bar{1}\text{)(132)}$&$\alpha ^2,-\alpha ^1,-\alpha ^3$\\
			\hline
			$\tilde{A}_{10}$ &66&$(\bar{2}\text{)(132)}$&$\alpha ^2,\alpha ^1,\alpha ^3$\\
			\hline
			$\tilde{A}_{11}$ &76&$(\bar{3}\text{)(132)}$&$-\alpha ^2,\alpha ^1,-\alpha ^3$\\
			\hline
			$\tilde{A}_{12}$ &86&$(\overline{123}\text{)(132)}$&$-\alpha ^2,-\alpha ^1,\alpha ^3$\\
			\hline
	\end{tabular}
	\quad
	\begin{tabular}{|c|c|c|c|}
		\hline
			Class & $b\nu$& $(BC_3^{b\nu})_\rI{}^{\rJ}$ & $i\tilde{\ell}_{IJ}^1, i\tilde{\ell}_{IJ}^2, i\tilde{\ell}_{IJ}^3 $\\
			\hline
			$\tilde{A}_{13}$ &12&$\text{(12)}$&$-\alpha ^3,\alpha ^1,\alpha ^2$\\
			\hline
			$\tilde{A}_{14}$ &22&$(\overline{12}\text{)(12)}$&$\alpha ^3,-\alpha ^1,\alpha ^2$\\
			\hline
			$\tilde{A}_{15}$ &32&$(\overline{13}\text{)(12)}$&$-\alpha ^3,-\alpha ^1,-\alpha ^2$\\
			\hline
			$\tilde{A}_{16}$ &42&$(\overline{23}\text{)(12)}$&$\alpha ^3,\alpha ^1,-\alpha ^2$\\
			\hline
			\hline
			$\tilde{A}_{17}$ &13&$\text{(13)}$&$\alpha ^2,-\alpha ^3,\alpha ^1$\\
			\hline
			$\tilde{A}_{18}$ &23&$(\overline{12}\text{)(13)}$&$\alpha ^2,\alpha ^3,-\alpha ^1$\\
			\hline
			$\tilde{A}_{19}$ &33&$(\overline{13}\text{)(13)}$&$-\alpha ^2,-\alpha ^3,-\alpha ^1$\\
			\hline
			$\tilde{A}_{20}$ &43&$(\overline{23}\text{)(13)}$&$-\alpha ^2,\alpha ^3,\alpha ^1$\\
			\hline
			\hline
			$\tilde{A}_{21}$ &14&$\text{(23)}$&$\alpha ^1,\alpha ^2,-\alpha ^3$\\
			\hline
			$\tilde{A}_{22}$ &24&$(\overline{12}\text{)(23)}$&$-\alpha ^1,\alpha ^2,\alpha ^3$\\
			\hline
			$\tilde{A}_{23}$ &34&$(\overline{13}\text{)(23)}$&$-\alpha ^1,-\alpha ^2,-\alpha ^3$\\
			\hline
			$\tilde{A}_{24}$ &44&$(\overline{23}\text{)(23)}$&$\alpha ^1,-\alpha ^2,\alpha ^3$\\
			\hline
	\end{tabular}
	\caption{The 24 $\tilde{\ell}$-equivalence classes that constitute the cis-tilde-equivalence class $\tilde{A}$ with $\chi_0 = +1$.  Shown are the values of $\tilde{\ell}$ that constitute $\brtV_{IJ}$ for $\tilde{\brL}_{\rI}^{b\nu}$. }
	\label{t:elltildecis}
\end{table}

\begin{table}[!h]
\centering
	\begin{tabular}{|c|c|c|c|}
	\hline
			Class & $b\nu$& $(BC_3^{b\nu})_\rI{}^{\rJ}$ & $i\tilde{\ell}_{IJ}^1, i\tilde{\ell}_{IJ}^2, i\tilde{\ell}_{IJ}^3 $\\
			\hline
			$\tilde{B}_{1}$ &  11&$\text{()}$&$\beta ^1,\beta ^2,-\beta ^3$\\
			\hline
			$\tilde{B}_{2}$ &21&$(\overline{12})$&$-\beta ^1,-\beta ^2,-\beta ^3$\\
			\hline
			$\tilde{B}_{3}$ &31&$(\overline{13})$&$-\beta ^1,\beta ^2,\beta ^3$\\
			\hline
			$\tilde{B}_{4}$ &41&$(\overline{23})$&$\beta ^1,-\beta ^2,\beta ^3$\\
			\hline
			\hline
			$\tilde{B}_{5}$ &15&$\text{(123)}$&$\beta ^2,-\beta ^3,\beta ^1$\\
			\hline
			$\tilde{B}_{6}$ &25&$(\overline{12}\text{)(123)}$&$-\beta ^2,-\beta ^3,-\beta ^1$\\
			\hline
			$\tilde{B}_{7}$ &35&$(\overline{13}\text{)(123)}$&$\beta ^2,\beta ^3,-\beta ^1$\\
			\hline
			$\tilde{B}_{8}$ &45&$(\overline{23}\text{)(123)}$&$-\beta ^2,\beta ^3,\beta ^1$\\
			\hline
			\hline
			$\tilde{B}_{9}$ &16&$\text{(132)}$&$-\beta ^3,\beta ^1,\beta ^2$\\
			\hline
			$\tilde{B}_{10}$ &26&$(\overline{12}\text{)(132)}$&$-\beta ^3,-\beta ^1,-\beta ^2$\\
			\hline
			$\tilde{B}_{11}$ &36&$(\overline{13}\text{)(132)}$&$\beta ^3,-\beta ^1,\beta ^2$\\
			\hline
			$\tilde{B}_{12}$ &46&$(\overline{23}\text{)(132)}$&$\beta ^3,\beta ^1,-\beta ^2$\\
			\hline
	\end{tabular}
	\quad
	\begin{tabular}{|c|c|c|c|}
		\hline
			Class & $b\nu$& $(BC_3^{b\nu})_\rI{}^{\rJ}$ & $i\tilde{\ell}_{IJ}^1, i\tilde{\ell}_{IJ}^2, i\tilde{\ell}_{IJ}^3 $\\
			\hline
			$\tilde{B}_{13}$ &52&$(\bar{1}\text{)(12)}$&$\beta ^2,-\beta ^1,-\beta ^3$\\
			\hline
			$\tilde{B}_{14}$ &62&$(\bar{2}\text{)(12)}$&$-\beta ^2,\beta ^1,-\beta ^3$\\
			\hline
			$\tilde{B}_{15}$ &72&$(\bar{3}\text{)(12)}$&$\beta ^2,\beta ^1,\beta ^3$\\
			\hline
			$\tilde{B}_{16}$ &82&$(\overline{123}\text{)(12)}$&$-\beta ^2,-\beta ^1,\beta ^3$\\
			\hline
			\hline
			$\tilde{B}_{17}$ &53&$(\bar{1}\text{)(13)}$&$-\beta ^3,\beta ^2,-\beta ^1$\\
			\hline
			$\tilde{B}_{18}$ &63&$(\bar{2}\text{)(13)}$&$-\beta ^3,-\beta ^2,\beta ^1$\\
			\hline
			$\tilde{B}_{19}$ &73&$(\bar{3}\text{)(13)}$&$\beta ^3,\beta ^2,\beta ^1$\\
			\hline
			$\tilde{B}_{20}$ &83&$(\overline{123}\text{)(13)}$&$\beta ^3,-\beta ^2,-\beta ^1$\\
			\hline
			\hline
			$\tilde{B}_{21}$ &54&$(\bar{1}\text{)(23)}$&$-\beta ^1,-\beta ^3,\beta ^2$\\
			\hline
			$\tilde{B}_{22}$ &64&$(\bar{2}\text{)(23)}$&$\beta ^1,-\beta ^3,-\beta ^2$\\
			\hline
			$\tilde{B}_{23}$ &74&$(\bar{3}\text{)(23)}$&$\beta ^1,\beta ^3,\beta ^2$\\
			\hline
			$\tilde{B}_{24}$ &84&$(\overline{123}\text{)(23)}$&$-\beta ^1,\beta ^3,-\beta ^2$\\
			\hline
	\end{tabular}
	\caption{The 24 $\tilde{\ell}$-equivalence classes that constitute the trans-tilde-equivalence class $\tilde{B}$ with $\chi_0 = -1$. Shown are the values of $\tilde{\ell}$ that constitute $\brtV_{IJ}$ for $\tilde{\brL}_{\rI}^{b\nu}$. }
	\label{t:elltildetrans}
\end{table}

We define the \emph{color-parity} of a given equivalence class as being odd (even) if there is an odd (even) number of total flips plus flops as listed in the $(BC_3^{b\nu})_\rI{}^{\rJ}$ column of Tables~\ref{t:ellcis}--\ref{t:elltildetrans}. Compare for instance the equivalence classes $\tilde{B}_{6}$ and $\tilde{B}_{16}$ in Table~\ref{t:elltildetrans}. The equivalence class $\tilde{B}_{6}$ has odd color-parity: two flips encoded by $(\overline{12})$ and three flops encoded by $(123)$ for a total of five flips plus flops. In contrast, the~equivalence class $\tilde{B}_{16}$ has even color-parity: three flips encode by ($\overline{123}$) and one flop encoded by $(12)$ for a total of four flips plus flops. Notice that the equivalence classes in each of the Tables~\ref{t:ellcis}--\ref{t:elltildetrans} are numbered such that Classes 1--12 have odd color-parity and Classes 13--24 have even color-parity. We also point out that the $\ell^{\hat{a}}_{\rm IJ}$ and $\tilde{\ell}^{\hat{a}}_{\rm IJ}$ for each equivalence class satisfy:
\begin{align}
	\ell^{\hat{a}}_{\rm IJ} = \frac{1}{2}\chi_0 \epsilon_{\rm IJ}{}^{\rm KL}\ell^{\hat{a}}_{\rm KL}~~~,~~~\tilde{\ell}^{\hat{a}}_{\rm IJ} = \frac{1}{2}\chi_0 \epsilon_{\rm IJ}{}^{\rm KL}\tilde{\ell}^{\hat{a}}_{\rm KL}~~~. 
\end{align}
This is due to the fact that classes $A$ and $\tilde{A}$ with $\chi_0 = +1$ have $\boldsymbol{l} \sim \bap$ and $\tilde{\boldsymbol{l}} \sim \bap$, classes $B$ and $\tilde{B}$ with $\chi_0 = -1$ have $\boldsymbol{l} \sim \bbt$ and $\tilde{\boldsymbol{l}} \sim \bbt$, and the fact that the $\bap$ and $\bbt$ matrices satisfy:
\begin{align}
	\alpha^{\hat{a}}_{\rm IJ} = +\frac{1}{2} \epsilon_{\rm IJ}{}^{\rm KL}\alpha^{\hat{a}}_{\rm KL}~~~,~~~\beta^{\hat{a}}_{\rm IJ} = -\frac{1}{2}\epsilon_{\rm IJ}{}^{\rm KL}\beta^{\hat{a}}_{\rm KL}~~~.
\end{align} 
The above property of the $\bap$ and $\bbt$ matrices can be seen from inspection of their explicit forms in Appendix~\ref{a:ab}.

On the 96 adinkras $(\rL_{\rI}^{b\nu})_i{}^{\hat{j}}$ and $(\rtL_{\rI}^{b\nu})_i{}^{\hat{j}}$ defined in Equations~(\ref{e:LLtequiv}), we perform $BC_4$ boson transformations generating 384 distinct adinkras within each $\ell$- and $\tilde{\ell}$-equivalence class:
\begin{align}\label{e:36864}
	(\rL_{\rI}^{\pm a\mu A b\nu})_i{}^{\hat{j}} = &(BC_4^{\pm a\mu A})_i{}^j(\rL_{\rI}^{b\nu})_j{}^{\hat{j}}~~~,~~~(\rtL_{\rI}^{\pm a\mu A b\nu})_i{}^{\hat{j}} = (BC_4^{\pm a\mu A})_i{}^j(\rtL_{\rI}^{b\nu})_j{}^{\hat{j}}~.
\end{align}
Equation~(\ref{e:36864}) therefore encodes \emph{all} 36,864 = $384 \times 96$ adinkras. The counting is summarized below:
\begin{table}[!h]
\centering
\begin{tabular}{ccccccccc}
index & $\pm$ & $a$ & $\mu$ & $A$ & $b$ & $\nu$& $\sim$ & product \\
\hline
count & 2 & 8 & 6 & 4 & 8 & 6 & 2 & 36,864
\end{tabular}
\end{table}

The representations we have used as examples throughout this paper, including their $\ell$- or $\tilde{\ell}$-equivalence classes, are:
\begin{subequations}
\begin{align}
	\brL_\rI^{(Q)} &= \brL_\rI^{+11111} \in B_1~~~, \\
	\tilde{\brL}_\rI^{(\tilde{Q})} &= \brL_\rI^{(trans)} = \tilde{\brL}_\rI^{+11111} \in \tilde{B}_1~~~,\\
	\brL_\rI^{(cis)} &= \tilde{\brL}_\rI^{+11171} \in \tilde{A}_3~~~,\\
	\brL_\rI^{(CM)} &= \brL_\rI^{+36251} \in A_1~~~,\\
	\brL_\rI^{(TM)} &= \tilde{\brL}_\rI^{+55321} \in \tilde{B}_2~~~,\\
	\brL_\rI^{(VM)} &= \tilde{\brL}_\rI^{-34241} \in \tilde{B}_4~~~.
\end{align}
\end{subequations}

\subsubsection{The One Billion, Three Hundred Fifty Eight Million, Nine Hundred Fifty Four Thousand, Four Hundred Ninety Six Gadgets, Revisited}
As first computed in~\cite{Gdgt2}, there are 1,358,954,496 = 36,864 $\times$ 36,864 possible gadget values amongst the 36,864 adinkras. The majority of these vanish, and both the non-vanishing and vanishing gadgets can now be easily understood in terms of $BC_4$ boson $\times$ $BC_3$ color. For instance, the 384 adinkras that constitute each $\ell$- or $\tilde{\ell}$-equivalence class have gadgets with others within their equivalence classes equal to one. Gadgets between different isomer- or isomer-tilde equivalence classes all vanish. 
\begin{subequations}\label{e:Gvanish}
\begin{align}
	\mathcal{G}[(B_n),(A_m)]=&\mathcal{G}[(B_n),(\tilde{B}_m)]=\mathcal{G}[(B_n),(\tilde{A}_m)]= 0 \\
	\mathcal{G}[(A_n),(\tilde{A}_m)]=&\mathcal{G}[(A_n),(\tilde{B}_m)]=\mathcal{G}[(\tilde{B}_n),(\tilde{A}_m)] = 0
\end{align} 
\end{subequations}

The only non-vanishing gadgets are those within the \emph{same} isomer or isomer-tilde equivalence class, and each takes the same matrix form where for instance the rows go over $A_m$ and the columns go over $A_n$: 
\begin{align*}
3\mathcal{G}[(A_m),(A_n)] =&
\text{\tiny $\left(
\begin{array}{cccccccccccccccccccccccc}
 3 & - & - & - & 0 & 0 & 0 & 0 & 0 & 0 & 0 & 0 & + & + & - & - & + & - & + & - & - & + & + & -
 \\
 - & 3 & - & - & 0 & 0 & 0 & 0 & 0 & 0 & 0 & 0 & + & + & - & - & - & + & - & + & + & - & - & +
 \\
 - & - & 3 & - & 0 & 0 & 0 & 0 & 0 & 0 & 0 & 0 & - & - & + & + & + & - & + & - & + & - & - & +
 \\
 - & - & - & 3 & 0 & 0 & 0 & 0 & 0 & 0 & 0 & 0 & - & - & + & + & - & + & - & + & - & + & + & -
 \\
 0 & 0 & 0 & 0 & 3 & - & - & - & 0 & 0 & 0 & 0 & + & - & + & - & - & + & + & - & + & + & - & -
 \\
 0 & 0 & 0 & 0 & - & 3 & - & - & 0 & 0 & 0 & 0 & - & + & - & + & + & - & - & + & + & + & - & -
 \\
 0 & 0 & 0 & 0 & - & - & 3 & - & 0 & 0 & 0 & 0 & + & - & + & - & + & - & - & + & - & - & + & +
 \\
 0 & 0 & 0 & 0 & - & - & - & 3 & 0 & 0 & 0 & 0 & - & + & - & + & - & + & + & - & - & - & + & +
 \\
 0 & 0 & 0 & 0 & 0 & 0 & 0 & 0 & 3 & - & - & - & - & + & + & - & + & + & - & - & + & - & + & -
 \\
 0 & 0 & 0 & 0 & 0 & 0 & 0 & 0 & - & 3 & - & - & + & - & - & + & + & + & - & - & - & + & - & +
 \\
 0 & 0 & 0 & 0 & 0 & 0 & 0 & 0 & - & - & 3 & - & + & - & - & + & - & - & + & + & + & - & + & -
 \\
 0 & 0 & 0 & 0 & 0 & 0 & 0 & 0 & - & - & - & 3 & - & + & + & - & - & - & + & + & - & + & - & +
 \\
 + & + & - & - & + & - & + & - & - & + & + & - & 3 & - & - & - & 0 & 0 & 0 & 0 & 0 & 0 & 0 & 0
 \\
 + & + & - & - & - & + & - & + & + & - & - & + & - & 3 & - & - & 0 & 0 & 0 & 0 & 0 & 0 & 0 & 0
 \\
 - & - & + & + & + & - & + & - & + & - & - & + & - & - & 3 & - & 0 & 0 & 0 & 0 & 0 & 0 & 0 & 0
 \\
 - & - & + & + & - & + & - & + & - & + & + & - & - & - & - & 3 & 0 & 0 & 0 & 0 & 0 & 0 & 0 & 0
 \\
 + & - & + & - & - & + & + & - & + & + & - & - & 0 & 0 & 0 & 0 & 3 & - & - & - & 0 & 0 & 0 & 0
 \\
 - & + & - & + & + & - & - & + & + & + & - & - & 0 & 0 & 0 & 0 & - & 3 & - & - & 0 & 0 & 0 & 0
 \\
 + & - & + & - & + & - & - & + & - & - & + & + & 0 & 0 & 0 & 0 & - & - & 3 & - & 0 & 0 & 0 & 0
 \\
 - & + & - & + & - & + & + & - & - & - & + & + & 0 & 0 & 0 & 0 & - & - & - & 3 & 0 & 0 & 0 & 0
 \\
 - & + & + & - & + & + & - & - & + & - & + & - & 0 & 0 & 0 & 0 & 0 & 0 & 0 & 0 & 3 & - & - & -
 \\
 + & - & - & + & + & + & - & - & - & + & - & + & 0 & 0 & 0 & 0 & 0 & 0 & 0 & 0 & - & 3 & - & -
 \\
 + & - & - & + & - & - & + & + & + & - & + & - & 0 & 0 & 0 & 0 & 0 & 0 & 0 & 0 & - & - & 3 & -
 \\
 - & + & + & - & - & - & + & + & - & + & - & + & 0 & 0 & 0 & 0 & 0 & 0 & 0 & 0 & - & - & - & 3
 \\
\end{array}
\right) $}
\end{align*}
\begin{align}\label{e:Gnonvanish}
	\mathcal{G}[(A_m),(A_n)]=&\mathcal{G}[(B_m),(B_n)]=\mathcal{G}[(\tilde{A}_m),(\tilde{A}_n)] =\mathcal{G}[(\tilde{B}_m),(\tilde{B}_n)]
\end{align}
The matrix above therefore encodes the possible gadget values within each isomer- and isomer-tilde-equivalence class and shows some interesting color-parity features within each of these classes. Recall from Tables~\ref{t:ellcis}--\ref{t:elltildetrans} that Class Numbers 1--12 (13--24) have odd (even) color-parity. Keeping this in mind while inspecting the block off-diagonal elements of the gadget matrix above, we notice that even and odd color-parity classes are \emph{never orthogonal}: they always have an inner product equal to either plus or minus one-third. In fact, the plus one-third gadget \emph{only} arises between classes of different color-parity. Inspecting instead the block diagonal elements, where the classes have the same color-parity, notice that different even (odd) flops are \emph{always orthogonal} and different flips paired with the same flop always have an inner product of minus one-third.

In~\cite{Gdgt2}, the 1,358,954,496 = 36,864 $\times$ 36,864 gadgets between all possible pairings of the 36,864 adinkras were exhaustively calculated. The counting of each is shown in Table~\ref{t:counting}. Notice that these counts add up to the total number of gadgets: 1,358,954,496. Also shown in Table~\ref{t:counting} is the number of times each value appears in each of the matrices in Equations~(\ref{e:Gvanish}) and~(\ref{e:Gnonvanish}).

\begin{table}[!htbp]
\centering
\begin{tabular}{|c|c|c|c|c|}
	$\mathcal{G}[(\mathcal{R}),(\mathcal{R}')]$ & count & \# entries in each matrix~(\ref{e:Gvanish}) & \# entries in each matrix~(\ref{e:Gnonvanish}) \\
	\hline
	0 & 1,132,462,080 & 576 & 192 \\
	-1/3& 127,401,984 & 0 & 216  \\
	+1/3 & 84,934,656 & 0 & 144   \\
	1 & 14,155,776 & 0 & 24 
\end{tabular}
\caption{The total count of each gadget value amongst all 1,358,954,496 gadgets and the number of times each value occurs in each matrix described by Eqs.~(\ref{e:Gvanish}) and~(\ref{e:Gnonvanish}).}
\label{t:counting}
\end{table}
Noting that there are $12 = 2 \times 6$ matrices described by Equation~(\ref{e:Gvanish}) (twice the number due to the symmetry of the gadget), four matrices described by Equation~(\ref{e:Gnonvanish}), and that each matrix entry has a degeneracy of 147,456 $= 384 \times 384$ (owing to there being 384 adinkras within each $\ell$- and $\tilde{\ell}$-equivalence class), we see that the count of each gadget value in Equations~(\ref{e:Gvanish}) and~(\ref{e:Gnonvanish}) matches that of~\cite{Gdgt2}:
\begin{subequations}
\begin{align}
	\text{count} =(\text{\# OF entries in}&\text{~each matrix (\ref{e:Gvanish})}\times 12~\text{matrices} \cr
	+\text{\# OF entries}&\text{~in each matrix (\ref{e:Gnonvanish}) } \times 4~\text{matrices} )\times\text{entry degeneracy} \\
	\mathcal{G}[(\mathcal{R}),(\mathcal{R}')] = 0:& \hspace*{20 pt} 1,132,462,080 = (576 \times 12 + 192 \times 4) \times 147,456 \\
	\mathcal{G}[(\mathcal{R}),(\mathcal{R}')] = -1/3:& \hspace*{20 pt} 127,401,984 = (0 \times 12 + 216 \times 4) \times 147,456 \\
	\mathcal{G}[(\mathcal{R}),(\mathcal{R}')] = +1/3:& \hspace*{20 pt} 84,934,656 = (0 \times 12 + 144 \times 4) \times 147,456 \\
	\mathcal{G}[(\mathcal{R}),(\mathcal{R}')] = 1:& \hspace*{20 pt} 14,155,776 = (0 \times 12 + 24 \times 4) \times 147,456
\end{align}
\end{subequations}

\newpage
\section{Moving Toward 1D, \texorpdfstring{$N$}{N}~= 4 Minimal Valises AND
\texorpdfstring{$BC_4$}{BC4}~Sigma-Models }\label{s:Weapon}

The previous sections have been devoted to constructing a streamlined
mathematical approach to sorting among the 36,864 adinkras that possess four 
colors, four closed nodes, and four open nodes. This is a problem in representation 
theory. At this point of our discussion, we will build on the previous sections' 
foundation to engage the application of this foundation to the construction
of 1D, $N$ = 4 non-linear sigma-models over the $BC_4$ Coxeter group.

\subsection{Review of the Discovery of Twisted Reps in Sigma-Models }\label{s:KahlerT}

For 4D, $\cal N$ = 1 $\s$-models \cite{KZ}, there are two ingredients, chiral superfields 
$\Phi{}^{\sc I}$ and a K\" ahler potential $K(\Phi{}^{\sc I}, {\Bar \Phi}{}^{\sc I}) $, which 
come together to define a dynamical system via the supersymmetrical action formula:
\be
S{}_{\s}^{(4D)}~=~\int d^4 x \, d^2 {\theta} \, d^2 {\Bar \theta}~K(\Phi{}^{\sc I}, 
{\Bar \Phi}{}^{\sc I})~~~.
\ee
This defines the most general possible 4D, $\cal N$ = 1 $\s$-model. For each
choice of $K(\Phi{}^{\sc I}, {\Bar \Phi}{}^{\sc I}) $ and choice of the range of the
superscript $\sc I$ on $\Phi{}^{\sc I}$, there is a model that is well defined. It is 
also the case that {{no}} other 4D, $\cal N$ = 1 $\s$-models exist. Therefore, there is a type of completeness description implicit in the fact that there is only one 4D, $N$ = 1 minimal superfield representation, which implies that a specification of $K$ completely describes the space of 4D, $N$ = 1 $\s$-models. This situation is what we refer to as ``control of the model space.''

It is simple to reduce the action above to one where only two of the four
spacetime manifold coordinates are retained. One is led to write:
\be
S{}_{\s}^{(2D)}~=~\int d^2 x \, d^2 {\theta} \, d^2 {\Bar \theta}~K(\Phi{}^{\sc I}, 
{\Bar \Phi}{}^{\sc I})~~~,
\ee
but control of the model space is lost. There is nothing wrong with the
action above. However, what changes is there exists another scalar supermultiplet
in 2D, $\cal N$ = 2 superspace that does not exist in 4D, $\cal N$ = 1
superspace. 

As was first shown in the works of \cite{Twstd1,Twstd2}, a distinct 2D, $\cal N$ = 2 
supersymmetric scalar multiplet, the so-called ``twisted chiral supermultiplet'' 
(denoted by $\chi{}^{\Hat {\sc I}}$ and where the range of the index ${ {\sc I}}$ 
may be different from that of the ${\Hat { I}}$ index) exists in the lower dimension. 
Thus, modifying the action to the form:
\be
S{}_{\s}^{(2D)}~=~\int d^2 x \, d^2 {\theta} \, d^2 {\Bar \theta}~K(\Phi{}^{\sc I}, 
{\Bar \Phi}{}^{\sc I}; \, \chi{}^{\Hat {\sc I}}, {\Bar \chi}{}^{\Hat {\sc I}})~~~,
\ee
with the inclusion of the twisted chiral supermultiplet restores control of the model
space for minimal off-shell representations. It can be seen that a K\" ahler-like 
potential $K$ still controls the geometry. In the case of the four-dimensional
$\cal N$ = 1 $\s$-model, this geometry describes a Riemannian K\" ahler manifold.
In the case of the complete two-dimensional $\cal N$ = 2 $\s$-model, this
geometry is a non-Riemannian bi-Hermitian manifold with torsion.

The distinction between two-dimensional $\cal N$ = 2 $\s$-models constructed
{{solely}} from chiral supermultiplets or {{solely}} from twisted chiral 
supermultiplets in comparison to two-dimensional $\cal N$ = 2 $\s$-models constructed
from {{both}} chiral supermultiplets {{and}} twisted chiral supermultiplets
arise from the representation theory fact that the two types of supermultiplets
are ``usefully inequivalent'' \cite{UI}.

When two-dimensional $\cal N$ = 2 supermultiplets are reduced to one-dimensional 
$N = 4$ supermultiplets, the distinction between the chiral supermultiplet ($CM$) and 
twisted chiral supermultiplet ($TCM$) can be seen in their gadget values~\cite{Gdgt2}:
\be\label{e:twistedgadget}
\eqalign{
&{ {\cal G}} \left[ (CM) \, , \, (CM) \right]~=~{ {\cal G}} \left[ (TCM) 
\, , \, (TCM) \right]~=~1~~~, \cr
&{~~~~~~~~~~~~~~~~~~~~~~} { {\cal G}} \left[ (CM) \, , \, (TCM) \right]~=~0
~~~~.
} \ee

As first discovered in~\cite{Twstd1,Twstd2} and later related to adinkras in 
~\cite{Gates:2014npa,Calkins:2014sma,Gdgt2}, the 2D $\mathcal{N} =2$ 
twisted chiral multiplet is the dimensionally-reduced 4D, $\mathcal{N}=1$ 
vector multiplet. We see in comparing Equation~(\ref{e:twistedgadget}) to 
Equation~(\ref{e:gadgetsCMTMVM}) that the gadget keeps track of this relationship 
between the twisted chiral and vector supermultiplets.

The two-dimensional $\s$-model actions can also be reduced to one-dimensional $\s$-model actions, 
\be\label{e:Ss1D}
S{}_{\s}^{(1D)}~=~\int d\t \, d^2 {\theta} \, d^2 {\Bar \theta}~K(\Phi{}^{\sc I}, 
{\Bar \Phi}{}^{\sc I} ; \, \chi{}^{\Hat {\sc I}}, {\Bar \chi}{}^{\Hat {\sc I}})~~~,
\ee
and the works of \cite{E5,E6,E7} in principle capture all of these 
(as we only consider valise supermultiplets, the chiral and twisted chiral superfields in~(\ref{e:Ss1D}) correspond to starting with their 4D progenitors where both auxiliary fields have been replaced by three-forms). However, here, it is
useful to recall the experience of the reduction from 4D, $\cal N$ = 1 $\s$-models
to 2D, $\cal N$ = 2 $\s$-models. The loss of control of the model space came
about because the actions and supermultiplets that appear in them are ``blind''
to the appearance of `new' supermultiplets that can result from the reduction process.

In order to demonstrate the loss once more, it is useful to show the reduction
from 2D, $\cal N$ = 2 supersymmetry to first consider the intermediate step where
we consider 2D, $\cal N$ = (4,0) supersymmetry. This will allow the explicit
demonstration of the emergence of new supermultiplets in the intermediate step. 
Thus, any further reduction to 1D, $\cal N$ = (4,0) supersymmetry must inherit 
these supermultiplets from 2D, $\cal N$ = (4,0) supersymmetry.

\noindent
{\subsection {2D, \texorpdfstring{$\cal N$}{N}~= (4,0) Supersymmetry Considerations}}

We introduce the bosonic coordinates for the worldsheet $\t$ and $\s$
assembled into light cone coordinates $\s^{\pp}$ and 
$\s^{\mm}$ such that:
\be
 d\s^{\pp} \, d\s^{\mm}~=~(d\t){}^2~-~(d\s)^2~~~.
\ee
For (4,0) superspace, four Grassmann coordinates correspond to 
the $+$ component of spinor helicity with regard to the worldsheet Lorentz group. 
Therefore, we have: 
\be
\zeta^{+ \,i } \,=\,
\left[\begin{array}{c}
\zeta^{+ \, 1}\\
\zeta^{+ \, 2} \\
\end{array}\right]~~.
\ee
As the Grassmann coordinates are complex, the ``isospin'' indices, denoted by
$i$, $j$, $\dots$ etc.) may be regarded as describing an internal su(2) symmetry.

Finally, we introduce the superspace ``covariant derivatives'':
\be
D_{+ i}~=~\left( \,
{ {\pa~~~~} \over {\pa \zeta^{+ \,i }}}~+~i \, \fracm 12 {\Bar \zeta}{}^{+}{}_{ \,i } \pa{}_{\pp}
\right) 
~~,~~
{\Bar D}_{+}^{ i}~=~\left( \,
{ {\pa~~~~} \over {\pa {\Bar \zeta}{}^{+}{}_{ \,i }}}~+~i \, \fracm 12 { \zeta}{}^{+ \,i }
{\pa}{}_{\pp} \right)~~~,
\ee
together with the light cone derivatives ${\pa}{}_{\pp}$ and ${\pa}{}_{\mm}$ to
describe the tangent space to the supermanifold. These definitions 
ensure the equations:
\be
\{ \, D_{+}{}_i~,~{\Bar D}_{+}^{ j} \, \}~=~i \, \d{}_i {}^j \, \pa_{\pp} ~~~.
\label{CC}
\ee

\noindent
{\subsection {Reviewing the Known 2D, \texorpdfstring{$\cal N$}{N}~= (4,0) Minimal Scalar Valise
Supermultiplets}}

Many years ago~\cite{Gates:1994bu,Dhanawittayapol:1996cr}, it was demonstrated that there is a minimum of 
four distinct (4,0) valise supermultiplets each containing four bosons and 
four fermions. Thus, one can introduce a ``representation label'' $({\cal R})$, which takes 
on four values denoted by SM-I, SM-II, SM-III, and SM-IV. The~field content 
of each is shown in Table~\ref{t:2DFieldContent}. All fields with 
two such indices are traceless. The bosons are ${\cal A},\, {\cal B}, \, \phi, 
\, \phi_i {}^j , \, {\cal A}_i $, and $ {\cal B}_i $, and of these, only $\phi$ and 
$\phi_i {}^j$ are real (or Hermitian). 
\begin{table}[!htbp]
\centering
\renewcommand\arraystretch{1.2}
\begin{tabular}{|c|c| }\hline
${\rm Multiplet}$  & ${\rm Field~Content}$  \\ \hline 
\hline
$~~~{\rm SM-I}~~~$ &  $~~({\cal A},~{\cal B}, ~ \psi^{- i} )~~$ \\ \hline
$~~~{\rm SM-II}~~~$ &  $~~(\phi,~\phi_i {}^j , ~ \l^- {}_i )~~$   \\ \hline
$~~~{\rm SM-III}~~~$ & $ ({\cal A}_i,~\rho^- , ~ \pi^-  )$   \\ \hline
$~~~{\rm SM-IV}~~~$ &  $ ({\cal B}_i,~\psi^- , ~ \psi^-_i {}^j  ) $   \\ \hline
\end{tabular}\caption{The field content of the 2D, $\mathcal{N}$ = (4,0) supermultiplets.}
\label{t:2DFieldContent}
\end{table}

Regarding the su(2) symmetry, the supercovariant derivatives, the bosonic 
fields, and the fermionic fields are distributed among different irreducible 
representations, as shown in Table~\ref{t:2Dspin}.
\begin{table}[!htbp]
\centering
\renewcommand\arraystretch{1.2}
\begin{tabular}{|c|c| }\hline
${\rm Quantity}$  & ${\rm su(2)-spin~J~value}$  \\ \hline 
\hline
${\cal A}, \, {\cal B} , \, {\phi} , \,  \rho^- , \,  \pi^- , \, \psi^- $ &  $0$   \\ \hline
$~~D_{+ i}, \, {\Bar D}_{+}^{i}, \, \psi^{- i},  \l^- {}_i~~$ &  $~~\fracm 12 ~~$ \\ \hline
$ \phi^-_i {}^j  , \, \psi^-_i {}^j $ &  $ 1 $   \\ \hline
\end{tabular}
\caption{The spin content of the 2D, $\mathcal{N}$ = (4,0) supermultiplets.}
\label{t:2Dspin}
\end{table}
It is noteworthy that the first two of these supermultiplets (i.e., SM-I
and SM-II) can be interpreted as arising from a dimensional reduction 
process applied to the well-known 4D, $\cal N$ = 1 ``chiral supermultiplet''
and ``vector supermultiplet'', respectively. With respect to the other two
supermultiplets, there are {{no}} discussions known to us that indicate 
they can arise {{solely}} as the results of such dimensional reductions.
The SM-III and SM-IV supermultiplets, however, are related respectively
to the SM-I and SM-II supermultiplets by a Klein transformation, where
bosonic fields and fermionic fields are exchanged one for the other.

These fields may be interpreted in two ways. In the first interpretation, all 
these are component fields that are functions solely of light cone coordinates 
$\s^{\pp}$ and $\s^{\mm}$. In the second interpretation, these are each
regarded as the lowest component of a corresponding superfield in the 
expansion over the basis of the superspace Grassmann coordinates.
 
The ``$D$-algebra'' or ``SUSY transformation law'' for each supermultiplet
is given in Equations~(\ref{TRsm1})--(\ref{TRsm4}), which follow.
\vskip 0.12in \noindent
$\text{SM-I}~{\rm {Supermultiplet:}}~({\cal A},~{\cal B},~\psi^{- i} )$ 
\be 
\eqalign{ 
D_{+ i} {\cal A}~=~& 2 C_{ij} \psi^- {}^{j} ,~~~~~~\,~~\,~~~~~~~~{\Bar D}_+
{}^i {\cal A}~=~0,~~~~~~~~~\cr
{\Bar D}_+ {}^i {\cal B}~=~& i 2 \psi^- {}^{i} ,~~~~~~~~~~~~~~~~~~~~\, 
D_{+ i} {\cal B}~=~0 ,~~~~~~\,~~~\cr
{\Bar D}_+ {}^i \psi^- {}^{j}~=~& i \fracm 12 \, C^{ij} \pa_{\pp} {\cal A} ,\,~~~~~~~~~~~
~\,\, \, D_{+ i} \psi^- {}^{j}~=~\fracm 12 \, \d_i {}^j \, \pa_{\pp} {\cal B},  ~~~\cr
{\Bar D}_+ {}^i {\Bar {\cal A}}~=~& - \, 2 C^{ij} {\Bar \psi}{}^- {}_{j} ,~~~~~
~~~~~~~{D}_+ {}_i {\Bar {\cal A}}~=~0 ,~~~~~~\,\,~~~\cr
{D}_+ {}_i {\Bar {\cal B}}~=~& i 2 {\Bar \psi}{}^- {}_{i} ,
~~~~~~~~~~~\,~~\, \,~~\,~\,~\, {\Bar D}_+ {}^i {\Bar {\cal B}}~=~0 ,~~~~~~~\,~~\cr
{D}_+ {}_i {\Bar \psi}{}^- {}_{j}~=~& - \, i \, \fracm 12 \, C_{ij} \, \pa_{\pp} {\Bar {\cal A}} 
,~~~~~~~~~\, {\Bar D}{}_{+}{}^{i} {\Bar \psi}^- {}_{j}~=~\d_i {}^j \fracm 12 \, 
 \pa_{\pp
 } {\Bar {\cal B}}, ~~~} \label{TRsm1}
\ee
\noindent
${\rm SM-II}~{\rm {Supermultiplet:}}~(\phi,~\phi_{j}{}^{k},~\lambda^- 
{}_j )$ 
\be
 \eqalign{ 
 D_{+ i}~\phi~=~&~i \lambda^- {}_i , 
~~~~~~~~~~~~~~~~~~~~~\,~~~{\Bar D}_{+}{}^{i}~\phi 
~=~i {\Bar \lambda}^- {}^i,~~~~{~~~~~~~~~~~~\,~~~~~~~~} \cr
 {~~~~~~~~~~~~} D_{+ i}~\phi_{j}{}^{k}~=~&~2 \d_i {}^{k} \lambda^- 
 {}_{j} - \d_{j} {}^{k} \lambda^- {}_{i} ,~~~~~~
{\Bar D}{}_{+}{}^{i}~\phi_{j}{}^{k}~=~- \,2 \d_j {}^{i} {\Bar \lambda}^- 
{}^{k} \,+\, \d_{j} {}^{k} {\Bar \lambda}{}^- {}^{i} ,~~~\cr
{D}_+ {}_i~{\Bar \lambda}^- {}^j~=~& \fracm 12 \, \d_j {}^i \pa_{\pp} 
\phi~-~i \fracm 12 \, \pa_{\pp} \phi_j{}^i,~~~~~~~D_{+ i}~\lambda^- {}_j~=~0
,~~~~~~~~~~~~~~~~~~~~~~~~~~~~\cr
{\Bar D}_+ {}^i~\lambda^- {}_j~=~& \fracm 12 \, \d_j {}^i \pa_{\pp} 
\phi~+~i \fracm 12 \, \pa_{\pp} \phi_j{}^i ,~\,~~~~~{\Bar D}_{+}{}^{i}~{\Bar \lambda}^- 
{}^j~=~0 ,~~~~~~~~~~~~~~~~~~~~~~~~~~~\,~\cr
\phi~=~& \phi^* ,~~~~~~\phi_{i}{}^{j}~=~(\phi_{j}{}^{i})^*,~~~~~~~~
\phi_{i}{}^{i}~=~0 ,~~~
} \label{TRsm2} \ee
 \noindent
${\rm SM-III}~{\rm {Supermultiplet:}}~({\cal A}_i,~\pi^-,~\rho^- )$ 
\be \eqalign{ 
D_{+ i} {\cal A}_j~=~& C_{ij} \pi^-,~~~~~~~~~~~~~~~{\Bar D}_+ {}^i 
{\cal A}_j~=~\d_j {}^i \rho^- ,~~~~~~~\cr
{\Bar D}_+ {}^i {\Bar {\cal A}}^j~=~& -\, C^{ij}{\Bar \pi}^- ,~~~~~\,~~~~~
D_{+ i} {\Bar {\cal A}}^j~=~-\, \d_i {}^j {\bar \rho}^- ,~~~\cr
D_{+ i} \rho^-~=~& i \, \pa_{\pp} {\cal A}_i ,~~~~~~~~~~~~~{\Bar 
D}_+ {}^i \rho^-~=~0 ,~~~~~~~~~~\,~~\cr
{\Bar D}_+ {}^i {\Bar \rho}^-~=~& -\, i \, \pa_{\pp} {\Bar {\cal A}}^i 
,~~~~~~~~~{D}_+ {}_i {\Bar \rho}^-~=~0 ,~~~~~~~~~~\,~~\cr
{\Bar D}_+ {}^i \pi^-~=~& i C^{ij} \, \pa_{\pp} {\cal A}_j,~~~~
~~~\, D_{+ i} \pi^-~=~0 ,~~~~~~~~~~\,~~\cr
{D}_+ {}_i {\Bar \pi}^-~=~&-\, i C_{ij} \, \pa_{\pp} {\Bar {\cal A}}^j , 
~~~~~~~\, {\Bar D}_{+}{}^{i} {\Bar \pi}^-~=~0 ,~~~~~~~~\,~~~~} \label{TRsm3}
\ee
 \noindent
${\rm SM-IV}~{\rm {Supermultiplet:}}~({\cal B}_i,~\psi^- ,~\psi^- 
{}_i {}^{j} )$ 
\be \eqalign{ {~~~~~}
{\Bar D}_+ {}^i {\cal B}_{j}~=~& \d_j {}^i \, \psi^-~+~i 2 \, \psi^- {}_j 
{}^{i} ,\,~~~~~~~~~~~~~~~~~\, D_{+ i} {\cal B}_{j}~=~0
,~~~~~~~~~~~~~~~~~~~~~~~~~~~~~\cr
{D}_+ {}_i {\Bar {\cal B}}^{j}~=~& -\, \d_i {}^j \, \psi^-~+~i 2 \, \psi^- {}_i 
{}^{j} ,~~~~\, {~~~~~~~~~} {\Bar D}_{+}{}^{i} {\Bar {\cal B}}^{j}~=~0
,~~~~~~~~~~~~~~~~~~~~~~~~~~~~~\cr
D_{+ i} \psi^-~=~& i \, \fracm 12 \, \pa_{\pp} {\cal B}_{i} ,~~~~~~~~{~~~~~~~\,~~}
~~~~~~\,~~~~\,~~D_{+ i} \psi^- {}_j {}^{k}~=~\fracm 12 \, \d_i {}^k \pa_{\pp 
} {\cal B}_{j} - \frac14 \d_j {}^{k} \, \pa_{\pp} {\cal B}_{i} ,~~~~\cr
{\Bar D}_{+}{}^{i} {\Bar \psi}^-~=~& -\, i \, \fracm 12 \, \pa_{\pp} {\Bar {\cal B}}^{i} 
,~~~~~~~~~~~~~{~~~~~~~\,~~}~~~
{\Bar D}_{+}{}^{i} \psi^- {}_j {}^{k}~=~\fracm 12 \, \d_j {}^i \, \pa_{\pp} {\Bar {\cal B
}}^{k} - \frac14 \d_j {}^{k} \, \pa_{\pp} {\Bar {\cal B}}^{i} ,~~~~\cr
&\psi{}^-~=~(\psi{}^- )^* ,~~~~~~\psi^- {}_i {}^{j}~=~(\psi^- {}_j {}^i)^* 
~~~,~~~\psi{}^-{}_{i}{}^{i}~=~0 .~~~} \label{TRsm4}
\ee
The fermionic holoraumy for each of these multiplets is given in Appendix~\ref{a:2DHoloraumy}.

\noindent
{\subsection {Uniformization via Real Formulations}}

In order to calculate the values of the first gadget, in the context of these $(4, \,0)$ supermultiplets 
described previously, it is necessary to convert all their descriptions into a real basis where the 
comparison process can be made in the simplest possible manner. In particular, the first step 
is to obtain the ``L-matrices'' and ``R-matrices'' associated with each of the four representations: SM-I, 
SM-II, SM-III, and SM-IV. As was emphasized in the Appendix of the work in~\cite{Gates:1995ch}, 
``L-matrices'' and ``R-matrices'' can be identified in dimensions greater than one. In this
particular example, the ``L-matrices'' and ``R-matrices'' were shown for superspaces with
three bosonic coordinates.

To begin, we note that for the operator $D_{+}{}_i$, we can write:
\be
D_{+}{}_i \,=\,
\left[\begin{array}{c}
D_{+}{}_1 \\
D_{+}{}_2 \\
\end{array}\right]~,~
D_{+}{}_1~=~ \, \left( \, {\rm D}_{+ \, {\Hat A}} \,-\, i \, {\rm D}_{+ \, {\Hat B}}
 \right)~,~D_{+}{}_2~=~ \, \left( \, {\rm D}_{+ \, {\Hat C}} \,-\, i \, 
 {\rm D}_{+ \, {\Hat D}} \right)~, \label{Maj1}
\ee
where the four ``supercovariant derivatives'' are defined with respect to the four 
real (Majorana) spinor coordinates for the $(4, \,0)$ superspace $\zeta^{+ \, {\Hat 
A}}$, $\zeta^{+ \, {\Hat B}}$, $\zeta^{+ \, {\Hat C}}$, and $\zeta^{+ \,D}$. It is important to note that the labels ${
\Hat A}$, ${\Hat B}$, ${\Hat C}$, and ${\Hat D}$ are fixed values, not indices
that take on different values.

Taking the 
complex conjugate of the results in (\ref{Maj1}), we find:
\be
{\Bar D}_{+}{}^i \,=\,
\left[\begin{array}{c}
{\Bar D}_{+}{}^1 \\
{\Bar D}_{+}{}^2 \\
\end{array}\right]~,~
{\Bar D}_{+}{}^1~=~ \, \left( \, {\rm D}_{+ \, {\Hat A}} \,+\, i \, {\rm D}_{+ \, {\Hat B}}
 \right)~,~{\Bar D}_{+}{}^2~=~ \, \left( \, {\rm D}_{+ \, {\Hat C}} \,+\, i \, 
 {\rm D}_{+ \, {\Hat D}} \right)~, \label{Maj2}
\ee
which imply also the validity of the complex conjugate version of (\ref{CC}). Together,
(\ref{Maj1}) and (\ref{Maj2}) imply:
\be \eqalign{
{\rm D}_{+ \, {\Hat A}}~&=~  \, \left( \, D_{+}{}_1~+~{\Bar D}_{+}{}^1 \, \right) ,~\,~~~\cr
{\rm D}_{+ \, {\Hat B}}~&=~i\,  \, \left( \, D_{+}{}_1~-~{\Bar D}_{+}{}^1 \, \right) ,~~~\cr
{\rm D}_{+ \, {\Hat C}}~&=~  \, \left( \, D_{+}{}_2~+~{\Bar D}_{+}{}^2 \, \right) ,~\,~~~\cr
{\rm D}_{+ \, {\Hat D}}~&=~i\,  \, \left( \, D_{+}{}_2~-~{\Bar D}_{+}{}^2 \, \right) ,~~~
} \label{RspnR} \ee
which 
in turn
allows the definition of a real (4, 0) superspace covariant derivative
through the equation:
\be \eqalign{
{\rm D}{}_{+{}_{\rm I}}~=~\left[\begin{array}{c}
{\rm D}_{+ \, {\Hat A}} \\
{\rm D}_{+ \, {\Hat B}} \\
{\rm D}_{+ \, {\Hat C}} \\
-{\rm D}_{+ \, {\Hat D}} \\
\end{array}\right]~=~
\left[\begin{array}{c}
\left( \, D_{+}{}_1 \,+\, {\Bar D}_{+}{}^1 \right) \\
i \, \left( \, D_{+}{}_1 \,-\, {\Bar D}_{+}{}^1 \right) \\
\left( \, D_{+}{}_2 \,+\, {\Bar D}_{+}{}^2 \right) \\
-i \, \left( \, D_{+}{}_2 \,-\, {\Bar D}_{+}{}^2 \right) 
\end{array}\right],~~~
} \label{Dreal} \ee
where subscript index I takes on the four fixed values ${\Hat A}$, $\dots$ ${\Hat D}$.
This definition implies that the equation:
\be
[ \, {\rm D}{}_{+{}_{\rm I}}~,~{\rm D}{}_{+{}_{\rm J}} \, \}~=~i \, 2 \, \d {}_{{}_{\rm I} \, 
{}_{\rm J}} \, \pa_{\pp} ,~~~\label{DeeEQ}
\ee
is satisfied. 

Let us also highlight that in (\ref{Dreal}), there is a notational device introduced. The quartet 
Majorana supercovariant derivative operator is denoted by ${\rm D}{}_{+ {}_{\rm I}}$, whereas 
the complex SU
(2)-doublet supercovariant derivative operator is denoted by the pair $(D_{+}
{}_i$, ${\Bar D}_{+}{}^i)$. Thus, Equation (\ref{Dreal}) solidifies the definition of the 
Majorana supercovariant derivative operator. The next step is also do this for all fields.

For the bosonic and fermionic fields in each supermultiplet, we define:
\be \eqalign{ {~~~}
 \Phi_{i}^{\rm {(SM-I)}}~=~ \fracm 12
 \left[\begin{array}{c}
 \left( \, {\cal B} \,+\, {\Bar {\cal B}} \, \right) \\
+ \, i\left( \, {\cal A} \,-\, {\Bar {\cal A}} \, \right) \\
-\, i \left( \, {\cal B} \,-\, {\Bar {\cal B}} \, \right) \\
 \left( \, {\cal A} \,+\, {\Bar {\cal A}} \, \right)
 \end{array}\right] 
,~~~~~~~~~
 \Psi_{+\, \hat k}^{\rm {(SM-I)}}~=~
 \left[\begin{array}{c}
( \, {\Bar \psi}^-{}_1 \,+\, \psi^-{}^1\, ) \\
i \, ( \, {\Bar \psi}^-{}_1 \,-\, \psi^-{}^1 \,) \\
( \, {\Bar \psi}^-{}_2 \,+\, \psi^-{}^2 \, ) \\
- i \, ( \, {\Bar \psi}^-{}_2 \,-\, \psi^-{}^2 \, ) 
\end{array}\right] ,~~}
\label{SM-Ips}
\ee
\be \eqalign{ {~~~}
 \Phi_{i}^{\rm {(SM-II)}}~=~
 \left[\begin{array}{c} \phi{}_1{}^1 \\
 \fracm 12 \, [ \phi{}_1{}^2 + ( \phi{}_1{}^2 )^* ] \\
 \phi \\
i\, \fracm 12 \, [ \phi{}_1{}^2 - ( \phi{}_1{}^2 )^* ] 
\end{array}\right] 
,~~~~~~
 \Psi_{+\, \hat k}^{\rm {(SM-II)}}~=~
 \left[\begin{array}{c}
( \, {\lambda}^-{}_1 \,+\, {\Bar \lambda}^-{}^1\, ) \\
-i \, ( \, { \lambda}^-{}_1 \,-\, {\Bar \lambda}^-{}^1 \,) \\
( \, {\lambda}^-{}_2 \,+\, {\Bar \lambda}^-{}^2 \, ) \\
i \, ( \, {\lambda}^-{}_2 \,-\, {\Bar \lambda}^-{}^2 \, ) 
\end{array}\right]
 ,{~~~}
} \label{SM-IIps}
\ee
\be \eqalign{ {~~~~}
\Phi_{i}^{\rm {(SM-III)}}~=~
 \left[\begin{array}{c}
( \, {\cal A}{}_1 \,+\, {\Bar {\cal A}}{}^1\, ) \\
i \, ( \, {\cal A}{}_1 \,-\, {\Bar {\cal A}}{}^1 \,) \\
( \, {\cal A}{}_2 \,+\, {\Bar {\cal A}}{}^2 \, ) \\
 i \, ( \, {\cal A}{}_2 \,-\, {\Bar {\cal A}}{}^2 \, ) 
\end{array}\right]
,~~~~~
 \Psi_{+\, \hat k}^{\rm {(SM-III)}}~=~
 \fracm 12
\left[\begin{array}{c}
\left( \, \pi^- \,+\, {\Bar \pi}^- \, \right) \\
i\left( \, \rho^- \,-\, {\Bar \rho}^- \, \right) \\
i \left( \, \pi^- \,-\, {\Bar \pi}^- \, \right) \\
\left( \, {\rho}^- \,+\, {\Bar \rho}^- \, \right)
\end{array}\right] 
,~~} \label{SM-IIips}
\ee
\be \eqalign{ {~~}
\Phi_{i}^{\rm {(SM-IV)}}~=~
\frac{1}{2}\left[\begin{array}{c}
( \, {\cal B}{}_1 \,+\, {\Bar {\cal B}}{}^1\, ) \\
i \, ( \, { \cal B}{}_1 \,-\, {\Bar {\cal B}}{}^1 \,) \\
( \, {\cal B}{}_2 \,+\, {\Bar {\cal B}}{}^2 \, ) \\
i \, ( \, {\cal B}{}_2 \,-\, {\Bar {\cal B}}{}^2 \, ) 
\end{array}\right]
,~~~~~
 \Psi_{+\, \hat k}^{\rm {(SM-IV)}}~=~
 \left[\begin{array}{c}
-\, 2 \psi^-{}_1{}^1 \\
 \, [ \psi^-{}_1{}^2 + ( \psi^-{}_1{}^2 )^* ] \\
 \psi^- \\
i\, \, [ \psi^-{}_1{}^2 - ( \psi^-{}_1{}^2 )^* ] 
\end{array}\right] .{~~~}
} \label{SM-IVps}
\ee

Therefore, when all of the bosons and fermions in Equations (\ref{TRsm1})--(\ref{TRsm4}) are expressed in terms of real quartets of functions as in 
(\ref{SM-Ips})--(\ref{SM-IVps}) and the Majorana supercovariant derivative
in (\ref{Dreal}) is used, they universally possess SUSY transformation 
laws in the form:
\be \eqalign{
{\rm D}{}_{+{}_{\rm I}} \Phi_{i}^{(\cal R)}~=~i \, \left( {\rm L}{}_{{}_{\rm I}}^{(\cal R)} \right) 
{}_{i \, {\hat k}} \, \, \Psi_{+\, \hat k}^{(\cal R)} ,~~~~~~
{\rm D}{}_{+{}_{\rm I}} \Psi_{+\, \hat k}^{(\cal R)}~=~\left( {\rm R}{}_{{}_{\rm I}}^{(\cal R)} \right)
{}_{{\hat k} \, i} \, \pa_{\pp} \, \Phi_{i}^{(\cal R)} .~~~
} \label{VH1}
\ee 
In the expression (\ref{VH1}), $\Phi_{i}^{(\cal R)}$ denotes the i$^{\text{th}}$ boson associated 
with the $({\cal R})$-th (4, 0) supermultiplet and $\Psi_{+ \, \hat k}^{(\cal R)} $ denotes 
the ${\hat k}$-th fermion associated with the $({\cal R})$-th supermultiplet, and
the explicit forms of the matrices $\brL_\rI {}^{(\cal R)}$ and $\brR_\rI {}^{(\cal R)}$
depend on the value of $({\cal R})$. For all representations, they satisfy:
\begin{align}\label{e:ortho}
	{\bm {\rm R}}_\rI^{(\cal R)} = ({\bm {\rm L}}_\rI^{(\cal R)})^{-1} = ({\bm {\rm L}}_\rI^{(\cal R)})^{T} .~~~
\end{align}
\begin{equation}
\begin{aligned}\label{e:LSMI}
	{\bm {\rm L}}_{1}^{(\rm SM-I)} = &\left(
\begin{array}{cccc}
 1 & 0 & 0 & 0 \\
 0 & 0 & 0 & -1 \\
 0 & 1 & 0 & 0 \\
 0 & 0 & -1 & 0 \\
\end{array}
\right)
	,~~~~~~
{\bm {\rm L}}_{2}^{(\rm SM-I)} =\left(
\begin{array}{cccc}
 0 & 1 & 0 & 0 \\
 0 & 0 & 1 & 0 \\
 -1 & 0 & 0 & 0 \\
 0 & 0 & 0 & -1 \\
\end{array}
\right)
,~~~~~~\cr
{\bm {\rm L}}_{3}^{(\rm SM-I)} =&\left(
\begin{array}{cccc}
 0 & 0 & 1 & 0 \\
 0 & -1 & 0 & 0 \\
 0 & 0 & 0 & -1 \\
 1 & 0 & 0 & 0 \\
\end{array}
\right)
,~~~~~~{\bm {\rm L}}_{4}^{(\rm SM-I)} =\left(
\begin{array}{cccc}
 0 & 0 & 0 & 1 \\
 1 & 0 & 0 & 0 \\
 0 & 0 & 1 & 0 \\
 0 & 1 & 0 & 0 \\
\end{array}
\right).~~~
\end{aligned}\end{equation}
%
\begin{equation}
\begin{aligned}\label{e:LSMII}
	{\bm {\rm L}}_{1}^{(\rm SM-II)} = &\left(
\begin{array}{cccc}
 0 & 1 & 0 & 0 \\
 0 & 0 & 0 & -1 \\
 1 & 0 & 0 & 0 \\
 0 & 0 & -1 & 0 \\
\end{array}
\right)
	,~~~~~~
{\bm {\rm L}}_{2}^{(\rm SM-II)} =\left(
\begin{array}{cccc}
 1 & 0 & 0 & 0 \\
 0 & 0 & 1 & 0 \\
 0 & -1 & 0 & 0 \\
 0 & 0 & 0 & -1 \\
\end{array}
\right)
,~~~~~~\cr
{\bm {\rm L}}_{3}^{(\rm SM-II)} =&\left(
\begin{array}{cccc}
 0 & 0 & 0 & 1 \\
 0 & 1 & 0 & 0 \\
 0 & 0 & 1 & 0 \\
 1 & 0 & 0 & 0 \\
\end{array}
\right)
,~~~~~~{\bm {\rm L}}_{4}^{(\rm SM-II)} =\left(
\begin{array}{cccc}
 0 & 0 & 1 & 0 \\
 -1 & 0 & 0 & 0 \\
 0 & 0 & 0 & -1 \\
 0 & 1 & 0 & 0 \\
\end{array}
\right).~~~
\end{aligned}\end{equation}
\begin{equation}
\begin{aligned}\label{e:LSMIII}
	{\bm {\rm L}}_{1}^{(\rm SM-III)} = &\left(
\begin{array}{cccc}
 0 & -1 & 0 & 0 \\
 0 & 0 & 0 & 1 \\
 -1 & 0 & 0 & 0 \\
 0 & 0 & -1 & 0 \\
\end{array}
\right)
	,~~~~~~
{\bm {\rm L}}_{2}^{(\rm SM-III)} =\left(
\begin{array}{cccc}
 0 & 0 & 0 & -1 \\
 0 & -1 & 0 & 0 \\
 0 & 0 & -1 & 0 \\
 1 & 0 & 0 & 0 \\
\end{array}
\right)
,~~~~~~\cr
{\bm {\rm L}}_{3}^{(\rm SM-III)} =&\left(
\begin{array}{cccc}
 1 & 0 & 0 & 0 \\
 0 & 0 & 1 & 0 \\
 0 & -1 & 0 & 0 \\
 0 & 0 & 0 & 1 \\
\end{array}
\right)
,~~~~~~{\bm {\rm L}}_{4}^{(\rm SM-III)} =\left(
\begin{array}{cccc}
 0 & 0 & -1 & 0 \\
 1 & 0 & 0 & 0 \\
 0 & 0 & 0 & 1 \\
 0 & 1 & 0 & 0 \\
\end{array}
\right).~~~
\end{aligned}\end{equation}
\begin{equation}
\begin{aligned}\label{e:LSMIV}
	{\bm {\rm L}}_{1}^{(\rm SM-IV)} = &\left(
\begin{array}{cccc}
 -1 & 0 & 0 & 0 \\
 0 & 0 & 1 & 0 \\
 0 & 1 & 0 & 0 \\
 0 & 0 & 0 & -1 \\
\end{array}
\right)
	,~~~~~~
{\bm {\rm L}}_{2}^{(\rm SM-IV)} =\left(
\begin{array}{cccc}
 0 & 0 & -1 & 0 \\
 -1 & 0 & 0 & 0 \\
 0 & 0 & 0 & 1 \\
 0 & 1 & 0 & 0 \\
\end{array}
\right)
,~~~~~~\cr
{\bm {\rm L}}_{3}^{(\rm SM-IV)} =&\left(
\begin{array}{cccc}
 0 & 1 & 0 & 0 \\
 0 & 0 & 0 & 1 \\
 1 & 0 & 0 & 0 \\
 0 & 0 & 1 & 0 \\
\end{array}
\right)
,~~~~~~{\bm {\rm L}}_{4}^{(\rm SM-IV)} =\left(
\begin{array}{cccc}
 0 & 0 & 0 & 1 \\
 0 & -1 & 0 & 0 \\
 0 & 0 & 1 & 0 \\
 -1 & 0 & 0 & 0 \\
\end{array}
\right).~~~
\end{aligned}\end{equation}

We have for $\chi_0$ values:
\begin{align}
\label{e:chi014}
	\chi_0^{(\rm SM-I)} =& \chi_0^{(\rm SM-IV)} = +1~~~\\
\label{e:chi023}
	\chi_0^{(\rm SM-II)} =& \chi_0^{(\rm SM-III)} = -1.~~~
\end{align}

Thus, by choosing to work in a real basis, the disparate forms of the field content
and transformation laws seen in (\ref{TRsm1})--(\ref{TRsm4}) have been subjected
to a ``uniformization.''

\noindent
{\subsection {2D, \texorpdfstring{$\cal N$}{N}~= (4, 0) Adinkra-Related Matrices and Gadget Values}}

With the results of the last subsection in hand, we can build on these and are in a 
position to calculate the fermionic holoraumy matrices defined by:
\be \eqalign{
[\, {\rm D}{}_{+{}_{\rm I}}~,~{\rm D}{}_{+{}_{\rm J}} \, ] \, \Psi_{+ \, \hat k}^{({\cal R})} 
~&=~2\, \left[ \, {\tilde{ \rm V}}^{({\cal R})}{}_{{}_{\rm I}}{}_{{}_{\rm J}} \, \right] {}_{{\hat 
k} \, {\hat \ell}} \, \pa_{\pp} \, \Psi_{+\, \hat \ell}^{({\cal R})} ~~~,
} \label{VH2}
\ee
where:
\be \eqalign{
\left[ \, {\tilde {\rm V}}{}_{\rI\rJ}^{({\cal R})} \,\right] {}_\hi{\,}^\hk~&=~-i\, \frac12 \, 
\left[ \, (\,{\rm R}_\rI {}^{({\cal R})} \,)_\hi{}^j\>(\, {\rm L}_\rJ {}^{({\cal R})} \,
)_j{}^\hk~-~(\,{\rm R}_\rJ {}^{({\cal R})} \, )_\hi{}^j\>(\,{\rm L}_\rI
{}^{({\cal R})} 
\,)_j{}^\hk \, \right]
~~.
} \label{VH4}
\ee
associated with each supermultiplet SM-I, SM-II, SM-III, and SM-IV. It can be seen that 
if one deletes the helicity label (i.e.,\ the + signs) from these and makes the replacement 
$\pa_{\pp}$ $\to$ $\pa_{\tau}$, then (\ref{DeeEQ})--(\ref{VH4}) take the exact form as the 
equations given for adinkras and 1D, $N$ = 4 supersymmetry~\cite{Gates:2014npa,Calkins:2014sma}. 
We~can decompose the $\brtV_{\rI\rJ}{}^{(\cal R)}$ in terms of $\ell_{\rm IJ}^{
\hat{a}}$ and $\tilde{\ell}^{\hat{a}(\cal R)}_{\rm IJ} $ as in Equation~(\ref{e:Vtilde}). In doing so, 
we find:
\begin{align}\label{e:ellSMI}
 i\ell^{\hat{a}(\rm SM-I)}_{\rm IJ} = - \a^{\hat{a}}_{\rm IJ}&,~~~~~~i\tilde{\ell}^{\hat{a
 }(\rm SM-I)}_{\rm IJ} = 0,~~~\\
 \label{e:ellSMII}
 i\ell^{\hat{a}(\rm SM-II)}_{\rm IJ} = 0&,~~~~~~i\tilde{\ell}^{\hat{a}(\rm SM-II)}_{
 \rm IJ} = (-1)^{\hat{a}+1} \b^{\hat{a}}_{\rm IJ},~~~~~~\text{no $\hat{a}$ sum},~~~\\
 \label{e:ellSMIII}
 i\ell^{\hat{a}(\rm SM-III)}_{\rm IJ} = 0 &,~~~~~~i\tilde{\ell}^{\hat{a}(\rm SM-III)}_{
 \rm IJ} = ( \b^{1}_{\rm IJ},-\b^{3}_{\rm IJ},-\b^{2}_{\rm IJ}),~~~\\
 \label{e:ellSMIV}
 i\ell^{\hat{a}(\rm SM-IV)}_{\rm IJ} = 0 &,~~~~~~i\tilde{\ell}^{\hat{a}(\rm SM-IV)}_{
 \rm IJ} = (-1)^{\hat{a}+1} \a^{\hat{a}}_{\rm IJ},~~~~~~\text{no $\hat{a}$ sum}.~~~
\end{align}

The matrix quantity $\brtV_{\rI\rJ}{}^{(\cal R)}$ defines the fermionic holoraumy
tensor for each supermultiplet, and once the matrices $\brL_\rI {}^{(\cal R)}$ and 
$\brR_\rI {}^{(\cal R)}$ are used to calculate it, we have a universal form of the holoraumy 
tensor for each supermultiplet. Denoting two of the $(4, \,0)$ supermultiplets by $({ {\cal 
R}})$ and $( {\cal R}^{\prime})$, along their holoraumy matrices $\brtV_{
\rI\rJ}{}^{(\cal R)}$ and $\brtV_{\rI\rJ}{}^{({\cal R}^{\prime})}$, the ``gadget 
value'' between the two representations is defined by the equation:
\be
{{\cal G}} [ ({ {\cal R}}) , ( {\cal R}^{\prime}) ]~=~\left[ {1 \over {48}} \right] \, 
\sum_{\rI , \, \rJ} \,{\rm {Tr}} \, \left[ \, \brtV_{
\rI\rJ}{}^{(\cal R)} \, \brtV_{\rI\rJ}{}^{({\cal R}^{\prime})} \right].~~~
\label{Gdgt1}
\ee
It is perhaps of note to observe that the discussion here is the first time that we have 
extended the concept of holoraumy into the realm of heterotic supersymmetry. The 
gadget values between the four supermultiplets can be represented by a 4 $\times$ 
4 matrix, and listing the order of the representation labels as (SM-I), (SM-II), (SM-III), and 
(SM-IV) for both rows and columns, we find:
\be
{{\cal G}} [ ({ {\cal R}}) , ( {\cal R}^{\prime}) ]~=~
\left[\begin{array}{cccc}
1 & 0 & 0 & 0\\
0 & 1 & \frac{1}{3} & 0\\
0 & \frac{1}{3} & 1 & 0\\
0 & 0 & 0 & 1 \\
\end{array}\right]
\label{Gdgt2}
\ee

The gadget values shown in the equation above explicitly show an important lesson about the reduction process. It can be seen that the value of + 1/3 appears. Recall that for the 2D, $\cal N$ = 2 supermultiplets, this value never occurs. However, in the case of 2D, $\cal N$ = (4,0) supermultiplets, this~gadget value readily appears.

As both the 2D, $\cal N$ = 2 supermultiplets and the 2D, $\cal N$ = (4,0) supermultiplets can be reduced to 1D, $N$ = 4 supermultiplets, this is an explicit demonstration that the model space of any and all possible 1D, $N$ = 4 non-linear $\s$-models must be larger than that found by dimensional reduction!

\noindent
{\subsection {Reducing to 1D, \texorpdfstring{$N$}{N} = 4 Superspace}}

The process of going from 2D, $\cal N$ = (4, 0) superspace to 1D, $N$ = 4 
superspace simply amounts to demanding that all fields depend solely on the 
temporal dimension of the higher dimension. Additionally, all the helicity indices 
are simply dropped from spinors, partial derivatives, etc., as there is no helicity in 
1D. However, in 1D, we are still able to maintain the distinction between bosons 
and fermions since the equations:
\be \eqalign{
\Phi_{i}^{(\cal R)} (\tau_1) \, \Phi_{j}^{(\cal R)} (\tau_2)~&=~+ \, \Phi_{j}^{(\cal R)} 
(\tau_2) \, \Phi_{i}^{(\cal R)} (\tau_1),~~~~\cr
 \Psi_{\hat k}^{(\cal R)} (\tau_1) \, \Psi_{\hat l}^{(\cal R)} (\tau_2) \, 
~&=~- \,  \Psi_{ \hat l}^{(\cal R)} (\tau_2) \, \Psi_{\hat k}^{(\cal R)} (\tau_1)
.~~~
} \label{Rings}
\ee 
must be valid in order to have well-defined component actions.

The main goal of this section was to provide a convincing argument that knowing
the standard supermultiplets associated with the reduction to two-dimensions of the
chiral and vector supermultiplet totally misses the appearance of two new 
supermultiplets that can be constructed on the basis of the use of the Klein
transformation. All of these supermultiplets are compactly described by quartets
of bosons and fermions that are expressed uniformly in the equation of (\ref{VH1}).

However, from our study of adinkras with four colors, we know that these four
supermultiplets are only four among a total of 36,864 such supermultiplets. This
raises questions.

How does one distinguish among the 36,864 such supermultiplets?

How does one describe a sigma-model built, not only on the basis of the four 
supermultiplets (SM-I), (SM-II), (SM-III), and (SM-IV), but one constructed from
any or all of these supermultiplets found by the study of adinkras?

The first of the questions is the one of classification, and the previous sections
deal with this issue and provide methods for classifying these 1D, N = 4 valise
supermultiplets. This was a primary motivation for the arguments developed.

The second question is equivalent to the question of providing a universal
formulation of {{all}} possible such sigma-models. A solution to this
will be presented in a subsequent section. This is very much a question of physics 
as descriptions of dynamical systems follow from such actions.

\newpage
\section{The Universal 1D, \texorpdfstring{$\bm N$}{N} = 4 Minimal Valise
\texorpdfstring{$\bm {BC_4}$}{BC4} Sigma-Model }\label{s:Kahler}

~~~~~
The enumeration of the 36,864 ${BC}_4$ adinkras implies 
that this constitutes a starting point where every possible 
linear minimal valise representation of 1D, $N$ = 4
supersymmetry has been made explicitly ``visible.'' Thus, if there is a way to
concatenate the standard 1D, $N$ = 4 superspace formalism together with the
$BC_4$-enumerated adinkra-based formalism, we are guaranteed to have 
complete control of the model space 1D, $N=4$ supersymmetric $\s$-models. 

The work in \cite{DFGHIL2} ensures the concatenation is possible. The key observations are:
\newline \noindent (a.)
The links in the adinkra use the $ {\rm D}{}_{{}_{{\rm I}}}$ symbol as their representation
and in \newline \noindent $~~~~~~~$
a traditional superfield these can be interpreted at the superspace super-
\newline \noindent $~~~~~~~$
covariant derivative.
\newline \noindent (b.)
Every bosonic and fermionic node of an adinkra may be regarded as the
 \newline \noindent $~~~~~~~$
lowest component of a corresponding superfield.
\newline \noindent 
These observations imply that one need only re-interpret the symbols used
for adinkra-based supermultiplets to obtain traditional superfield equations.

In particular, this implies that a $\s$-model action formula involving superfields obtained 
from $BC_4$ adinkras via the prescription above must take the form:
\be
S{}_{\s}^{(BC_4)}~=~\fracm 1{4!} \, \int d\t \, \epsilon{}^{{\rm I}_1 \, } {}^{{\rm I}_2 
\, } {}^{{\rm I}_3 \, } {}^{{\rm I}_4} {\rm D}{}_{{}_{{\rm I}_1}} {\rm D}{}_{{}_{{\rm 
I}_2}} {\rm D}{}_{{}_{{\rm I}_3}} {\rm D}{}_{{}_{{\rm I}_4}} K ( \Phi_{i}^{({\cal R})}) {\Big |}
\ee
where $ \Phi_{i_1}^{({\cal R}_1)}$ is
a choice of the scalar superfields associated with any representation $({\cal R})$ among
the 36,864 $BC_4$-based valise adinkras, and the notation $ { |}$ as usual means
first perform all the indicated differentiations followed by taking the limit as all superspace
Grassmann coordinates are set to zero. Now, on these superfields, we have the following
equations.
\be \eqalign{
{\rm D}{}_{{}_{{\rm I}_1}} \Phi_{i_1}^{({\cal R}_1)}~&=~i \, \left( {\rm L}{}_{{}_{{\rm I}_1}}^{({
\cal R}_1)} \right) {}_{i_1 \, {\hat k}_1} \, \, \Psi_{{\hat k}_1}^{({\cal R}_1)}
, \cr
{\rm D}{}_{{}_{{\rm I}_3}} {\rm D}{}_{{}_{{\rm I}_4}} \Phi_{i_1}^{({\cal R}_1)}~&=~i\,
 \, \left( {\rm L}{}_{{}_{{\rm I}_4}}^{({\cal R}_1)} {\rm R}{}_{{}_{{\rm I}_3}}^{({\cal R}_1)}
 \right) {}_{i_1 \, {\ell}_1} \, \pa_{\tau} \, \Phi_{{\ell}_1}^{({\cal R}_1)}
, \cr
{\rm D}{}_{{}_{{\rm I}_2}} {\rm D}{}_{{}_{{\rm I}_3}} {\rm D}{}_{{}_{{\rm I}_4}} \Phi_{
i_1}^{({\cal R}_1)}~&=~- \, \left( {\rm L}{}_{{}_{{\rm I}_4}}^{({\cal R}_1)} {\rm R}{}_{{}_{
{\rm I}_3}}^{({\cal R}_1)}{\rm L}{}_{{}_{{\rm I}_2}}^{({\cal R}_1)} \right) 
{}_{i_1 \, {\Hat \ell}_1} \, \pa_{\tau} \, \Psi_{{\Hat \ell}_1}^{({\cal R}_1)}
, \cr
{\rm D}{}_{{}_{{\rm I}_1}} {\rm D}{}_{{}_{{\rm I}_2}} {\rm D}{}_{{}_{{\rm I}_3}} {\rm D}{}_{
{}_{{\rm I}_4}} \Phi_{i_1}^{({\cal R}_1)}~&=~- \, \left( {\rm L}{}_{{}_{{\rm I}_4}}^{({\cal 
R}_1)} {\rm R}{}_{{}_{{\rm I}_3}}^{({\cal R}_1)} {\rm L}{}_{{}_{{\rm I}_2}}^{({\cal R}_1)} 
{\rm R}{}_{{}_{{\rm I}_1}}^{({\cal R}_1)} \right) {}_{i_1 \, { \ell}_1} \, \pa_{\tau}^2 \,
 \Phi_{{ \ell}_1}^{({\cal R}_1)},
}  \ee
and after performing all the differentiations shown in the action formula, this yields
the form of the final action. This can be simplified if we express it in terms of four
coefficients denoted by: 
$~$
\newline $~~~~~$
(a.) ${\rm B}{}_{{j}_1  \,  {j}_2  } [ ({\cal R}_1) : ({\cal R}_2) ]$ - a metric 
for the bosonic kinetic energy terms, 
$~$
\newline $~~~~~$
(b.) $ {\rm F} {}_{{\hat k}_1  \,  {\hat k}_2  } [ ({\cal R}_1) : ({\cal R}_2) ] $ 
 - a metric for the fermionic kinetic energy terms, 
$~$
\newline $~~~~~$
(c.)
$ f_{{\hat k}_1  \,  {\hat k}_2  \, {j}_3 } [ ({\cal R}_1) : ({\cal R}_2) : ({\cal R}_3) ]$ 
 - a coupling of bosons and quadratic in $~$
\newline $~~~~~~~~~~~~~~~~~~~~~~~~~~~~~~~~~~~~~~~~~~~~~~~~$
 fermions, and
$~$
\newline $~~~~~$
(d.)
$ f_{{\hat k}_1 \,  {\hat k}_2  \, {\hat 
k}_3 \, {\hat k}_4} [({\cal R}_1) : ({\cal R}_2) : ({\cal R}_3) : ({\cal R}_4) ]$ 
 - a coupling of bosons and $~$
\newline $~~~~~~~~~~~~~~~~~~~~~~~~~~~~~~~~~~~~~~~~~~~~~~~~$
quartic in fermions. \newline These allow the action to be expressed as:
\be\label{e:1DKahler2}
\eqalign{ {~~}
S{}_{\s}^{(BC_4)}~\equiv~&\int d \t {\Big [}~\fracm 12\, {\rm B}{}_{{j}_1 \, {j}_2 } [
({\cal R}_1) : ({\cal R}_2) ] \, [ \pa_{\t} {\Phi}{{}^{j_1 ({\cal R}_1)}} \, ] \, [ \, \pa_{\t} 
{\Phi}{{}^{j_2 ({\cal R}_2)}} \, ] 
\cr
&~~~~~~~~~+~
 \, i\, \fracm 12\, {\rm F}{}_{{\hat k}_1 \, {\hat 
k}_2 } [ ({\cal R}_1) : ({\cal R}_2) ] \, \Psi_{{\hat k}_1}^{({\cal R}_1)} \, \pa_{\t} 
\Psi_{{\hat k}_2}^{({\cal R}_2)}\,
\cr
&~~~~~~~~~\,+~f_{{\hat k}_1 \, {\hat k}_2 \, {j}_3 } [ ({\cal R}_1) : ({\cal R}_2) : 
({\cal R}_3) ] \Psi_{{\hat k}_1}^{({\cal R}_1)}\, \Psi_{{\hat k}_2}^{({\cal R}_2)}\, 
 \,[ \pa_{\t} {\Phi}{{}^{j_3 ({\cal R}_3)}} \, ] \cr
&~~~~~~~~~\,+~f_{{\hat k}_1 \, {\hat k}_2 \, {\hat k}_3 \, {\hat k}_4} [
({\cal R}_1) : ({\cal R}_2) : ({\cal R}_3) : ({\cal R}_4) ]
\Psi_{{\hat k}_1}^{({\cal R}_1)} \, \Psi_{{\hat k}_2}^{({\cal R}_2)}\,
\Psi_{{\hat k}_3}^{({\cal R}_3)} \, \Psi_{{\hat k}_4}^{({\cal R}_4)}~~
{\Big ]}. }  \ee
The definitions of these coefficients follow from comparing 
Equation~(\ref{e:1DKahler2}) to Equation~(\ref{e:1DKahler1}).
\be\label{e:1DKahler1}
\eqalign{ {~~~~~}
{\rm B}{}_{{j}_1 \, {j}_2 }~&=~
 \epsilon{}^{{\rm I}_1 \, } {}^{{\rm I}_2 \, } {}^{{\rm I}_3 \, } 
{}^{{\rm I}_4} {\big [ } \, \fracm 1{4} \,
K_{ {{{}^{, i_1 ({\cal R}_1) \, ,i_2 ({\cal R}_2)}}{}} } ( {\rm V}{}_{{
}_{{\rm I}_1}}{}_{{}_{{\rm I}_2}} {}^{({\cal R}_1)} ) {}^{i_1}{}_{j_1} \, ( {\rm V}{}_{
 {}_{{\rm I}_3}}{}_{{}_{{\rm I}_4}} {}^{({\cal R}_2)} ) {}^{i_2}{}_{j_2} 
 \cr
&~~~~~~~~~~~~~~~~~-~\fracm 1{12} \,
 K_{ {{{}^{, i_1 ({\cal R}_1) \, ,j_2 ({\cal R}_2)}}{}} } ( {\rm V}{
 }_{{}_{{\rm I}_1}}{}_{{}_{{\rm I}_2}} {}^{({\cal R}_1)} {\rm V}{}_{{}_{{\rm I}_3}}{
 }_{{}_{{\rm I}_4}} {}^{({\cal R}_1)} ) {}^{i_1}{}_{j_1} \, {\big ]},
 \cr
 &\cr
 {\rm F}{}_{{\hat k}_1 \, {\hat 
k}_2 }~&= i\, \fracm 1{3} \, 
 \epsilon{}^{{\rm I}_1 \, } {}^{{\rm I}_2 \, } {}^{{\rm I}_3 \, } {}^{{\rm I}_4} 
K_{ {{{}^{, i_1 ({\cal R}_1) \, ,i_2 ({\cal R}_2)}}{}} } ( {\rm L}{}_{{}_{{\rm I}_1}}^{({\cal 
R}_1)} ) {}_{i_1 \, {\hat k}_1} \, ( {\rm V}{}_{{}_{{\rm I}_3}}{}_{{}_{{\rm I}_4}} {}^{(
{\cal R}_2)} {\rm L}{}_{{}_{{\rm I}_2}}^{({\cal R}_2)} ) {}_{i_2 \, {\hat k}_2},
 \cr
 &~~\cr
 f_{{\hat k}_1 \, {\hat k}_2 \, {j}_3 }
~&= -~\,\fracm 1{4} \, \epsilon{}^{{\rm I}_1 \, } {}^{{\rm I}_2 \, } {}^{{\rm I}_3 \, } {
}^{{\rm I}_4} K_{ {{{}^{, i_1 ({\cal R}_1) \, , i_2 ({\cal R}_2) \, , i_3 ({\cal R}_3) }}}} 
( {\rm L}{}_{{}_{{\rm I}_1}}^{({\cal R}_1)} ) {}_{i_1 {\hat k}_1} ( {\rm L}{}_{{}_{{\rm 
I}_2}}^{({\cal R}_2)} ) {}_{i_2 {\hat k}_2} ( {\rm V}{}_{{}_{{\rm I}_3}}{}_{{}_{{\rm I}_4}} {}^{({\cal 
R}_3)} ) {}^{i_3}{}_{j_3}, 
 \cr 
 &  \cr
f_{{\hat k}_1 \, {\hat k}_2 \, {\hat k}_3 \, {\hat k}_4} 
~&=
\fracm 1{4!} \, \epsilon{}^{{\rm I}_1 \, } {}^{{\rm I}_2 \, }{}^{{\rm I}_3 \, } {}^{{\rm I}_4}
K_{ {{{}^{, i_1 ({\cal R}_1) \, , i_2 ({\cal R}_2) \, , i_3 ({\cal R}_3) \, , i_4 ({\cal R
}_4) }}}} \, \times \cr
&{~~~~~~~~~~~~~~~~~~}
( {\rm L}{}_{{}_{{\rm I}_1}}^{({\cal R}_1)} ) {}_{i_1 \, {\hat k}_1} 
( {\rm L}{}_{{}_{{\rm I}_2}}^{({\cal R}_2)} ) {}_{i_2 \, {\hat k}_2} 
( {\rm L}{}_{{}_{{\rm I}_3}}^{({\cal R}_3)} ) {}_{i_3 \, {\hat k}_3} 
( {\rm L}{}_{{}_{{\rm I}_4}}^{({\cal R}_4)} ) {}_{i_4 \, {\hat k}_4} 
}~~~. \ee
All repeated indices in these equations (including the representation indices) 
are to be summed over all possible values that occur in K\" ahler-like potential. 
In these expression, we have used the notations
\be
K_{ {{{}^{, i_1 ({\cal R}_1)}}{}} }~=~
\left[ \, { {\pa{} K} \over {{\pa {\Phi}{{}^{i_1 ({\cal R}_1)}}} 
}} \, \right] 
\ee
\be
K_{ {{{}^{, i_1 ({\cal R}_1) \, i_2 ({\cal R}_2)}}} }~=~
\left[ \, { {\pa{}^2 K} \over {{\pa {\Phi}{{}^{i_1 ({\cal R}_1)}}} \, {\pa {\Phi}{{}^{i_2 ({\cal R}_2)}}}
}} \, \right] 
\ee
\be
K_{ {{{}^{, i_1 ({\cal R}_1) \, i_2 ({\cal R}_2) \, i_3 ({\cal R}_3)}}} }~=~
\left[ \, { {\pa{}^3 K} \over {{\pa {\Phi}{{}^{i_1 ({\cal R}_1)}}} \, {\pa {\Phi}{{}^{i_2 ({\cal R}_2)}}}
 \, {\pa {\Phi}{{}^{i_2 ({\cal R}_2)}}}
}} \, \right] 
\ee
\be
K_{ {{{}^{, i_1 ({\cal R}_1) \, i_2 ({\cal R}_2) \, i_3 ({\cal R}_3) \, i_4 ({\cal R}_4)}}} }~=~
\left[ \, { {\pa{}^4 K} \over {{\pa {\Phi}{{}^{i_1 ({\cal R}_1)}}} \, {\pa {\Phi}{{}^{i_2 ({\cal R}_2)}}}
 \, {\pa {\Phi}{{}^{i_3 ({\cal R}_3)}}} \, {\pa {\Phi}{{}^{i_4 ({\cal R}_4)}}}
}} \, \right] 
\ee
to denote the derivatives of the K\" ahler-like potential.

To show this yields the usual component level free action for the 
36,864 valise adinkras described in the previous sections, Equation~(\ref{e:L0}),
one should take any superfields and choose $K$ to be the quadratic function:
\be
K~=~\frac{1}{16} \chi_0^{({\cal R})}\, \delta{}_{i \, j} \sum_{({\cal R})} \, 
{\Phi}{{}^{i ({\cal R})}} \, {\Phi}{{}^{j ({\cal R})}}
\ee
(where $\cal R$ can be any of the shown adinkras) of the bosonic nodal fields for 
each of the adinkras. For~this choice of the
K\" ahler-like potential, the geometry of the $\s$-model is flat. If additional terms
higher order in the fields, including even powers of fermion superfields, are included,
then the geometry associated with the K\" ahler-like potential becomes non-trivial.

The coefficient functions denoted as ${\rm B}{}_{{j}_1 \, {j}_2 } [({\cal R}_1) : ({\cal 
R}_2) ]$ and $ {\rm F}{}_{{\hat k}_1 \, {\hat k}_2 } [({\cal R}_1) : ({\cal R}_2) ]$ define 
the metrics respectively on the spaces of component bosons and fermions of the 
supermultiplets. In a similar manner comparing $ f_{{\hat k}_1 \, {\hat k}_2 \, {j}_3 } 
[({\cal R}_1) : ({\cal R}_2) : ({\cal R}_3) ]$ to familiar such terms of supersymmetric
$\s$-models, the obvious interpretation of this term is that it defines an affine connection
on the space of fermions. Finally, making a similar comparison leads to the conclusion
that $f_{{\hat k}_1 \, {\hat k}_2 \, {\hat k}_3 \, {\hat k}_4} [({\cal R}_1) : ({\cal R}_2) : 
({\cal R}_3) : ({\cal R}_4)]$ should describe the curvature tensor of the $\s$-models.
For the choice where $K ( \Phi_{i}^{({\cal R})})$ is restricted to solely quadratic
terms, $ f_{{\hat k}_1 \, {\hat k}_2 \, {j}_3 } [({\cal R}_1) : ({\cal R}_2) : ({\cal R}_3) ]$
and $f_{{\hat k}_1 \, {\hat k}_2 \, {\hat k}_3 \, {\hat k}_4} [({\cal R}_1) : ({\cal R}_2) : 
({\cal R}_3) : ({\cal R}_4)]$ vanish.

The most important points to take away from this discussion is the forms of the action 
and the related equations of motion depend on two distinct types of data. One of 
these is the form of the K\" ahler-like potential K. The other is the representation of 
the valises that are chosen to appear in the $\s$-model. In the action formula, these
representation data are contained in the V-matrices and L-matrices.

\section{Conclusions}
Adinkras are useful and interesting tools with which to study supersymmetry. Perhaps the most useful aspect is the ability to encode information about higher dimensional supersymmetry. To refine holographic procedures, a clear description of equivalent adinkras is necessary. This paper used the gadget to provide a descriptions of 96 such equivalence classes: the $\brtV$-equivalence classes of four-color, four-boson, and four-fermion adinkras. These are equivalence classes of 48 $BC_3$ color transformations (signed three-permutations) of two $\brtV$-inequivalent quaternion adinkras. Each equivalence class was shown to contain 384 adinkras; hence, the 96 equivalence classes contain all 36,864 $= 384 \times 96$ four-color, four-boson, and four-node adinkras. 

These equivalence classes serve to elucidate some of the mysteries of the gadget seen in~\cite{Gdgt2}. For instance, we showed that the plus one-third gadget only arises between equivalence classes of different color-parity. We also found correlations between the frequencies of the gadget values and the color-parities of the equivalence classes. The gadget value of zero occurs most frequently because it is the only gadget value that occurs between adinkras of different isomer- and isomer-tilde-equivalence classes. Within each isomer- and isomer-tilde-equivalence class, color-parity seems to control why the minus-one third gadget appears more frequently than the plus one-third gadget: both appear equally between adinkras of different color-parity, but only the minus one-third gadget appears between adinkras of the same color-parity. The gadget value of one is the least frequent, as it can only occur between adinkras of the same $\brtV$-equivalence class.

Another important result of this paper is Equation~(\ref{e:36864}), which compactly encodes \emph{all 36,864 four-color, four-boson, and four-fermion valise adinkras}. Furthermore, the utility of the gadget in distinguishing multiplets related by dimensional reduction was demonstrated. Specifically, the twisted chiral multiplet and the vector multiplet were used as an example. Dimensional reductions of actions were discussed in terms of K\"ahler-like potentials; technology we plan to use to advance SUSY holography. Future work will focus on utilizing equivalence classes to develop holographic techniques that we plan to extend to adinkras with more supercharges such as 4D, $\mathcal{N}=4$ super Yang--Mills theory and 10D and 11D supergravity.

The discussion under (\ref{Gdgt2}) is a demonstration of the presence of SUSY holography once more. This provides a beautiful physics reason for why the gadget value of +1/3 is so distinctive from the other gadget values of $-$1/3, 0, 1. The latter occur in the one-dimensional shadows of supermultiplets all the way up to 4D, $\cal N$ = 1 theories, but the former only occurs in the one-dimensional shadows of supermultiplets up to 2D, $\cal N$ =(2,0). The ability of adinkras and their holoraumy to keep track of this level of subtlety should be convincing to any skeptic who doubts the efficacy of this approach to the study of supersymmetric representation theory.

\vspace{.05in}
 \begin{center}
 \parbox{5in}{{``Equality is not in regarding different things similarly, equality is in regarding different things differently.''}---Tom Robbins}
 \end{center} 

\section*{Acknowledgments}
$~~~~$ 
This work was partially supported by the National Science Foundation Grant PHY-1315155. This
research was also supported in part by the University of Maryland Center for String \& Particle Theory (CSPT). K.S. would also like to thank Pepperdine University for hospitality and travel funds and Northwest Missouri State University for computing equipment and travel funds that supported this work. Additional acknowledgment is given by L.\ Kang to the CSPT, for his participation in the 2017 SSTPRS (Student Summer Theoretical Physics Research Session) program.

\appendix
\section{\texorpdfstring{$\bap$}{a} and \texorpdfstring{$\bbt$}{b} Matrices}
The $\bap$ and $\bbt$ matrices used in this paper are:
\label{a:ab}
\begin{align*}
	\bap^1 =& \left(
\begin{array}{cccc}
 0 & 0 & 0 & -i \\
 0 & 0 & -i & 0 \\
 0 & i & 0 & 0 \\
 i & 0 & 0 & 0 \\
\end{array}
\right),~~~
	\bap^2 = \left(
\begin{array}{cccc}
 0 & -i & 0 & 0 \\
 i & 0 & 0 & 0 \\
 0 & 0 & 0 & -i \\
 0 & 0 & i & 0 \\
\end{array}
\right) ,~~~
	\bap^3 = \left(
\begin{array}{cccc}
 0 & 0 & -i & 0 \\
 0 & 0 & 0 & i \\
 i & 0 & 0 & 0 \\
 0 & -i & 0 & 0 \\
\end{array}
\right) 
\end{align*}

\begin{align*}
	\bbt^1 = & \left(
\begin{array}{cccc}
 0 & 0 & 0 & -i \\
 0 & 0 & i & 0 \\
 0 & -i & 0 & 0 \\
 i & 0 & 0 & 0 \\
\end{array}
\right),~~~
	\bbt^2 = \left(
\begin{array}{cccc}
 0 & 0 & -i & 0 \\
 0 & 0 & 0 & -i \\
 i & 0 & 0 & 0 \\
 0 & i & 0 & 0 \\
\end{array}
\right),~~~
	\bbt^3 = \left(
\begin{array}{cccc}
 0 & -i & 0 & 0 \\
 i & 0 & 0 & 0 \\
 0 & 0 & 0 & i \\
 0 & 0 & -i & 0 \\
\end{array}
\right).
\end{align*}
 terms of tensor products of Pauli spin matrices $\sigma^i$ and the $2\times2$ identity matrix $I_2$, this can be written as:
\begin{align*}
	\bap^1 =& \sigma^2 \otimes \sigma^1,~~~\bap^2 = I_2 \otimes \sigma^2,~~~\bap^3 = \sigma^2 \otimes \sigma^3 \\
	\bbt^1 =& \sigma^1 \otimes \sigma^2,~~~\bbt^2 = \sigma^2 \otimes I_2,~~~\bbt^3 = \sigma^3 \otimes \sigma^2 .
\end{align*}
These matrices form two mutually-commuting su(2) algebras:
\begin{align*}
	[ \bap^{\hat{a}}, \bap^{\hat{b}} ] = & 2 i \epsilon^{\hat{a}\hat{b}\hat{c}} \bap^{\hat{c}},~~~[ \bbt^{\hat{a}}, \bbt^{\hat{b}} ] = 2 i \epsilon^{\hat{a}\hat{b}\hat{c}} \bbt^{\hat{c}},~~~[ \bap^{\hat{a}}, \bbt^{\hat{b}} ] = 0.
\end{align*}
Owing to the algebra above, the $\bap$ and $\bbt$ matrices satisfy the trace orthogonality relationship: \begin{equation}
\begin{aligned}\label{e:abtraces}
	Tr( \bap^{\hat{a}} \bbt^{\hat{b}} ) =& 0 \cr
  Tr(\bap^{\hat{a}} \bap^{\hat{b}}) = & Tr( \bbt^{\hat{a}} \bbt^{\hat{b}}) = 4\delta^{\hat{a}\hat{b}}~~~.
\end{aligned}\end{equation}

\newpage
\section{Explicit Matrix Forms for Flips and Flops}\label{a:FlipFlop}
Since flip matrices with Boolean codes greater than seven are simply the negative of a corresponding matrix with codes less than seven, it is only necessary to list Boolean matrices less than eight:
\begin{align}
	() =& \left(
\begin{array}{cccc}
 1 & 0 & 0 & 0 \\
 0 & 1 & 0 & 0 \\
 0 & 0 & 1 & 0 \\
 0 & 0 & 0 & 1 \\
\end{array}
\right) ,~~~
		(\overline{12}) = \left(
\begin{array}{cccc}
 -1 & 0 & 0 & 0 \\
 0 & -1 & 0 & 0 \\
 0 & 0 & 1 & 0 \\
 0 & 0 & 0 & 1 \\
\end{array}
\right),~~~
\end{align}
\begin{align}
		(\overline{13}) = \left(
\begin{array}{cccc}
 -1 & 0 & 0 & 0 \\
 0 & 1 & 0 & 0 \\
 0 & 0 & -1 & 0 \\
 0 & 0 & 0 & 1 \\
\end{array}
\right) ,~~~
		(\overline{23}) =& \left(
\begin{array}{cccc}
 1 & 0 & 0 & 0 \\
 0 & -1 & 0 & 0 \\
 0 & 0 & -1 & 0 \\
 0 & 0 & 0 & 1 \\
\end{array}
\right),
\end{align}
\begin{align}
		(\overline{1}) = \left(
\begin{array}{cccc}
 -1 & 0 & 0 & 0 \\
 0 & 1 & 0 & 0 \\
 0 & 0 & 1 & 0 \\
 0 & 0 & 0 & 1 \\
\end{array}
\right),~~~
		(\overline{2}) =& \left(
\begin{array}{cccc}
 1 & 0 & 0 & 0 \\
 0 & -1 & 0 & 0 \\
 0 & 0 & 1 & 0 \\
 0 & 0 & 0 & 1 \\
\end{array}
\right),
\end{align}
\begin{align}
(\overline{3}) =& \left(
\begin{array}{cccc}
 1 & 0 & 0 & 0 \\
 0 & 1 & 0 & 0 \\
 0 & 0 & -1 & 0 \\
 0 & 0 & 0 & 1 \\
\end{array}
\right),~~~
	(\overline{123}) = \left(
\begin{array}{cccc}
 -1 & 0 & 0 & 0 \\
 0 & -1 & 0 & 0 \\
 0 & 0 & -1 & 0 \\
 0 & 0 & 0 & 1 \\
\end{array}
\right) 
\end{align}

Flop matrices with their corresponding $S_4$ code (matrix row and column indices suppressed) are given by:\begin{equation}
\begin{aligned}
	() =&\left(
\begin{array}{cccc}
 1 & 0 & 0 & 0 \\
 0 & 1 & 0 & 0 \\
 0 & 0 & 1 & 0 \\
 0 & 0 & 0 & 1 \\
\end{array}
\right),~~~
	(12)(34) = \left(
\begin{array}{cccc}
 0 & 1 & 0 & 0 \\
 1 & 0 & 0 & 0 \\
 0 & 0 & 0 & 1 \\
 0 & 0 & 1 & 0 \\
\end{array}
\right) ,~~~\cr
	(13)(24) =& \left(
\begin{array}{cccc}
 0 & 0 & 1 & 0 \\
 0 & 0 & 0 & 1 \\
 1 & 0 & 0 & 0 \\
 0 & 1 & 0 & 0 \\
\end{array}
\right),~~~
	(14)(23) = \left(
\begin{array}{cccc}
 0 & 0 & 0 & 1 \\
 0 & 0 & 1 & 0 \\
 0 & 1 & 0 & 0 \\
 1 & 0 & 0 & 0 \\
\end{array}
\right),~~~
\end{aligned}\end{equation}\begin{equation}
\begin{aligned}
	(12) =&\left(
\begin{array}{cccc}
 0 & 1 & 0 & 0 \\
 1 & 0 & 0 & 0 \\
 0 & 0 & 1 & 0 \\
 0 & 0 & 0 & 1 \\
\end{array}
\right),~~~
	(34) = \left(
\begin{array}{cccc}
 1 & 0 & 0 & 0 \\
 0 & 1 & 0 & 0 \\
 0 & 0 & 0 & 1 \\
 0 & 0 & 1 & 0 \\
\end{array}
\right),~~~\cr
	(1324) =& \left(
\begin{array}{cccc}
 0 & 0 & 0 & 1 \\
 0 & 0 & 1 & 0 \\
 1 & 0 & 0 & 0 \\
 0 & 1 & 0 & 0 \\
\end{array}
\right),~~~
	(1423) = \left(
\begin{array}{cccc}
 0 & 0 & 1 & 0 \\
 0 & 0 & 0 & 1 \\
 0 & 1 & 0 & 0 \\
 1 & 0 & 0 & 0 \\
\end{array}
\right),~~~
\end{aligned}\end{equation}\begin{equation}
\begin{aligned}
	(13) =&\left(
\begin{array}{cccc}
 0 & 0 & 1 & 0 \\
 0 & 1 & 0 & 0 \\
 1 & 0 & 0 & 0 \\
 0 & 0 & 0 & 1 \\
\end{array}
\right),~~~
	(1234) = \left(
\begin{array}{cccc}
 0 & 0 & 0 & 1 \\
 1 & 0 & 0 & 0 \\
 0 & 1 & 0 & 0 \\
 0 & 0 & 1 & 0 \\
\end{array}
\right),~~~\cr
	(24) =& \left(
\begin{array}{cccc}
 1 & 0 & 0 & 0 \\
 0 & 0 & 0 & 1 \\
 0 & 0 & 1 & 0 \\
 0 & 1 & 0 & 0 \\
\end{array}
\right),~~~
	(1432) = \left(
\begin{array}{cccc}
 0 & 1 & 0 & 0 \\
 0 & 0 & 1 & 0 \\
 0 & 0 & 0 & 1 \\
 1 & 0 & 0 & 0 \\
\end{array}
\right),~~~
\end{aligned}\end{equation}\begin{equation}
\begin{aligned}
	(23) =&\left(
\begin{array}{cccc}
 1 & 0 & 0 & 0 \\
 0 & 0 & 1 & 0 \\
 0 & 1 & 0 & 0 \\
 0 & 0 & 0 & 1 \\
\end{array}
\right),~~~
	(1342) = \left(
\begin{array}{cccc}
 0 & 1 & 0 & 0 \\
 0 & 0 & 0 & 1 \\
 1 & 0 & 0 & 0 \\
 0 & 0 & 1 & 0 \\
\end{array}
\right),~~~\cr
	(1243) =& \left(
\begin{array}{cccc}
 0 & 0 & 1 & 0 \\
 1 & 0 & 0 & 0 \\
 0 & 0 & 0 & 1 \\
 0 & 1 & 0 & 0 \\
\end{array}
\right),~~~
	(14) = \left(
\begin{array}{cccc}
 0 & 0 & 0 & 1 \\
 0 & 1 & 0 & 0 \\
 0 & 0 & 1 & 0 \\
 1 & 0 & 0 & 0 \\
\end{array}
\right),~~~
\end{aligned}\end{equation}\begin{equation}
\begin{aligned}
	(123) =&\left(
\begin{array}{cccc}
 0 & 0 & 1 & 0 \\
 1 & 0 & 0 & 0 \\
 0 & 1 & 0 & 0 \\
 0 & 0 & 0 & 1 \\
\end{array}
\right),~~~
	(134) = \left(
\begin{array}{cccc}
 0 & 0 & 0 & 1 \\
 0 & 1 & 0 & 0 \\
 1 & 0 & 0 & 0 \\
 0 & 0 & 1 & 0 \\
\end{array}
\right),~~~\cr
	(243) =& \left(
\begin{array}{cccc}
 1 & 0 & 0 & 0 \\
 0 & 0 & 1 & 0 \\
 0 & 0 & 0 & 1 \\
 0 & 1 & 0 & 0 \\
\end{array}
\right),~~~
	(142) = \left(
\begin{array}{cccc}
 0 & 1 & 0 & 0 \\
 0 & 0 & 0 & 1 \\
 0 & 0 & 1 & 0 \\
 1 & 0 & 0 & 0 \\
\end{array}
\right),~~~
\end{aligned}\end{equation}\begin{equation}
\begin{aligned}
	(132) =&\left(
\begin{array}{cccc}
 0 & 1 & 0 & 0 \\
 0 & 0 & 1 & 0 \\
 1 & 0 & 0 & 0 \\
 0 & 0 & 0 & 1 \\
\end{array}
\right),~~~
	(234) = \left(
\begin{array}{cccc}
 1 & 0 & 0 & 0 \\
 0 & 0 & 0 & 1 \\
 0 & 1 & 0 & 0 \\
 0 & 0 & 1 & 0 \\
\end{array}
\right),~~~\cr
	(124) =& \left(
\begin{array}{cccc}
 0 & 0 & 0 & 1 \\
 1 & 0 & 0 & 0 \\
 0 & 0 & 1 & 0 \\
 0 & 1 & 0 & 0 \\
\end{array}
\right),~~~
	(143) = \left(
\begin{array}{cccc}
 0 & 0 & 1 & 0 \\
 0 & 1 & 0 & 0 \\
 0 & 0 & 0 & 1 \\
 1 & 0 & 0 & 0 \\
\end{array}
\right).
\end{aligned}\end{equation}

\section{Fermionic Holoraumy and 2D, \texorpdfstring{$\cal N$}{N} = (4,0) Minimal Scalar Valise Supermultiplets}\label{a:2DHoloraumy} 

In the text of this work, we calculated the holoraumy associated with the
2D, $\cal N$ = (4,0) supermultiplets by going to a real basis. Of course, this can
also be done in the initial su(2) basis used to describe the SM-I, $\dots$, SM-IV 
representations. In other words, it is necessary to calculate the effects of the 
three operators:
\be \eqalign{ {~~~}
{\cal O}{}_1~\equiv~\left[ \, D_{+ i}~,~D_{+ j}  \, \right],~~
{\cal O}{}_2~\equiv~\left[ \, D_{+ i}~,~{\Bar D}_+ {}^j  \, \right],~~
{\cal O}{}_3~\equiv~\left[ \, {\Bar D}_+ {}^i~,~ {\Bar D}_+ {}^j \, \right] 
,~
} \label{H1} \ee
on all of the fermions in each supermultiplet. Upon completion of these calculations,
we find the results shown in the equations of (\ref{H2})--(\ref{H5}).
\vskip0.1in
\noindent
$ {\rm {2D}},~{\cal N}~$=$~$(4,0)$~{\rm SM-I}~{\rm {Supermultiplet }}~{\rm {Holoraumy}} $ 
\be \eqalign{ {~~~~}
\left[ \, D_{+ i},~D_{+ j}  \, \right] \psi{}^{- k}~&=~0,~~\cr
\left[ \, D_{+ i},~{\Bar D}_+ {}^j  \, \right] \psi{}^{- k}~&=~
 -\, i \,
      ({\vec \s}){}_i{}^j \, {\bm \cdot} \, ({\vec \s}){}_l{}^k \, \pa_{\pp} \psi^- {}^{l} 
,~~\cr
\left[ \, {\Bar D}_+ {}^i ,~ {\Bar D}_+ {}^j \, \right] 
 \psi{}^{- k}~&=~0,~
} \label{H2} \ee
\vskip0.1in
\noindent
$ {\rm {2D}},~{\cal N}~$=$~$(4,0)$~{\rm SM-II}~{\rm {Supermultiplet }}~{\rm {Holoraumy}} $ 
\be \eqalign{ {~~~~}
\left[ \, D_{+ i},~D_{+ j}  \, \right] \lambda{}^{-}{}_{ k}~&=~0,~~\cr
\left[ \, D_{+ i},~{\Bar D}_+ {}^j  \, \right]  \lambda{}^{-}{}_{ k}~&=~
 i \, \d_i {}^j \, \pa_{\pp} \lambda^- {}_k
,~~\cr
\left[ \, {\Bar D}_+ {}^i,~ {\Bar D}_+ {}^j \, \right] \lambda{}^{-}{}_{ k}
~&=~-\, i \, 2 \, C{}^{i j} \, C{}_{ k l} \, \pa_{\pp} {\Bar \lambda}^- {}^l 
,~
} \label{H3} \ee
 \vskip0.1in
\noindent
$ {\rm {2D}},~{\cal N}~$=$~$(4,0)$~{\rm SM-III}~{\rm {Supermultiplet }}~{\rm {Holoraumy}} $ 
\be \eqalign{ {~~~~}
\left[ \, D_{+ i},~D_{+ j}  \, \right] \pi{}^{-}~&=~0~~,~~\cr
\left[ \, D_{+ i},~{\Bar D}_+ {}^j  \, \right] \pi{}^{-}~&=~i \, \d_i {}^j \pa_{\pp} \pi{}^{-} 
,~~\cr
\left[ \, {\Bar D}_+ {}^i ,~ {\Bar D}_+ {}^j \, \right] 
 \pi{}^{- }~&=~- \, i \, 2 \, C^{i j} \, \pa_{\pp} \rho{}^-
,~~\cr
\left[ \, D_{+ i},~D_{+ j}  \, \right] \rho{}^{-}~&=~i 2 \, C_{ij} \, , \pa_{\pp} \pi^-,~~\cr
\left[ \, D_{+ i},~{\Bar D}_+ {}^j  \, \right] \rho{}^{-}~&=~-i \, \d_i {}^j \, \pa_{\pp} \rho{}^{-} 
,~~\cr
\left[ \, {\Bar D}_+ {}^i ,~ {\Bar D}_+ {}^j \, \right] 
 \rho{}^{- }~&=~0,~~
} \label{H4} \ee
 \vskip0.1in
\noindent
$ {\rm {2D}},~{\cal N}~$=$~$(4,0)$~{\rm SM-IV}~{\rm {Supermultiplet }}~{\rm {Holoraumy}} $ 
\be \eqalign{ {~~~~}
\left[ \, D_{+ i},~D_{+ j}  \, \right] \psi{}^{- }~&=~0,~~\cr
\left[ \, D_{+ i},~{\Bar D}_+ {}^j  \, \right] \psi{}^{- }~&=~2 \, (\pa_{\pp } \, 
\psi^- {}_i {}^{j}),~~\cr
\left[ \, {\Bar D}_+ {}^i,~ {\Bar D}_+ {}^j \, \right] 
 \psi{}^{- }~&=~0,~\cr
\left[ \, D_{+ i},~D_{+ j}  \, \right] \psi{}^{-}{}_{k}{}^{l}~&=~0,~~\cr
\left[ \, D_{+ i},~{\Bar D}_+ {}^j  \, \right] \psi{}^{-}{}_{k}{}^{l}~&=~
 i \, \fracm 12 (\vec \s){}_i{}^j \cdot \left[ \, (\vec \s){}_k{}^p ( \pa_{\pp } \psi^- {}_p {}^{l} )
\,-\, ( \pa_{\pp } \psi^- {}_k {}^{p} ) (\vec \s){}_p{}^l \, 
 \,+\, i\, (\vec \s){}_k{}^l \, \pa_{\pp } \psi^- \right],~~\cr
\left[ \, {\Bar D}_+ {}^i,~ {\Bar D}_+ {}^j \, \right] 
 \psi{}^{-}{}_{k}{}^{l}~&=~0,~
} \label{H5} \ee

\bibliographystyle{utphys}
\bibliography{Bibliography}  
       
\end{document}